\pdfoutput=1
\documentclass[11pt,twoside,a4paper,cmspaper,final,collab]{cms-tdr}
\def\svnVersion{dc09198}\def\svnDate{2024/08/06}\def\cmsCernNoTag{CERN-EP-2024-008}\def\cmsCernDate{\today}\def\cmsMessage{Published in Physical Review D as \href{http://dx.doi.org/10.1103/PhysRevD.110.032007}{\doi{10.1103/PhysRevD.110.032007}.}}
\begin{document}\cmsNoteHeader{EXO-21-008}

\newlength\cmsTabSkip\setlength{\cmsTabSkip}{1ex}

\DeclareRobustCommand{\PS}{{\HepParticle{S}{}{}}\xspace}
\DeclareRobustCommand{\Pomega}{{\HepSusyParticle{\upomega}{}{}}\xspace}
\DeclareRobustCommand{\PPsi}{{\HepParticle{\Uppsi}{}{}}\xspace}
\DeclareRobustCommand{\Peta}{{\HepSusyParticle{\PGh}{}{}}\xspace}

\newcommand{\ddbar}{\ensuremath{\PQd\PAQd}\xspace}
\newcommand{\PKpKm}{\ensuremath{\PKp\PKm}\xspace}
\newcommand{\PKzKz}{\ensuremath{\PKz\PKz}\xspace}

\newcommand{\htoss}{\ensuremath{\PH\to\PS\PS}\xspace}
\newcommand{\brhtoss}{\ensuremath{\mathcal{B}(\htoss)}\xspace}

\newcommand{\htopsipsi}{\ensuremath{\PH\to\PPsi\PPsi}\xspace}
\newcommand{\brhtopsipsi}{\ensuremath{\mathcal{B}(\htopsipsi)}\xspace}
\newcommand{\nhits}{\ensuremath{N_\text{hits}}\xspace}
\newcommand{\nclusters}{\ensuremath{N_\text{clusters}}\xspace}
\newcommand{\deltaphi}{\ensuremath{\Delta\phi{\mathrm{(}\ptvecmiss\mathrm{, cluster)}}}\xspace}
\ifthenelse{\boolean{cms@external}}{\providecommand{\cmsLeft}{upper\xspace}}{\providecommand{\cmsLeft}{left\xspace}}
\ifthenelse{\boolean{cms@external}}{\providecommand{\cmsRight}{lower\xspace}}{\providecommand{\cmsRight}{right\xspace}}
\providecommand{\cmsTable}[1]{\resizebox{\columnwidth}{!}{#1}}

\cmsNoteHeader{EXO-21-008}

\title{Search for long-lived particles decaying in the CMS muon detectors in proton-proton collisions at \texorpdfstring{$\sqrt{s}=13\TeV$}{sqrt(s)=13 TeV}}

\date{\today}
\abstract{
  A search for long-lived particles (LLPs) decaying in the CMS muon detectors is presented.
  A data sample of proton-proton collisions at $\sqrt{s}=13\TeV$ corresponding to an integrated luminosity of 138\fbinv, recorded at the LHC in 2016--2018, is used.
  The decays of LLPs are reconstructed as high multiplicity clusters of hits in the muon detectors.
  In the context of twin Higgs models, the search is sensitive to LLP masses from 0.4 to 55\GeV and a broad range of LLP decay modes, including decays to hadrons, \PGt leptons, electrons, or photons.
  No excess of events above the standard model background is observed.
  The most stringent limits to date from LHC data are set on the branching fraction of the Higgs boson decay to a pair of LLPs with masses below 10\GeV.
  This search also provides the best limits for various intervals of LLP proper decay length and mass.
  Finally, this search sets the first limits at the LHC on a dark quantum chromodynamic sector whose particles couple to the Higgs boson through gluon, Higgs boson, photon, vector, and dark-photon portals, and is sensitive to branching fractions of the Higgs boson to dark quarks as low as $2 \times 10^{-3}$.
}

\hypersetup{
  pdfauthor={CMS Collaboration},
  pdftitle={Search for long-lived particles decaying in the CMS muon detectors in proton-proton collisions at sqrt(s)=13 TeV},
  pdfsubject={CMS},
  pdfkeywords={CMS, long-lived particles, muon detector}
}

\maketitle

\section{Introduction}
\label{sec:intro}
Many extensions of the standard model (SM) predict the existence of neutral, weakly coupled particles that have long proper lifetimes.
These long-lived particles (LLPs) naturally arise in a broad range of models beyond the SM including supersymmetry (SUSY)~\cite{Giudice:2004tc,Hewett:2004nw,ArkaniHamed:2004yi,Gambino:2005eh,Arvanitaki:2012ps,ArkaniHamed:2012gw,Fayet:1974pd,Farrar:1978xj,Weinberg:1981wj,Barbier:2004ez,GIUDICE1999419,Meade:2008wd,Buican:2008ws,Fan:2011yu,Fan:2012jf}, hidden valley scenarios~\cite{Strassler:2006im,Strassler:2006ri,Han:2007ae}, inelastic dark matter~\cite{PhysRevD.64.043502}, and twin Higgs models~\cite{Chacko:2005pe,Curtin:2015fna,Cheng:2015buv}.

In this paper, we describe a search at the CERN LHC that uses the CMS muon detectors as a sampling calorimeter to identify particle showers produced by LLP decays.
This analysis extends and improves the techniques deployed in a previous CMS publication~\cite{CMS:2021juv}.
The search is based on proton-proton collision data collected at a center-of-mass energy of 13\TeV during 2016--2018 at the LHC, corresponding to an integrated luminosity of 138\fbinv.
The CMS muon detectors are composed of gaseous detector chambers interleaved with steel layers of the magnet flux-return yoke.
Decays of LLPs in the muon detectors induce hadronic and electromagnetic showers, giving rise to a large multiplicity of hits in a localized detector region, referred to as a muon detector shower (MDS) object.
The hadron calorimeter (HCAL), solenoid magnet, and steel flux-return yoke together provide 12--27 nuclear interaction lengths of shielding~\cite{Layter:343814, CMS:2008xjf}, which, together with explicit vetoes on the inner detector activity due to prompt SM particles, strongly suppress particle showers from jets that are not fully contained within the calorimeters' volume (punch-through).
An LLP produced with a large Lorentz boost and decaying after it has traversed the calorimeter systems may produce large missing transverse momentum because its momentum is not properly measured or associated with a reconstructed particle.
Therefore, the analyzed data are required to have a magnitude of the missing transverse momentum vector above 200\GeV.

This search is sensitive to the production of single or multiple LLPs decaying to final states including hadrons, tau leptons, electrons, or photons.
The LLPs decaying to muons very rarely produce a particle shower, as muons lose energy mostly through ionization and will generally not be detected by this search.
While this search is sensitive to many models predicting LLPs, we interpret the results in two separate benchmark scenarios.
The first is a simplified model motivated by the twin Higgs scenario~\cite{Chacko:2005pe,Curtin:2015fna,Cheng:2015buv} where the SM Higgs boson (\PH) decays to a pair of neutral long-lived scalars (\PS), each of which decays in turn to a pair of fermions or a pair of photons.
We search for long-lived scalars with masses between 0.4 and 55\GeV in a wide range of decay modes, including decays resulting primarily in hadronic showers ($\bbbar$, $\ddbar$, $\PKpKm$, $\PKzKz$, and $\Pgpp\Pgpm$), decays resulting primarily in electromagnetic showers ($\Pgpz\Pgpz$, $\Pgg\Pgg$, and \EE), and decays to $\PGtp\PGtm$, which may result in hadronic or electromagnetic showers.
The most stringent previous constraints for mean proper decay lengths $c\tau < 0.3\unit{m}$ are based on a search for displaced jets in the CMS tracker~\cite{Sirunyan:2020cao}.
For $c\tau > 0.3\unit{m}$, a search for displaced vertices in the ATLAS muon spectrometer~\cite{Aaboud:2018aqj, ATLAS:2022gbw} and the CMS search using the endcap muon detectors~\cite{CMS:2021juv} set the most stringent previous limits.

\begin{figure}[h!]
  \centering
  \includegraphics[width=0.6\columnwidth]{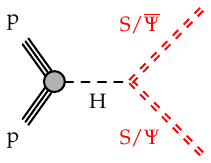}
  \caption{
    Diagram representing the twin Higgs and dark-shower models.
    The SM Higgs boson (\PH) decays to a pair of neutral long-lived scalars (\PS) in the twin Higgs model or to a pair of dark-sector quarks (\PPsi) in the dark shower model.
  }
  \label{fig:feynman_diagram}
\end{figure}

We also interpret the search results in terms of a set of ``dark shower'' models with perturbative parton showers~\cite{Knapen:2021eip}.
We consider production of an SM Higgs boson that decays to a pair of dark-sector quarks, each of which hadronizes into a dark shower consisting of short- and long-lived dark-sector mesons (scalar or vector) that eventually decay back to SM particles through portals that couple the dark sector to the SM.
Depending on the symmetries and decay portal, the proper lifetime of the dark mesons and the final-state SM particles can vary, resulting in a wide range of dark-shower signatures.
We interpret the search results in a framework for long-lived states with masses between 2 and 20\GeV and five different decay portals~\cite{Knapen:2021eip}, namely the gluon portal producing hadron-rich showers, the photon portal with photon showers, the vector portal with semi-visible jets including a mixture of invisible and visible dark mesons, the Higgs boson portal with heavy-flavor-rich showers, and the dark-photon portal with lepton-rich showers.
This is the first search at the LHC with an interpretation in this framework.
A diagram representing both benchmark models is shown in Fig.~\ref{fig:feynman_diagram}.

There are two key advantages of the LLP search strategy presented in this paper over searches that employ displaced vertices.
\begin{enumerate}
  \item[(i)] The absorbers in front of the muon detectors act as shielding material to maintain a sufficiently low level of background for the detection of a single LLP decay.
    This level of background rejection could only be achieved in current hadronically decaying displaced-vertex searches by requiring the detection of two LLP decays.
  \item[(ii)] The MDS signature is sensitive to the LLP energy only and insensitive to its mass, rendering this search equally sensitive to all LLP masses considered, as shown in Section~\ref{sec:results}.
    In contrast, the vertex reconstruction efficiency in a displaced-vertex search tends to decrease with the LLP mass because of the increasingly smaller opening angles.
\end{enumerate}
Because of these advantages, the signal acceptance and sensitivity are improved relative to those of the analyses leading to the previous best results~\cite{Sirunyan:2020cao, Aaboud:2018aqj, ATLAS:2022gbw} for all LLP masses and proper lifetimes.

This paper is organized as follows.
We briefly describe the CMS detector in Section~\ref{sec:detector}.
Section~\ref{sec:samples} provides a summary of the simulated samples used in the analysis.
The reconstruction of the final state objects is discussed in Sections~\ref{sec:object} and~\ref{sec:mds}, respectively.
The event selection is described in Section~\ref{sec:eventsel}.
The background estimation methods are detailed in Section~\ref{sec:background}.
The signal modeling and systematic uncertainties are discussed in Sections~\ref{sec:systematics}.
We report and interpret the results in Section~\ref{sec:results}.
Finally, a summary is given in Section~\ref{sec:summary}.
The tabulated results are provided in the HEPData record for this analysis~\cite{hepdata}.

\section{The CMS detector}
\label{sec:detector}

The central feature of the CMS detector is a superconducting solenoid of 6\unit{m} internal diameter, providing a magnetic field of 3.8\unit{T}.
Within the solenoid volume are a silicon pixel and strip tracker, a lead tungstate crystal electromagnetic calorimeter (ECAL), and a brass and scintillator HCAL, each composed of a barrel and two endcap sections.
Forward calorimeters extend the pseudorapidity ($\eta$) coverage provided by the barrel and endcap detectors.
Muons are identified in gas-ionization detectors embedded in the steel flux-return yoke outside the solenoid using three technologies: drift tubes (DTs) in the barrel, cathode strip chambers (CSCs) in the endcaps, and resistive-plate chambers (RPCs) in the barrel and endcaps.
A more detailed description of the CMS detector, together with a definition of the coordinate system used and the relevant kinematic variables, can be found in Ref.~\cite{CMS:2008xjf}.

The DT and CSC detectors covering the barrel and endcaps, respectively, play critical roles in the search described in this paper.
The barrel DT detectors cover a region of $\abs{\eta} < 1.2$ and are organized in four concentric cylindrical layers (``stations'') around the beamline and five wheels along the beamline axis ($z$), as illustrated in Fig.~\ref{fig:cluster_eff_2d}.
The four DT stations labeled MB1 to MB4 are located approximately 4, 5, 6, and 7\unit{m} away from the interaction point radially ($r$) and interleaved with the layers of the steel flux-return yoke.
The DT chambers in a particular station and wheel are arranged into rings, where the chambers are split into 12 $\phi$-sectors, each subtending $30^\circ$ in $\phi$.
Each DT station contains 8--12 layers of DT cells.
As charged particles traverse the DT stations, they ionize the gas and produce charges that drift to the anode wire at the center of each of the DT cells.
A signal pulse measured at the anode wire is recorded as a hit.
The first three stations, each containing twelve layers of DT cells, are each arranged in three groups of four staggered layers called ``superlayers'' (SLs).
The innermost and outermost SLs measure the hit coordinate in the $r$--$\phi$ plane (where $\phi$ is the azimuthal
angle in radians), and the central SL measures the position in the $z$ direction, along the beamline.
The fourth station only contains two SLs measuring the hit position in the $r$--$\phi$ plane.
Individual hits have a resolution of about 600\mum, which can be improved to about 260\mum by combining information from hits in the same SL~\cite{Chatrchyan:1223944}.

The CSC detectors cover a region of $0.9 < \abs{\eta} < 2.4$ and comprise four stations in each endcap.
The four CSC stations labeled ME1 to ME4 are located approximately 7.0, 8.0, 9.5, and 10.5\unit{m} away from the interaction point along the beamline axis on both ends of the detector, and are interleaved between steel absorbers.
In the $r$ direction, each station is composed of two or three rings, labeled as ME1/n-ME4/n, where integer n increases with the radial distance from the beam line.
Each chamber is composed of six thin layers containing cathode strips along the radial direction and anode wires perpendicular to the strips.
Charged particles traversing the chambers ionize the gas.
The resulting electrons are accelerated towards the anode wires producing an avalanche, while the positive ions travel to the opposite end and induce signals in the cathode strips.
By combining the information from signals on the anode wires and the cathode strips of each layer, the space and time coordinates of each hit can be determined with a resolution of 400--500\mum and 5\unit{ns}~\cite{Sirunyan:2018fpa,Chatrchyan:2009af}.

Each RPC consists of two parallel high-resistivity plastic plates enclosing a volume of gas that operates in avalanche mode in order to provide a fast response.
Signals collected from the readout strips measure the position and time of a muon hit.
The RPCs' excellent time resolution (1\unit{ns}) enables pairing the muon hits to the LHC bunch crossing without ambiguities.
Six layers of RPCs are located in the barrel and three in the endcaps, covering a region of $\abs{\eta}<1.6$.

Events of interest are selected using a two-tiered trigger system.
The first level, composed of custom hardware processors, uses information from the calorimeters and muon detectors to select events at a rate of around 100\unit{kHz} within a fixed latency of about 4\mus~\cite{CMS:2020cmk}.
The second level, known as the high-level trigger, consists of a farm of processors running a version of the full event reconstruction software optimized for fast processing and reduces the event rate to around 1\unit{kHz} before data storage~\cite{CMS:2016ngn}.

\section{Simulated samples}
\label{sec:samples}

The simulated $\htoss$ signal samples are generated at next-to-leading order (NLO) with the \POWHEG2.0~\cite{Nason:2004rx,Frixione:2007vw,Alioli:2010xd,Re:2010bp} generator for the five main Higgs production processes: gluon fusion, vector boson fusion, associated production with a \PZ or \PW boson, and associated production with a pair of top quarks.
The Higgs boson mass is set to 125\GeV, while the \PS mass ($m_\PS$) is set to 0.4, 1.0, 1.5, 3.0, 7.0, 15.0, 40.0, or 55.0\GeV.
The proper decay length is set to various values ranging between 10\unit{mm} and 100\unit{m}, resulting in detectable decays in the muon detectors at decay lengths of 4 to 11\unit{m} in the lab frame.
We consider decays to \bbbar, \ddbar, \PKpKm, \PKzKz, and $\Pgpp\Pgpm$, hereafter referred to as the fully hadronic decay modes, decays to $\Pgpz\Pgpz$, $\Pgg\Pgg$, and \EE, referred to as the fully electromagnetic decay modes, and decays to $\PGtp\PGtm$.
Di-muon decays are not considered, since muons typically do not create high hit-multiplicity showers in the muon detectors.
These specific decay modes are selected because they represent the dominant decay modes for Higgs-like scalar particles with mass in various ranges~\cite{Bezrukov:2013fca,Winkler:2018qyg}.
For mass below 0.2\GeV, the \EE and $\Pgg\Pgg$ decay modes are dominant; between 0.3 and 1\GeV, the $\Pgpp\Pgpm$ and $\Pgpz\Pgpz$ decay modes are dominant; between 1 and 2\GeV, the \PKpKm and \PKzKz decay modes are dominant; and above 2\GeV the $\Pgg\Pgg$ and $\Pq\Pq$ decay modes are dominant.
The analysis has some sensitivity below 0.4\GeV owing to the calorimetric nature of the signature.
However, because of the limitations in the signal simulation, the analysis sensitivity below 0.4\GeV is not quantified.

The dark shower signal models are generated similarly through Higgs boson production at NLO with \POWHEG2.0 and include only the dominant gluon fusion production mode.
The Higgs boson mass is set to 125\GeV.
The Higgs boson decay and the phenomenology of the dark showers are generated following the tools and theory assumptions presented in Ref.~\cite{Knapen:2021eip}, using the \textsc{pythia}8~\cite{Sjostrand:2014zea} hidden-valley module~\cite{Carloni:2011kk,Carloni:2010tw}.
In the generation, the dark sector is reduced to a single dark quark (\PPsi), vector meson (\Pomega), and scalar meson (\Peta), and there are three dark quantum chromodynamic (QCD) colors.

As mentioned in Section~\ref{sec:intro}, we generate signal samples for five different decay portals.
In the vector portal, \Pomega is long-lived and couples to SM particles, while \Peta is invisible.
For the dark-photon portal, \Peta decays into a pair of long-lived dark photons with masses equal to $0.4 m_{\Peta}$, which then each decay into SM particles.
For all other portals, \Peta is long-lived and couples to SM particles, while \Pomega is invisible.
The LLP mass is varied between 2 and 20\GeV.
The minimum LLP mass for each portal is motivated by the minimal ultraviolet completion and fine-tuning considerations discussed in Ref.~\cite{Knapen:2021eip}.
The proper decay length is varied between 1\unit{mm} and 10\unit{m}.
Characteristics of the models are the \Pomega to \Peta meson mass ratio, $\xi_{\omega}$, and the ratio of the dark sector QCD scale to the \Peta mass, $\xi_{\Lambda}$.
We consider three sets of numerical values: $(\xi_{\omega}, \xi_{\Lambda}) = (2.5, 2.5)$, $(2.5, 1.0)$, and $(1.0, 1.0)$.
These three hierarchies represent regimes where \Pomega and \Peta mesons are both produced and \Pomega can decay to a pair of \Peta mesons, only \Peta mesons are produced, and \Pomega and \Peta mesons are both produced but \Pomega cannot decay to a pair of \Peta mesons, respectively.
The three sets of values can be set for all portals, except for the vector portal, where only \Pomega couples to SM particles, so only $(\xi_{\omega}, \xi_{\Lambda}) = (1.0, 1.0)$ creates reconstructable signature.
These distinct scenarios present a wide range of signatures with different LLP multiplicities, visible decay product multiplicities, and missing transverse momenta (\ptvecmiss).

For both signal models, parton showering, hadronization, and the underlying event are modeled by \textsc{pythia}8.205 (8.230)~\cite{Sjostrand:2014zea} with parameters set by the CUETP8M1~\cite{Khachatryan:2015pea} (CP5~\cite{Sirunyan:2019dfx}) tune used for samples simulating the 2016 (2017 and 2018) data set conditions.
The NNPDF3.0~\cite{Ball:2014uwa} (3.1~\cite{Ball:2017nwa}) parton distribution functions (PDFs) are used in the generation of all simulated samples.
At least $7\times10^6$ events were generated for each simulated sample.
The \GEANTfour~\cite{geant4} toolkit is used to model the response of the CMS detector.
Simulated minimum-bias events are mixed with the hard interactions in simulated events to reproduce the effect of additional proton-proton interactions within the same or neighboring bunch crossings (pileup).
Events are weighted such that the distribution of the number of interactions per bunch crossing agrees with that observed during each data-taking period.

\section{Event reconstruction}
\label{sec:object}

A particle-flow (PF) algorithm~\cite{CMS:2017yfk} aims to reconstruct and identify each individual particle in an event, with an optimized combination of information from the various elements of the CMS detector.
The energy of photons is obtained from the ECAL measurement.
The energy of electrons is determined from a combination of the electron momentum at the primary interaction vertex as determined by the tracker, the energy of the corresponding ECAL cluster, and the energy sum of all photons spatially compatible with originating from the electron track, hence considered as bremsstrahlung.
The energy of muons is obtained from the curvature of the corresponding track.
The energy of charged hadrons is determined from a combination of their momentum measured in the tracker and the matching ECAL and HCAL energy deposits, corrected for the response function of the calorimeters to hadronic showers.
Finally, the energy of neutral hadrons is obtained from the corresponding corrected ECAL and HCAL energies.

Muons are measured in the range $\abs{\eta} < 2.4$.
The efficiency to reconstruct and identify muons with transverse momentum $\pt>20\GeV$ is greater than 96\%~\cite{Sirunyan:2018fpa,Sirunyan:2019yvv}.
Matching muons to tracks measured in the silicon tracker results in a relative \pt resolution of 1\% in the barrel and 3\% in the endcaps for muons with \pt up to 100\GeV, and better than 7\% in the barrel for muons with \pt up to 1\TeV~\cite{Sirunyan:2018fpa}.
Energetic muons producing electromagnetic radiation may generate MDS clusters.
In this search, reconstructed muons are used to veto these clusters to suppress background from muon bremsstrahlung.
Depending on the search categories, different muon \pt and identification requirements are applied.

For each event, hadronic jets are clustered from these reconstructed particles using the infrared- and collinear-safe anti-\kt algorithm~\cite{Cacciari:2008gp, Cacciari:2011ma} with a distance parameter of 0.4.
The jet momentum is determined as the vectorial sum of all particle momenta in the jet, and is found from simulation to be, on average, within 5--10\% of the true momentum over the whole \pt spectrum and detector acceptance~\cite{CMS:2016lmd}.
Pileup interactions produce additional tracks and calorimetric energy deposits that contribute to the jet momentum.
To mitigate this effect, charged particles identified to be originating from pileup vertices are discarded and an offset correction is applied to correct for remaining contributions.
Jet energy corrections are derived from simulation to bring the measured response of jets to that of particle-level jets on average.
In situ measurements of the momentum balance in dijet, $\text{photon} + \text{jet}$, $\PZ + \text{jet}$, and multijet events are used to account for any residual differences in the jet energy scale between data and simulation~\cite{CMS:2016lmd}.
Additional selection criteria, detailed in Section~\ref{sec:eventsel}, are applied to each jet to remove jets potentially dominated by anomalous contributions from various subdetector components or reconstruction failures~\cite{Chatrchyan:2011tn}.

The vector \ptvecmiss is computed as the negative vector \pt sum of all the PF candidates in an event, and its magnitude is denoted as \ptmiss.
The \ptvecmiss is modified to account for corrections to the energy scale of the reconstructed jets in the event~\cite{CMS:2019ctu}.

\section{Muon detector showers}
\label{sec:mds}

When charged particles traverse the CSCs, signal pulses are collected on the anode wires and the cathode strips.
The signals from the wire groups are combined with signals from the cathode strips to form a point on a two-dimensional plane in each chamber layer called a CSC hit.
When charged particles traverse the DT chambers, signal pulses are collected on the anode wires at the center of the DT cells.
Since the DTs only provide measurement in either the $\phi$ or $z$ dimension, the DT hit position is assumed to be at the center of each DT chamber in the orthogonal direction.
For LLPs that decay within or just in front of the muon system, the material in the iron return yoke structure will induce a hadronic or electromagnetic shower, creating a geometrically localized and isolated cluster of signal hits in the muon detectors.

Because the CSC and DT hits contain different information, they are handled separately.
The CSC and DT hits are clustered in $\eta$ and $\phi$ using the \textsc{dbscan} algorithm~\cite{dbscan}, which groups hits in high-density regions into clusters.
A minimum of 50 hits and a distance parameter $\Delta R_{\mathrm{cluster}} = \sqrt{\smash[b]{(\Delta\eta)^2+(\Delta\phi)^2}}$ of 0.2 is used.
A minimum ionizing muon is expected to produce a maximum of 24 or 44 hits in the CSC or DT detectors, respectively.
We choose a minimum of 50 hits to explicitly avoid misidentifying minimum ionizing muons as an MDS cluster object.
A spatial position is associated with each cluster by taking the geometric center of the hits in the cluster.
From this, we can calculate the $\eta$ and $\phi$ coordinates of each cluster.
Nearby clusters are merged if they satisfy $\Delta R < 0.6$.
This is repeated until all clusters within an event are isolated.
This merging procedure ensures that clusters coming from the same source are reconstructed as one object.
In the overlap region of the muon detectors, with $0.9 < \abs{\eta} < 1.2$, if both CSC and DT clusters are reconstructed with $\Delta R < 0.4$, the CSC clusters are given precedence because the CSC cluster response is larger, and the corresponding DT clusters are removed.

\begin{figure}[h!]
  \centering
  \includegraphics[width=0.9\columnwidth]{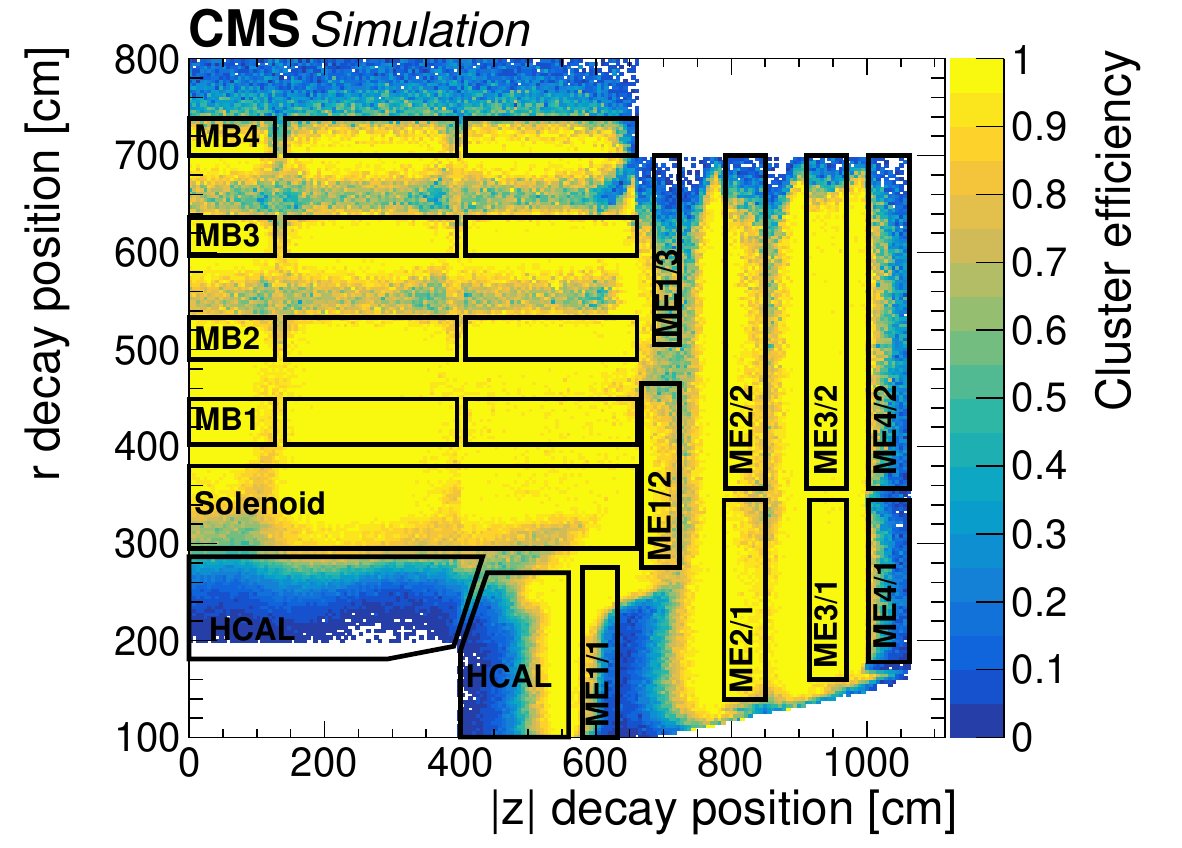}
  \caption{
    The cluster reconstruction efficiency, including both DT and CSC clusters, as a function of the simulated $r$ and $\abs{z}$ decay positions of the particle \PS decaying to $\ddbar$ in events with $\ptmiss > 200\GeV$, for a mass of 40\GeV and a range of $c\tau$ values uniformly distributed between 1 and 10\unit{m}.
    The cluster reconstruction efficiency appears to be nonzero beyond MB4 because the MB4 chambers are staggered so that the outer radius of the CMS detector ranges from 738 to 800\unit{cm}.
    The barrel and endcap muon stations are drawn as black boxes and labeled by their station names.
    The region between labeled sections are mostly steel return yoke.
  }
  \label{fig:cluster_eff_2d}
\end{figure}

\begin{figure}[h!]
  \centering
  \includegraphics[width=0.45\textwidth]{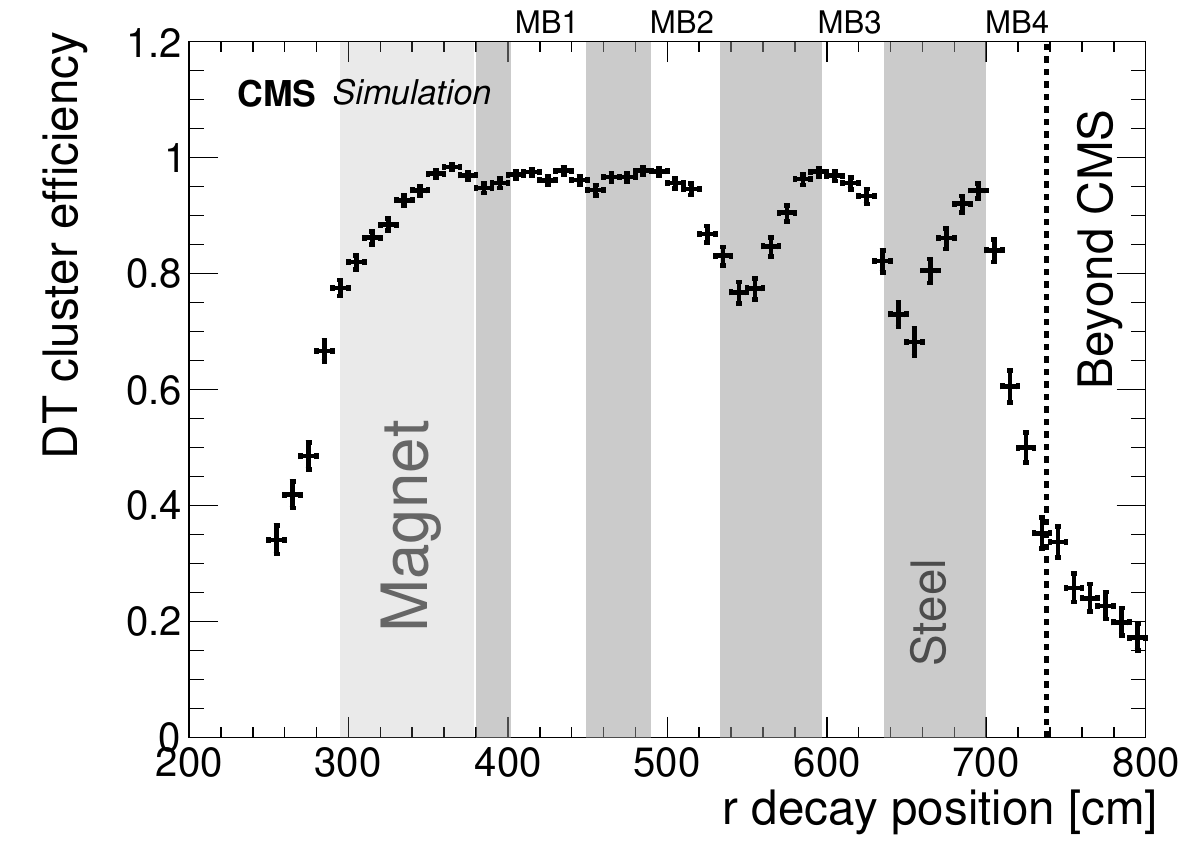}
  \includegraphics[width=0.45\textwidth]{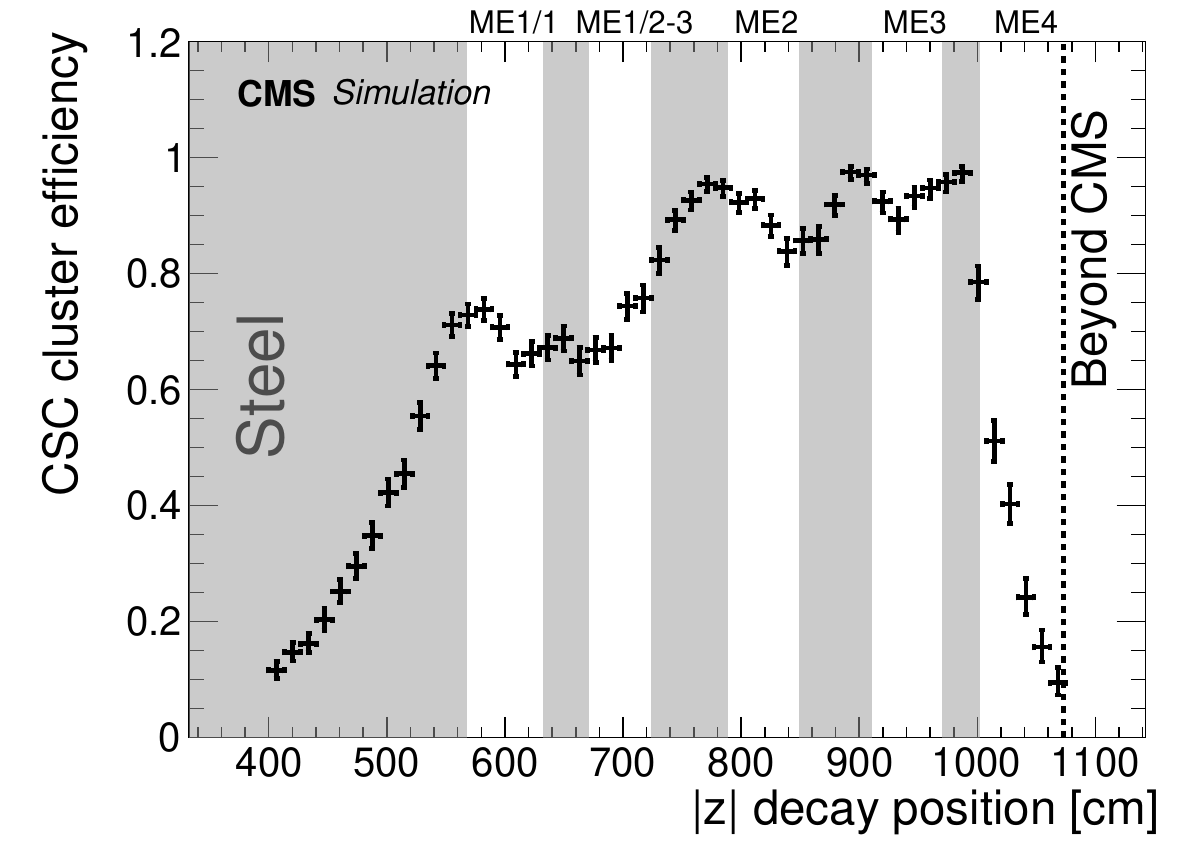}
  \caption{
    The DT (\cmsLeft) and CSC (\cmsRight) cluster reconstruction efficiency as a function of the simulated $r$ or $\abs{z}$ decay positions of \PS decaying to $\ddbar$ in events with $\ptmiss > 200\GeV$, for a mass of 40\GeV and a range of $c\tau$ values between 1 and 10\unit{m}.
    The DT cluster reconstruction efficiency is shown for events where the LLP decay occurs at $\abs{z}< 700\unit{cm}$.
    The DT cluster reconstruction efficiency appears to be nonzero beyond MB4 because the MB4 chambers are staggered so that the outer radius of the CMS detector ranges from 738 to 800\unit{cm}.
    The CSC cluster reconstruction efficiency is shown for events where the LLP decay occurs at $\abs{r}< 700\unit{cm}$ and $\abs{\eta}< 2.6$.
    The clusters are selected from signal events satisfying the $\ptmiss > 200\GeV$ requirement.
    Regions occupied by steel shielding are shaded in gray.
  }
  \label{fig:cluster_eff_1d}
\end{figure}

The simulated cluster-reconstruction efficiency, including both DT and CSC clusters, as a function of the generated $r$ and $\abs{z}$ decay positions of the particle \PS is shown in Fig.~\ref{fig:cluster_eff_2d}.
The DT (CSC) cluster reconstruction efficiency is shown separately as a function of the generated $r$ ($\abs{z}$) in Fig.~\ref{fig:cluster_eff_1d}.
The efficiencies are shown for events satisfying $\ptmiss > 200\GeV$, as required in the search described in Section~\ref{sec:eventsel} and depends strongly on the LLP decay position.
The efficiency is highest when the LLP decays near the edges of the shielding absorber material, where there is enough material to induce the shower, but not so much that it stops the shower secondary particles.
Decays that occur at the beginning or just before a thick section of absorber material will have a reduced efficiency because a fraction of the particle shower will be absorbed by the material before it can be detected.
The effect appears as dips in the efficiency in Fig.~\ref{fig:cluster_eff_1d} in the regions labeled MB2, MB3, and ME2.
The efficiency decreases to zero when the decay occurs near the end of the last stations in MB4 or ME4 because there is an insufficient amount of absorber material to induce any particle shower.

The cluster reconstruction efficiency also depends on whether the LLP decays to hadrons or to electrons or photons.
In general, hadronic showers have higher efficiency compared to electromagnetic showers because they are more likely to penetrate through the steel in between the muon stations.
Showers induced by electromagnetic decays generally occupy just one station and are stopped by the steel between the stations.
When the LLP decays close to or in the CSCs, defined as the union of the two regions: (i) $500 < \abs{z} < 661\unit{cm}$, $r <  270\unit{cm}$, and $\abs{\eta}< 2.4$, and (ii) $660 < \abs{z} < 1100\unit{cm}$, $r <  695.5\unit{cm}$, and $\abs{\eta}<2.4$, the inclusive CSC cluster reconstruction efficiency is approximately 80, 55, and 35\% for fully hadronic, $\PGtp\PGtm$, and fully leptonic decays, respectively.
When the LLP decays close to or in the DTs, defined as the region $380 < r < 738\unit{cm}$ and $\abs{z} <  661\unit{cm}$, the inclusive DT cluster reconstruction efficiency is approximately 80, 60, and 45\% for fully hadronic, $\PGtp\PGtm$, and fully leptonic decays, respectively.

The accuracy of the simulation modeling of the cluster reconstruction efficiency has been studied using a $\PZ\to\MM$ data sample, where clusters are produced when one of the muons undergoes bremsstrahlung and the associated photon produces an electromagnetic shower.
The difference between data and simulation is propagated as a systematic uncertainty in the cluster reconstruction efficiency, as detailed in Section~\ref{sec:systematics}.

\section{Search strategy and event selection}
\label{sec:eventsel}

An LLP, like the neutral scalar or the dark meson, that decays after it has traversed the calorimeter systems may produce large $\ptmiss$ because its momentum is not properly measured or associated with a particle by the PF algorithm.
We exploit this feature by analyzing the data collected by triggering on events with online $\ptmiss > 120\GeV$~\cite{CMS:2016ngn}, and subsequently requiring offline $\ptmiss > 200\GeV$, both computed from PF candidates, to ensure that the selected events are well above the trigger threshold, where the high-level trigger efficiency is effectively constant.
The trigger efficiency is about 95\% and the signal acceptance for the offline \ptmiss requirement is of the order of 1\% for both models.
We require at least one jet with $\pt > 30\GeV$ and $\abs{\eta} < 2.4$ passing the ``Tight Lepton Veto'' jet identification criteria~\cite{JetID}, because the \ptmiss requirement imposes an implicit large boost requirement on the Higgs boson that ensures that the Higgs boson is always produced together with a jet from initial-state radiation.
To suppress noncollision backgrounds, we apply beam-halo filters that remove events containing beam-halo muons and calorimeter noise filters~\cite{Chatrchyan:2011tn}.
The event-level selections are kept minimal to be as model independent as possible, and the efficiency of these selections is about 95\%.

\begin{figure}[h!]
  \centering
  \includegraphics[width=0.8\columnwidth]{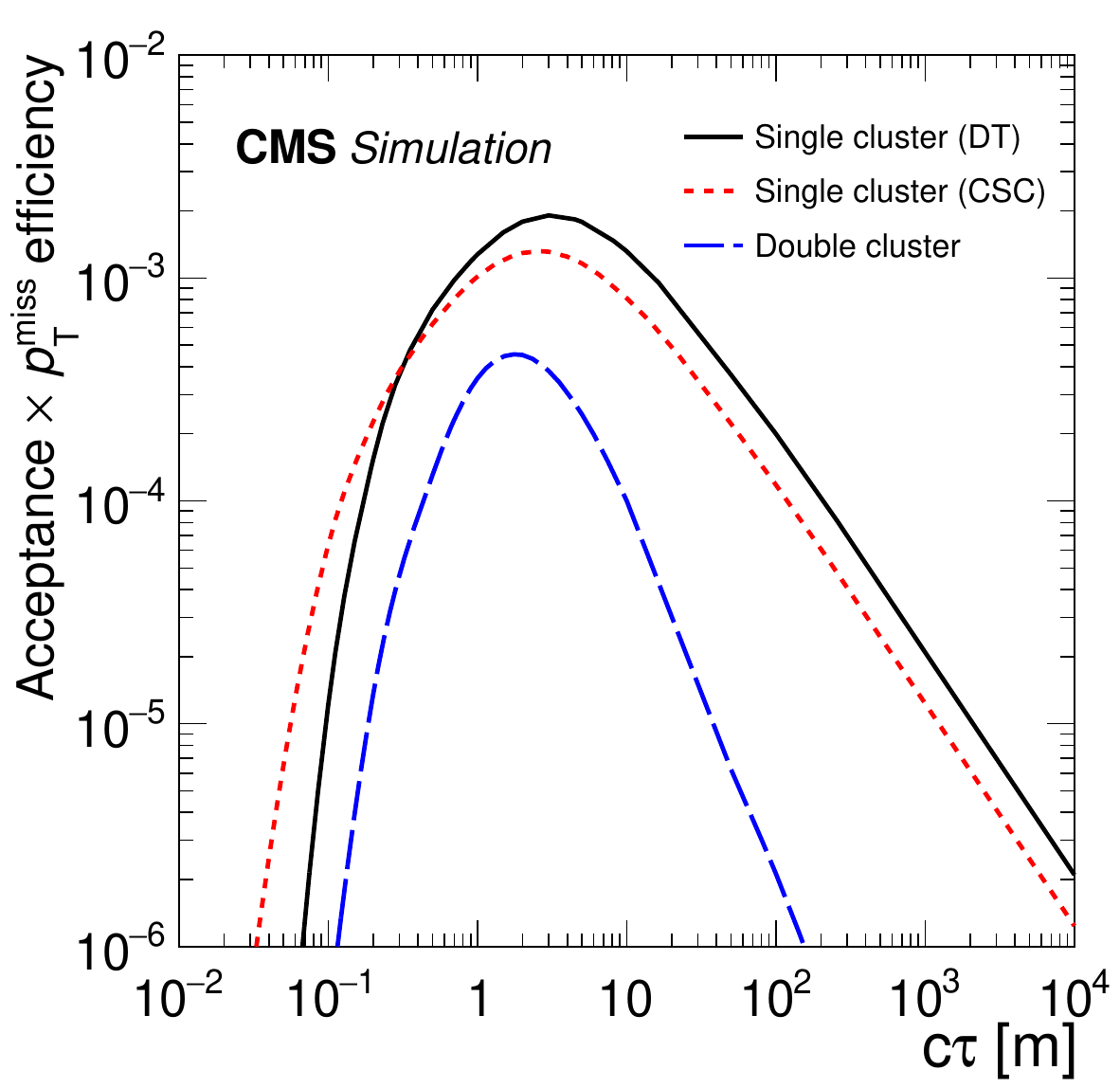}
  \caption{
    The geometric acceptance multiplied by the efficiency of the $\ptmiss > 200\GeV$ selection, as a function of the proper decay length $c\tau$ for a scalar particle \PS with a mass of 40\GeV.
  }
  \label{fig:acceptance}
\end{figure}

After the event-level selections, the events are separated into three mutually exclusive categories based on the number and location of the clusters: events (1) with two clusters in the muon detectors, (2) with exactly one CSC cluster and no DT cluster, and (3) with exactly one DT cluster and no CSC cluster.
Events with two clusters are further categorized into categories with two CSC clusters, two DT clusters, and one CSC and one DT cluster.
The definition of the category with exactly one CSC cluster is based on a previous search using the endcap muon detectors~\cite{CMS:2021juv} with a few changes, such as explicitly excluding overlapping double-cluster events and loosening the event-level selections to be consistent with those in other categories.
The geometric acceptance multiplied by the efficiency of the \ptmiss selection for each category is shown in Fig.~\ref{fig:acceptance}.

The main SM backgrounds are similar among the three categories and include punch-through jets, muons that undergo bremsstrahlung, and isolated hadrons from pileup, recoils, or underlying events.
To suppress background from punch-through jets or muon bremsstrahlung, we reject CSC and DT clusters that have a jet or muon within $\Delta R < 0.4$ in all categories.
However, depending on the category, the \pt thresholds and identification requirements of jets and muons are different.
Furthermore, additional tighter vetoes are applied to the single-cluster categories to reject background, as discussed in the following subsections.

  {\tolerance=800
    Additionally, the angular difference (\deltaphi) between the \ptvecmiss and the cluster location is used as a discriminating variable in all three categories.
    For signal, \deltaphi peaks near zero because the large \ptmiss requirement tends to select events where the Higgs boson has a large momentum and is nearly collinear with the \PS.
    For the backgrounds, \deltaphi is uniformly distributed because the cluster and \ptvecmiss are independent.
    The exact threshold on the variable is different for the barrel and endcap, and looser for the double-cluster category, as detailed in the following subsections.
    \par}

Finally, the number of hits in the clusters (\nhits) is used to discriminate signal and background.
Signal clusters tend to have large \nhits, while background clusters are expected to have small \nhits.
The threshold on \nhits is determined separately in different categories by optimizing the expected limits in each category.
At the determined thresholds, both the signal and background estimates are not sensitive to the thresholds, as the distributions for signals are relatively flat, while the background estimate is close to zero near the optimal point.
The exact threshold on \nhits is different for the barrel and endcap, and looser for the double-cluster category, as detailed in the following subsections that describe the detailed event selections for each category.

\subsection{Double clusters}

The double-cluster category includes events containing two clusters satisfying the selection criteria described below.
We ensure that different cluster objects are spatially separated from each other by requiring $\Delta R > 0.4$.
Events are separated into three categories depending on whether there are two DT clusters, two CSC clusters, or one CSC and one DT cluster present.
Requiring two muon system clusters significantly reduces the expected background, so the selection requirements in the double-cluster category are much looser compared to the single cluster categories.

The CSC clusters are rejected if any jet with $\pt > 30\GeV$ or a global muon~\cite{Sirunyan:2018fpa}, built by matching muon tracks in the muon detectors and in the tracker, with $\pt > 30\GeV$ is found within $\Delta R < 0.4$.
Similarly, DT clusters are rejected if any jet with $\pt > 50\GeV$ or a muon passing loose identification criteria~\cite{Sirunyan:2018fpa, Sirunyan:2019yvv} with $\pt > 10\GeV$ is found within $\Delta R < 0.4$.

We veto CSC clusters that are entirely contained in the innermost rings of the ME1 station (ME1/1) and veto DT clusters that have more than 90\% of the hits contained in the innermost station (MB1), both of which have the least absorber material between them and the interaction point, and larger background contamination.

We further reject CSC clusters produced by adjacent bunch crossings, known as out-of-time (OOT) pileup, by requiring that the cluster time ($t_\text{cluster}$) is consistent with an in-time interaction ($-5.0 < t_\text{cluster} <  12.5$\unit{ns}).
The cluster time is defined as the average time of the hits in the cluster relative to the LHC clock and corrected for the particle time-of-flight from the interaction point to the respective muon detector chambers assuming speed of light propagation.
An asymmetric time window is used to capture signal clusters with longer delays from slower-moving LLPs.
To reject clusters composed of hits from multiple bunch crossings, the root-mean-square spead of the hit times of each cluster is required to be less than 20\unit{ns}.

Tracks from muons in the barrel are likely to deposit a similar number of hits in all four DT stations, while showers from LLP decays are likely to have hits concentrated in one or two stations.
Therefore, to reject DT clusters from muon bremsstrahlung we veto clusters that contain hits in all four stations and that have a ratio of the minimum to maximum number of hits per station less than 0.4.

Cosmic ray muons produce hits in both the upper and lower hemispheres of the muon barrel system.
To suppress this background, we reject DT clusters if there are at least six segments, which are straight-line tracks built within each DT chamber, and at least one segment in every station found in the opposite hemisphere ($\abs{\Delta\phi} > 2$) from the cluster.
In addition, cosmic ray muon showers produce hits in multiple regions of the CMS detector.
Thus, we reject any event in which more than a quarter of the DT or CSC rings, consisting of chambers with the same $r$ and $z$ coordinates, contain 50 or more hits.
Finally, we require $\deltaphi < 1.0$~(1.2) for CSC (DT) clusters.

For signal events with two clusters, the $\Delta R$ between two LLPs, thus between the clusters, is typically around 1.
For background, the $\Delta R$ between clusters is usually large ($\Delta R > 2$) because the two clusters generally come from separate processes, especially for the CSC-CSC and DT-CSC categories.
Therefore, we require the $\Delta R$ between the two clusters to be less than 2.0~(2.5) for the CSC-CSC~(DT-CSC) subcategory.
There is already an implicit $\Delta\phi$ selection between the two clusters by requiring both clusters to pass the \deltaphi selection.
The $\Delta R$ selection additionally requires the two clusters to be close in $\eta$.

Finally, \nhits is used to discriminate signal and background.
We require $\nhits \geq 100$ (80) for CSC (DT) clusters.

\subsection{Single CSC cluster}
\label{ssec:strategy:single_csc}
The single-CSC-cluster category includes events in which only one LLP decay produces a displaced cluster in the endcap muon system.
In this category, the expected background yield is significantly higher than in the double-cluster category, so we apply much tighter cluster veto requirements to achieve the same near-zero background level.

The cluster veto requirements are the same as in Ref.~\cite{CMS:2021juv}.
Clusters that have a jet with $\pt > 10\GeV$ or a muon with $\pt > 20\GeV$ within $\Delta R < 0.4$ are rejected.
We veto clusters that have any hits in the two innermost rings of the ME1 station (ME1/1 and ME1/2), which have the least absorber material between them and the interaction point, or match any hit (with $\Delta R(\text{cluster, hit})<0.4$) in the RPCs located immediately next to ME1/2.
In the region where the barrel and endcap muon detectors overlap ($0.9 < \abs{\eta} < 1.2$), any cluster matched to any track segment reconstructed in the innermost station of the DT detectors (MB1), or any hit in the RPCs situated in front of and behind MB1 matched to within $\Delta R(\text{cluster, segment or hit})<0.4$, is vetoed.
We reject clusters with $\abs{\eta} > 2.0$ to suppress the muon bremsstrahlung background that evaded the muon veto because of the decreasing muon reconstruction and identification efficiencies at larger $\abs{\eta}$.

As in the double-cluster category, events in which more than a quarter of the DT or CSC rings contain 50 or more hits are rejected.

\begin{figure}[h!]
  \centering
  \includegraphics[width=0.8\columnwidth]{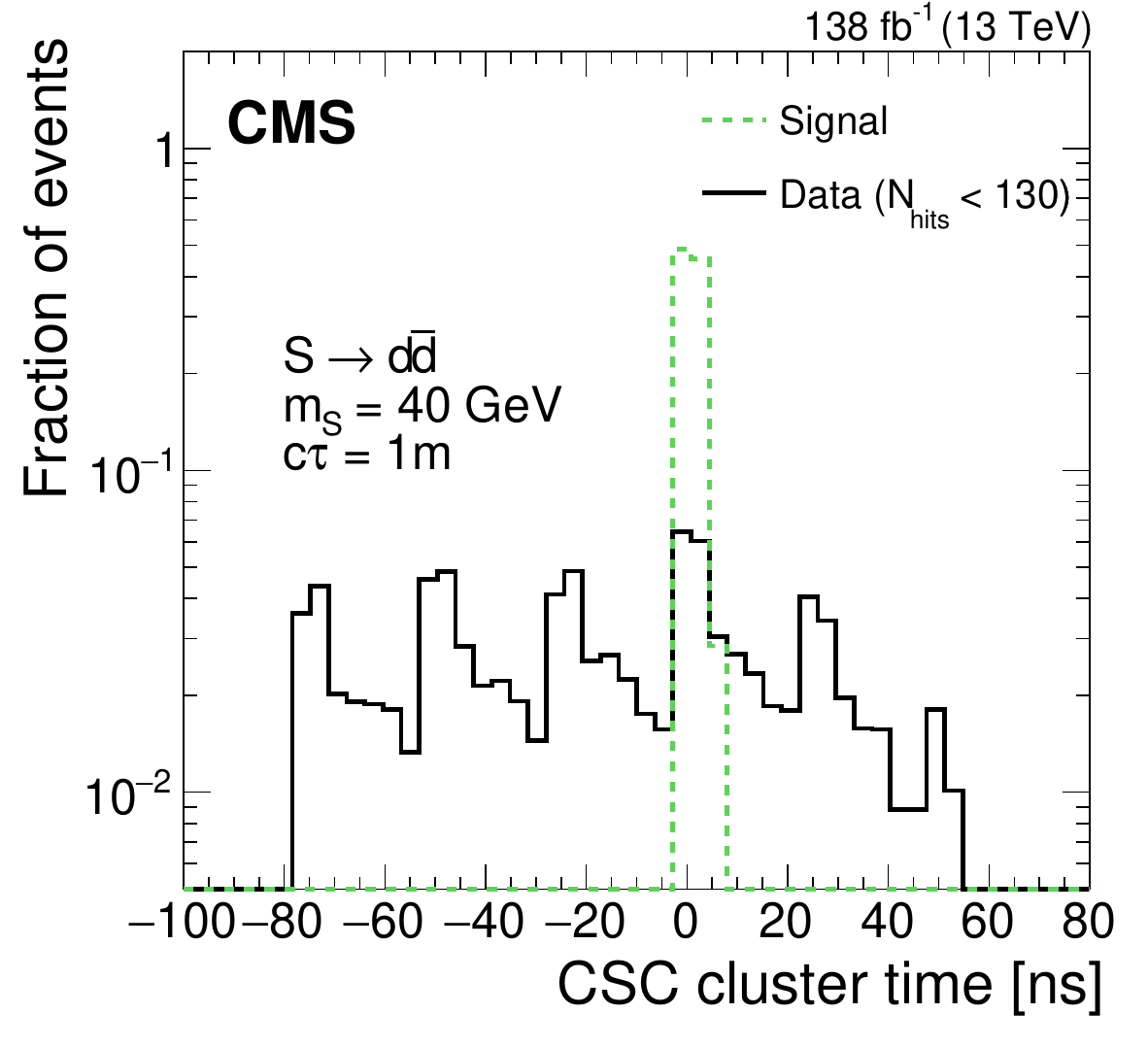}
  \caption{
    Distributions of the cluster time ($t_\text{cluster}$) for signal, where \PS decays to $\ddbar$ with a proper decay length $c\tau$ of 1\unit{m} and mass of 40\GeV, and for a background-enriched sample in data selected by inverting the \nhits requirement.
  }
  \label{fig:csc_strategy_clusterTime}
\end{figure}

After the veto requirements are applied, the dominant background source consists of decays of SM LLPs, which are predominantly produced by pileup interactions and are independent of the primary interaction that yielded the large \ptmiss.
These pileup interactions may be in-time or OOT with the primary interaction, as shown in Fig.~\ref{fig:csc_strategy_clusterTime}.
Clusters produced by OOT pileup are rejected by requiring $-5.0 < t_\text{cluster} < 12.5$\unit{ns}, as in the double-cluster category.
The time window requirement suppresses the background by a factor of 5.
Similarly, the root-mean-square spread of the hit times of each cluster is required to be less than 20\unit{ns}.

It was observed that clusters from all background processes occur more often at larger values of $\abs{\eta}$, as the effectiveness of the jet and muon vetoes decreases because of decreasing reconstruction efficiencies.
Signal clusters often occupy more than one CSC station ($N_{\text{stations}}>1$) and occur more frequently in stations farther away from the primary interaction point.
To distinguish signal and background clusters, a cluster identification algorithm was devised that makes more restrictive $\abs{\eta}$ requirements for clusters that occupy only one CSC station ($N_{\text{stations}}=1$) and are closer to the primary interaction point.
The $\abs{\eta}$ requirements are:
\begin{itemize}
  \item $\abs{\eta} < 1.9$ if $N_{\text{stations}} > 1$,
  \item $\abs{\eta} < 1.8$ if $N_{\text{stations}} = 1$ and the cluster is in station 4,
  \item $\abs{\eta} < 1.6$ if $N_{\text{stations}} = 1$ and the cluster is in station 3 or station 2, and
  \item $\abs{\eta} < 1.1$ if $N_{\text{stations}} = 1$ and the cluster is in station 1 because of an implicit selection from the ME1/1 and ME1/2 vetoes.
\end{itemize}

The cluster identification algorithm has $\sim$80\% efficiency for simulated clusters originating from \PS decays, and suppresses the background by a factor of 3.
The events that pass the cluster identification criteria are used to define the search region (or signal region, SR), and those that fail are used as an additional in-time validation region (VR).

\begin{figure}[h!]
  \centering
  \includegraphics[width=0.45\textwidth]{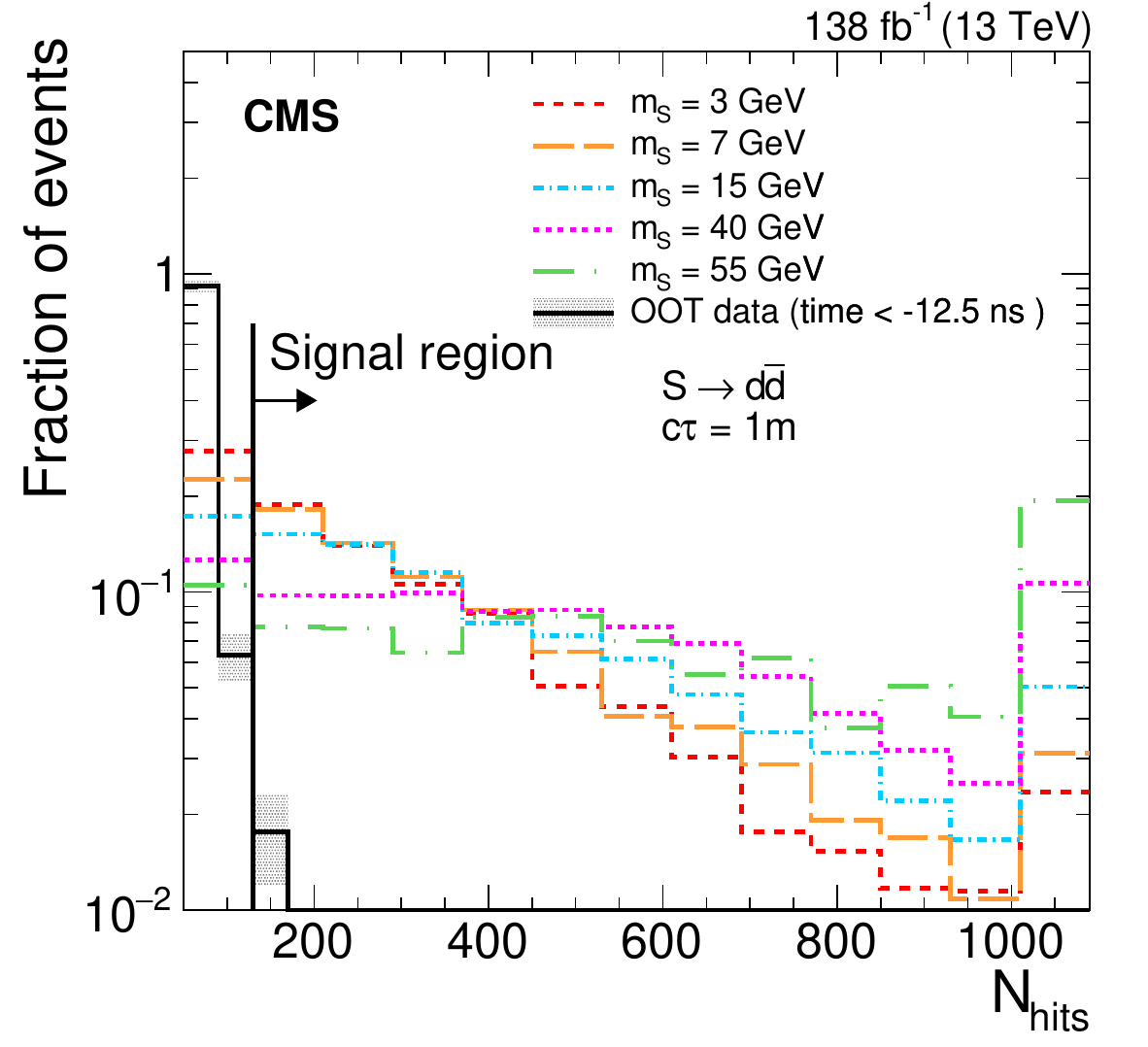}
  \includegraphics[width=0.45\textwidth]{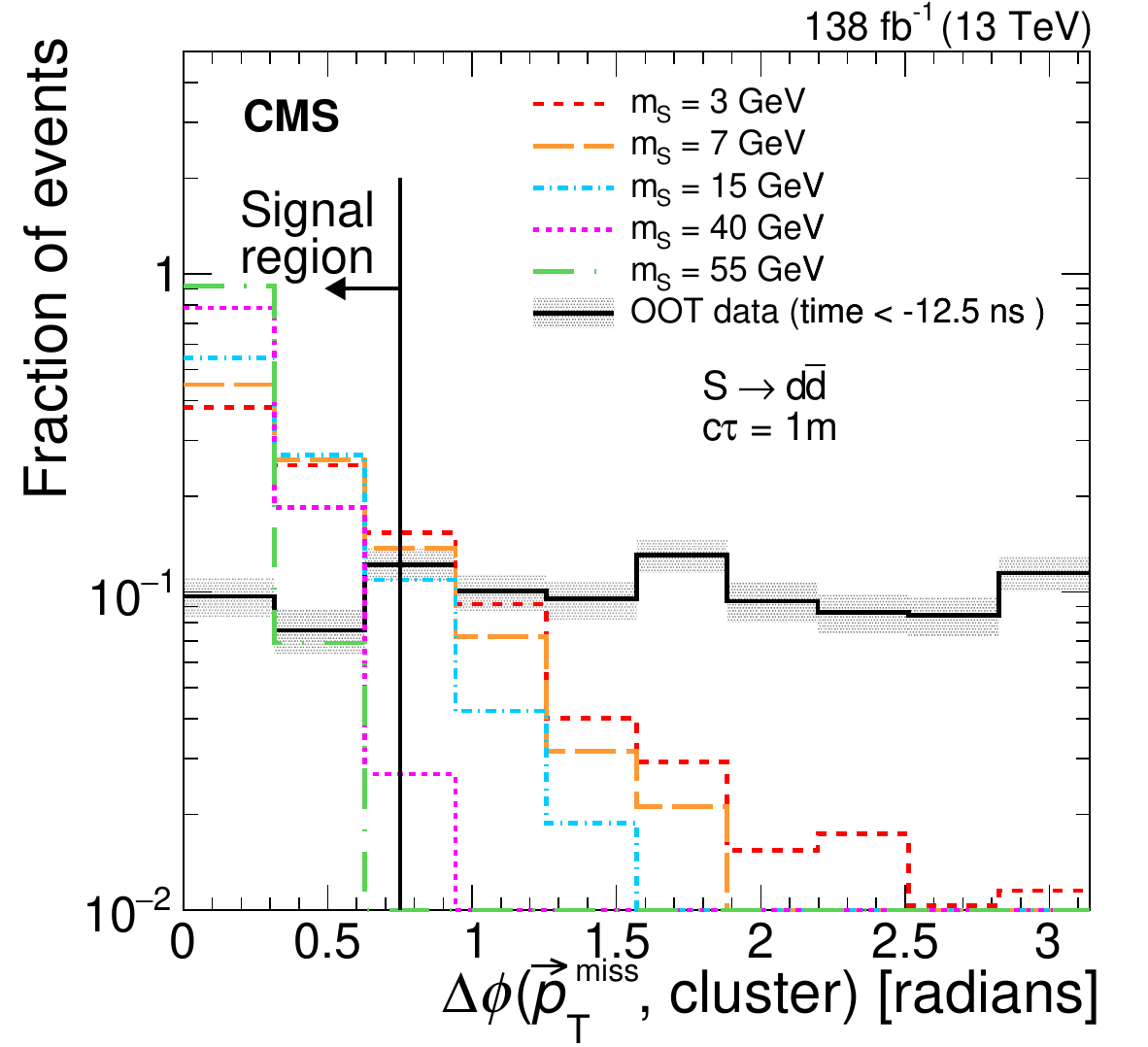}
  \caption{
    The distributions of \nhits (\cmsLeft) and  \deltaphi (\cmsRight) for single CSC clusters are shown for \PS decaying to \ddbar for a proper decay length of 1\unit{m} and various masses compared to the OOT background ($t_\text{cluster} < -12.5\unit{ns}$).
    The OOT background is representative of the overall background shape, because the background passing all the selections described above is dominated by pileup and underlying events.
    The shaded bands show the statistical uncertainty in the background.
  }
  \label{fig:csc_strategy_nrechits_dphi}
\end{figure}

Both \nhits and \deltaphi are used to discriminate signal and background.
We require $\deltaphi < 0.75$ and $\nhits > 130$.
The signal and background shapes of the two discriminants are shown in Fig.~\ref{fig:csc_strategy_nrechits_dphi}.
For the backgrounds, \deltaphi is independent of \nhits, as shown in Section~\ref{sec:background}, enabling the use of the ABCD method to predict the background yield in the signal-enriched bin.

\subsection{Single DT cluster}

The single-DT-cluster category targets events in which only one LLP decay produces a displaced cluster in the barrel muon system.
Events passing the selection criteria for the double-cluster or single-CSC-cluster categories are not considered in this category, to give precedence to the category with higher sensitivity (double cluster) and to minimize differences with the previously published search.

  {\tolerance=800
    First, to remove high-$\ptmiss$ events due to mismeasured jets, we require the minimum of $\abs{\Delta\phi(\text{jet}, \ptvecmiss)}$ over all the jets with $\pt > 30\GeV$ to be greater than 0.6.
    This requirement reduces the background from SM events composed uniquely of jets produced through the strong interaction, referred to as QCD multijet events, and is only applied to the single-DT-cluster category because this category is dominated by the punch-through jet background. \par}

We veto clusters that have a jet with $\pt > 10\GeV$ or a muon with $\pt > 10\GeV$ passing loose identification criteria~\cite{ Sirunyan:2018fpa, Sirunyan:2019yvv} within $\Delta R < 0.4$.
In addition, we reject clusters that are within $\Delta R < 1.2$ from the leading-\pt jet.
Furthermore, DT clusters that are within $\Delta R < 0.4$ of two or more hits in the innermost station MB1 are rejected.
Additionally, clusters with maximum hit counts in MB3 or MB4 are rejected if they are within $\Delta R < 0.4$ of two or more MB2 hits.
Each cluster is associated with one of the five wheels based on the average $z$ position of its hits.
To reject clusters from noise in the DTs, we require clusters to be matched to at least 1 RPC hit from the same wheel and within $\Delta\phi < 0.5$.

To suppress background from cosmic ray muons, we veto clusters that have more than 8 hits in MB1 within $\Delta\phi < \pi/4$ in any of the wheels adjacent to the wheel containing the hits of the cluster.
In addition, we veto clusters with maximum hit counts in MB3 and MB4 that have more than 8 hits in MB2 within $\Delta\phi < \pi/4$ in any adjacent wheel.
Furthermore, we look for DT segments that are far from the clusters with $\Delta R > 0.4$ in the upper or lower hemisphere of the DT wheels.
We veto the cluster if more than 14 segments are found in either hemisphere or more than 10 segments are found in both hemispheres.

\begin{figure}[h!]
  \centering
  \includegraphics[width=0.45\textwidth]{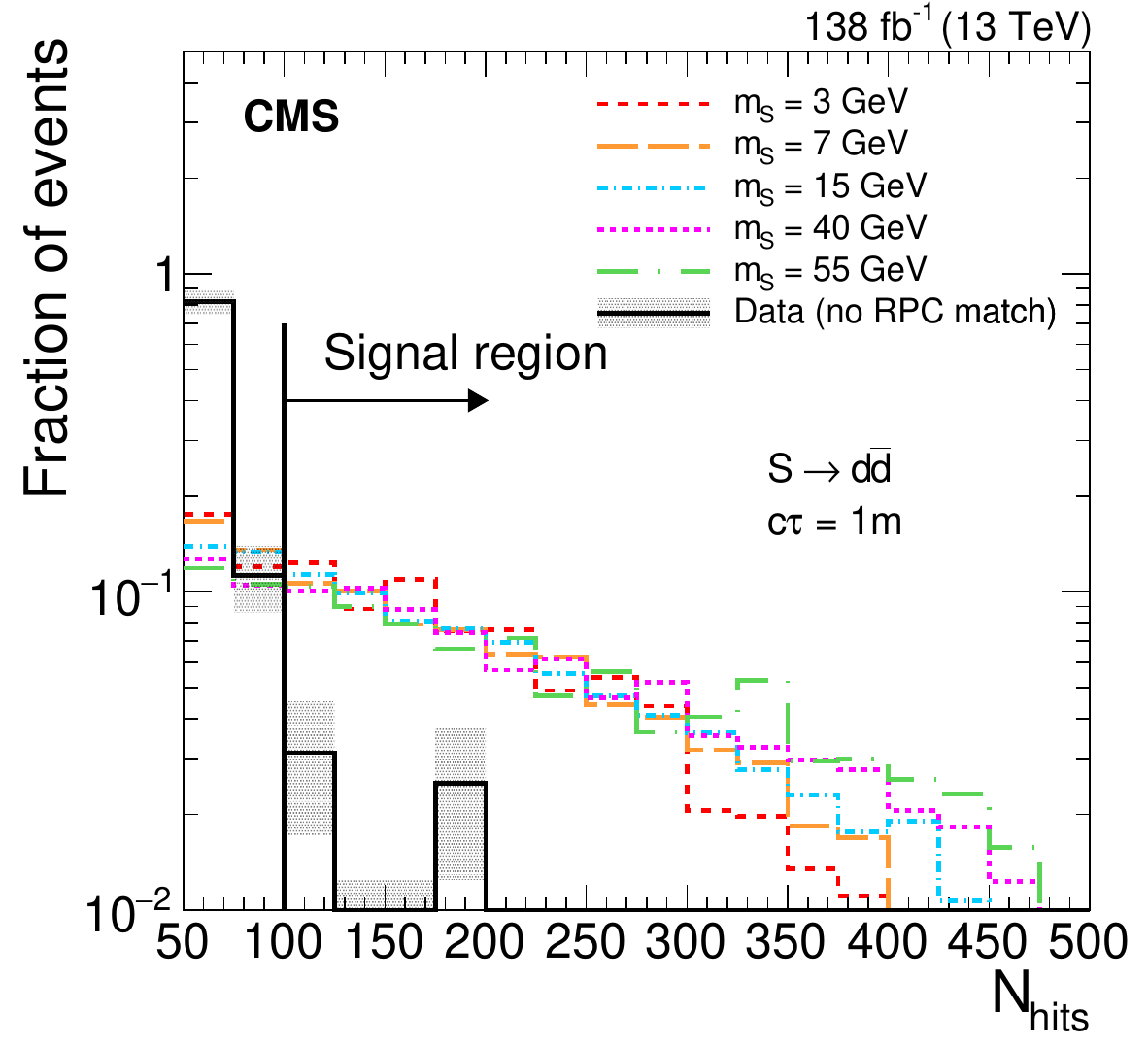}
  \includegraphics[width=0.45\textwidth]{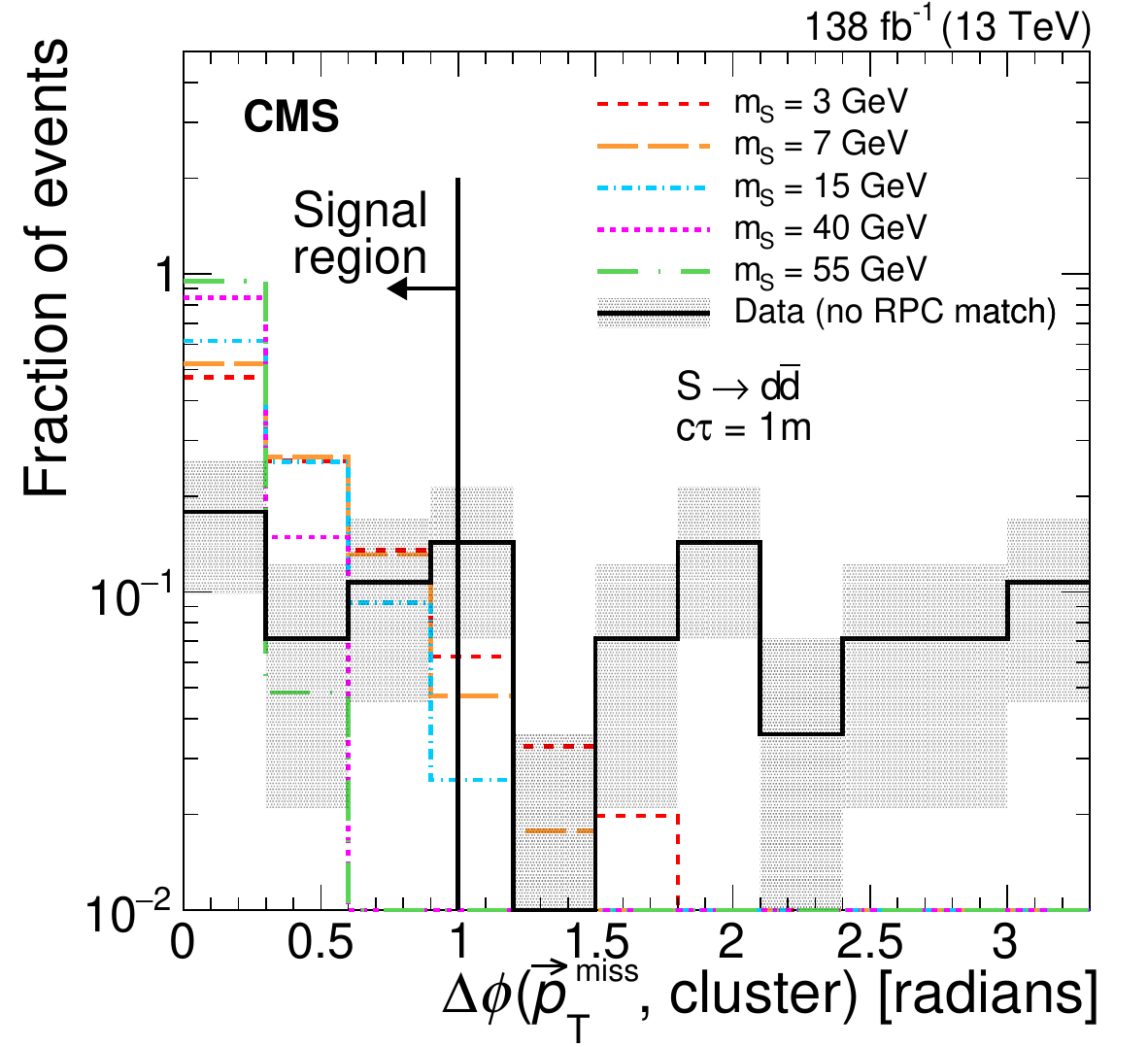}
  \caption{
    The distributions of \nhits (\cmsLeft) and \deltaphi (\cmsRight) for DT clusters are shown for \PS decaying to $\ddbar$ for a proper decay length of 1\unit{m} and various masses compared to the shape of background in a selection in which the cluster is not matched to any RPC hit and passes all other selections.
    The background is dominated by clusters from noise and low-\pt particles.
    The shaded bands show the statistical uncertainty in the background.
  }
  \label{fig:dt_strategy_nrechits_dphi}
\end{figure}

As in the single-CSC-cluster sub-category, \nhits and \deltaphi are used to discriminate signal and background.
We require $\deltaphi < 1$ and $\nhits > 100$.
The signal and background distributions of the two discriminants are shown in Fig.~\ref{fig:dt_strategy_nrechits_dphi}.

Finally, the DT clusters are categorized into 3 exclusive categories according to the station that contains the most hits: MB2, MB3, or MB4.
These categories have different background compositions, where the punch-through jet background is more prominent in the stations that are closer to the interaction point.

\section{Background estimation}
\label{sec:background}

The ABCD method~\cite{Kasieczka:2020pil} based on control samples in data is used for background estimation for all three categories.
The ABCD method requires two variables that discriminate signal and background and are independent of one another for the background.
Two separate requirements, one on each variable, partition the two-dimensional space into four bins, A, B, C, and D, where bin A contains events that pass both signal-like requirements, events in bins B and D only pass one of the requirements, and events in bin C pass neither requirement.
Because of the independence of the two variables, the expected background event rate in the signal-enriched bin A is related to the other three bins by $\lambda_\mathrm{A} = (\lambda_\mathrm{B}\lambda_\mathrm{D})/\lambda_\mathrm{C}$, where $\lambda_\mathrm{X}$ is the expected background event rate in each bin X.
In the double-cluster categories, the two variables that are used are the \nhits of each of the clusters, while in the single DT and CSC cluster categories, the two variables are \nhits and \deltaphi.
To account for a potential signal contribution to bins A, B, and C, a binned maximum likelihood fit is performed simultaneously in the four bins, with a common signal strength parameter scaling the signal yields in each bin.
The background component of the fit is constrained to obey the ABCD relationship.

In addition to the background predicted by the ABCD method, the single-DT-cluster category estimates the punch-through jet background separately, since it is non-negligible and does not have independent \nhits and \deltaphi variables, while the punch-through jet background in the other categories is negligible.
Other noncollision backgrounds, including cosmic ray muons, have been suppressed by dedicated filters described in Section~\ref{sec:eventsel}.
Their contributions are negligible in the SR after these dedicated filters.

The following subsections describe the main background component for each category, the background estimation method, and its validation.

\subsection{Double clusters}

\begin{figure}[h!]
  \centering
  \includegraphics[width=0.45\textwidth]{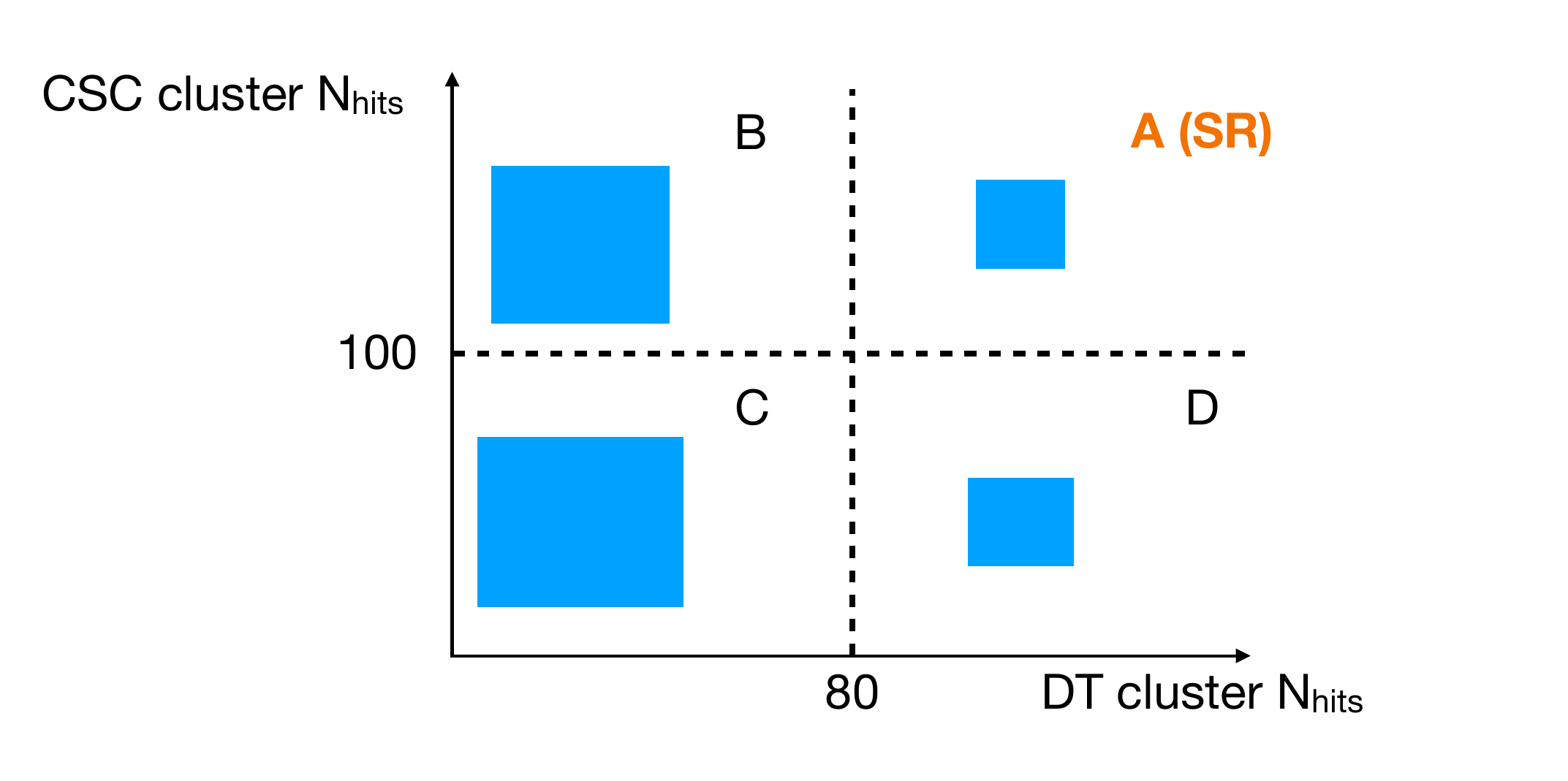}
  \includegraphics[width=0.45\textwidth]{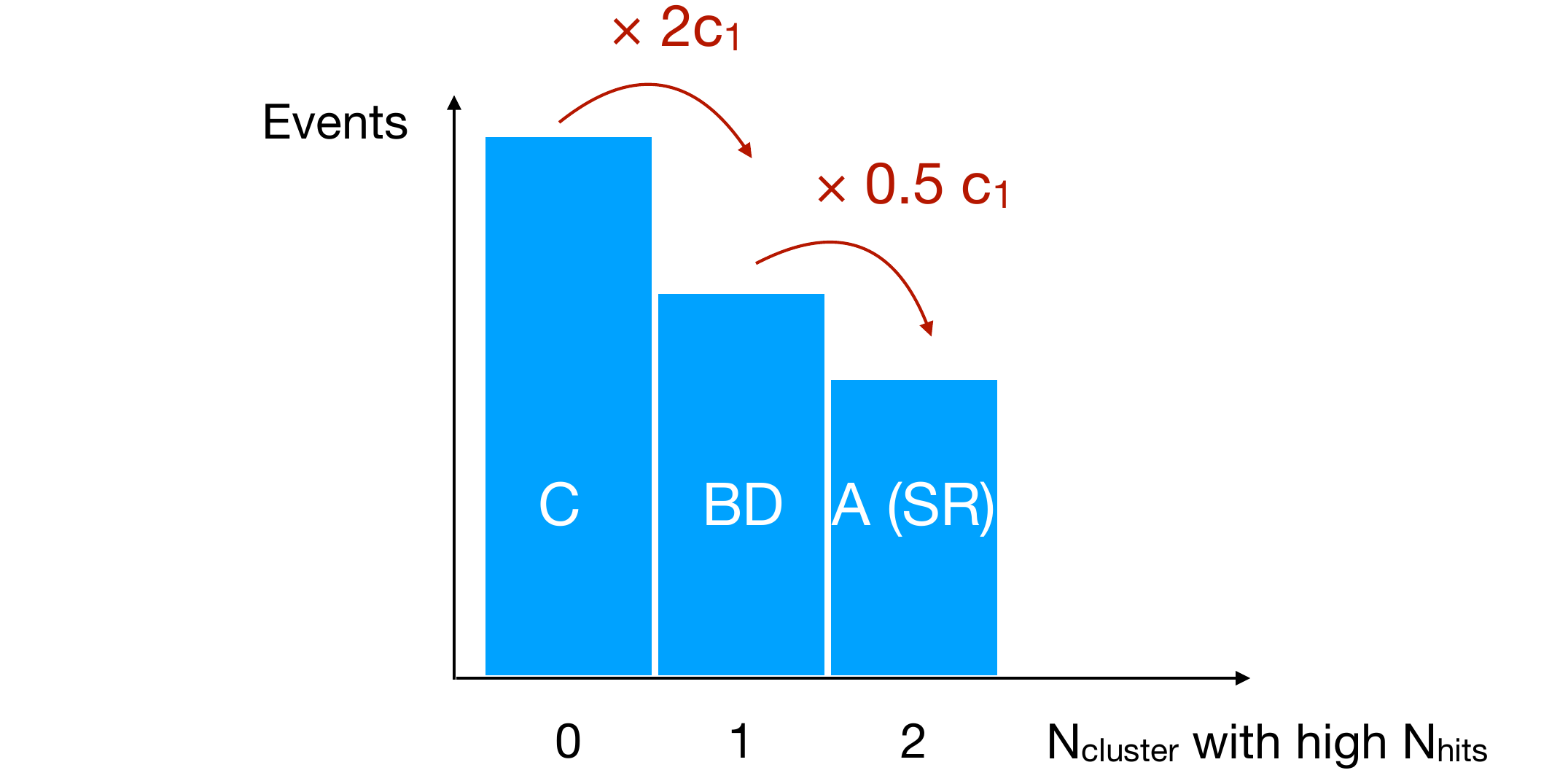}
  \caption{
    Diagrams illustrating the ABCD plane for the DT-CSC category (\cmsLeft), and for the DT-DT and CSC-CSC categories (\cmsRight).
    The variable $c_1$ is the pass-fail ratio of the \nhits selection for the background clusters.
    Bin A is the signal region (SR) for all categories.
    The size of the blue boxes on the left represents the approximate size of the expected background yield in each bin.
  }
  \label{fig:double_abcd}
\end{figure}

For the DT-CSC category, the two independent discriminating variables are the \nhits of the DT and CSC cluster, respectively.
The four bins comprise events with clusters having the following properties, as illustrated schematically in Fig.~\ref{fig:double_abcd} (upper):

\begin{itemize}
  \item Bin A: CSC cluster with $\nhits > 100$ and DT cluster with $\nhits > 80$;
  \item Bin B: CSC cluster with $\nhits > 100$ and DT cluster with $\nhits \leq 80$;
  \item Bin C: CSC cluster with $\nhits \leq 100$ and DT cluster with $\nhits \leq 80$; and
  \item Bin D: CSC cluster with $\nhits \leq 100$ and DT cluster with $\nhits > 80$.
\end{itemize}

For the CSC-CSC and DT-DT categories, the two variables are symmetric, so we combine bins B and D and define the combined expected background rate as $\lambda_\mathrm{BD}$.
Bins A, BD, and C contain events with 2, 1, and 0 clusters passing the \nhits selection, respectively, as illustrated in Fig.~\ref{fig:double_abcd} (lower).
The expected background yield in the signal-enriched bin A is related to the other two bins as $\lambda_\mathrm{A} = (\lambda_\mathrm{BD}/2)^2/\lambda_\mathrm{C}$.

To validate the background estimation method, we define two VRs: the inverted-\nhits region and the inverted-\deltaphi region.
The inverted-\nhits VR is defined by inverting the \nhits requirements for both clusters, while maintaining all the other cluster-level selections.
The \nhits threshold used in the inverted-\nhits VR is 70 (80) for DT (CSC) clusters.
Similarly, the inverted-\deltaphi VR is defined by inverting the \deltaphi requirement of both clusters while maintaining all the other cluster-level selections.
To probe signal-like events in the inverted-\deltaphi VR, we additionally require $\Delta\phi\text{(cluster1,cluster2)}<2$ for the DT-DT category.
For the CSC-CSC and DT-CSC categories, the two clusters are close to each other because of the $\Delta R\text{(cluster1,cluster2)}$ requirement.
The \deltaphi VR allows us to test for any nonnegligible backgrounds at high \nhits that cannot be accessed in the inverted-\nhits VR.
The background estimate agrees with the observed number of background events in both VRs and all three categories, as shown in Table~\ref{tab:double_bkg_validation}.

\begin{table*}[htb]
  \centering
  \caption{
    Validation of the ABCD method for the double-cluster category in both VRs.
    The uncertainty in the prediction is the statistical uncertainty propagated from bins B, C, and D or bins BD and C.
    The expected background event rate in bin A ($\lambda_\mathrm{A}$) and the background event rate in bins B, C, D, BD, and A ($N_\mathrm{B}$, $N_\mathrm{C}$, $N_\mathrm{D}$, $N_\mathrm{BD}$, $N_\mathrm{A}$) are shown.
  }
  \label{tab:double_bkg_validation}
  \begin{scotch}{l l l l l l D{+}{\,\pm\,}{3,3} l}
    Category                 & Validation region  & $N_\mathrm{B}$ & $N_\mathrm{C}$ & $N_\mathrm{D}$ & $N_\mathrm{BD}$ & \multicolumn{1}{c}{$\lambda_\mathrm{A}$} & $N_\mathrm{A}$ \\
    \hline
    \multirow{2}{*}{DT-DT}   & Inverted \deltaphi & \NA            & 11             & \NA            & 1               & 0.02+0.05      & 0              \\
    & Inverted \nhits    & \NA            & \phantom{1}2              & \NA            & 1               & 0.12+0.27      & 0              \\ 
    [\cmsTabSkip]
    \multirow{2}{*}{CSC-CSC} & Inverted \deltaphi & \NA            & \phantom{1}8              & \NA            & 2               & 0.12+0.18      & 0              \\
    & Inverted \nhits    & \NA            & \phantom{1}4              & \NA            & 2               & 0.25+0.38      & 0              \\ 
    [\cmsTabSkip]
    \multirow{2}{*}{DT-CSC}  & Inverted \deltaphi & 0              & 19             & 3              & \NA             & 0+0.3      & 0              \\
    & Inverted \nhits    & 2              & 11             & 1              & \NA             & 0.18+0.23      & 0         \\     
  \end{scotch}
\end{table*}

\subsection{Single CSC cluster}

\begin{figure}[h!]
  \centering
  \includegraphics[width=0.8\columnwidth]{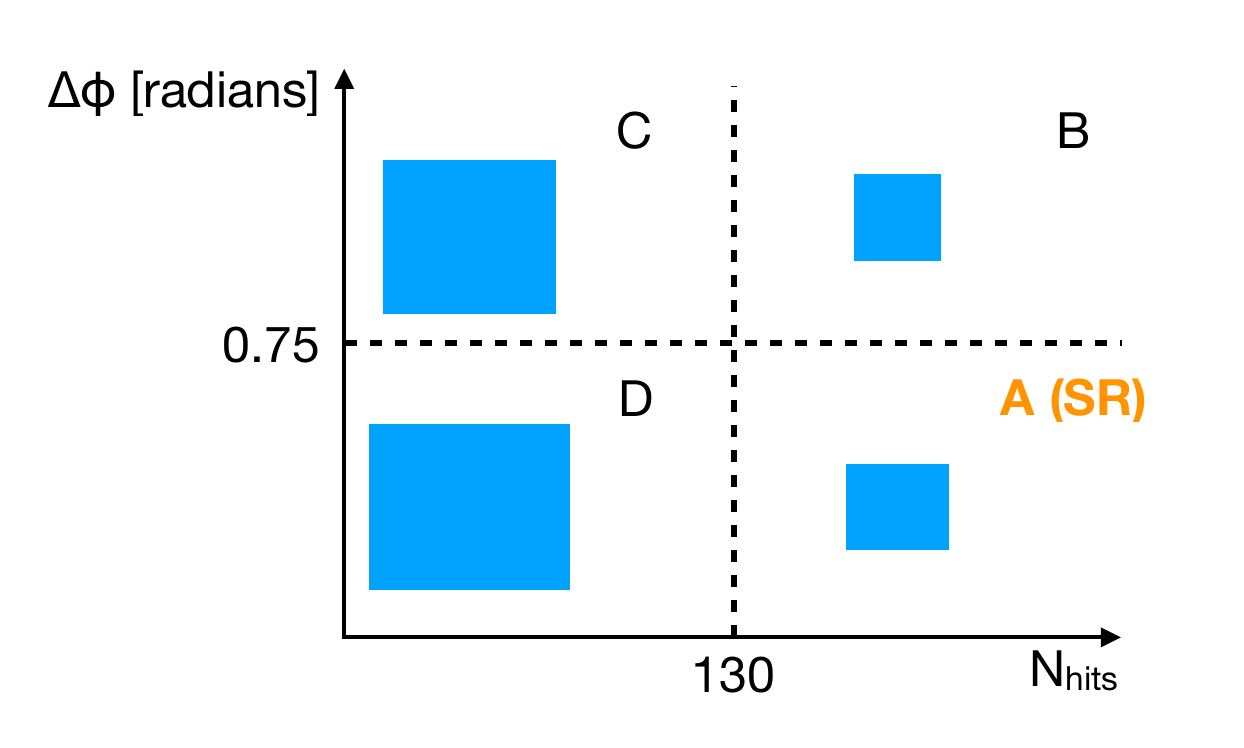}
  \caption{
    Diagram illustrating the ABCD plane for the single-CSC-cluster category, where bin A is the signal region (SR).
  }
  \label{fig:single_abcd}
\end{figure}

For the single-CSC-cluster category, the two discriminating variables are \deltaphi and \nhits.
For the backgrounds, \deltaphi is independent of \nhits enabling the use of the ABCD method.
As illustrated schematically in Fig.~\ref{fig:single_abcd}, the four bins comprise events with clusters having the following properties:

\begin{itemize}
  \item Bin A: $\deltaphi < 0.75$ and $\nhits > 130$;
  \item Bin B: $\deltaphi \ge 0.75$ and $\nhits > 130$;
  \item Bin C: $\deltaphi \ge 0.75$ and $\nhits \le 130$; and
  \item Bin D: $\deltaphi < 0.75$ and $\nhits \le 130$.
\end{itemize}

The background estimation procedure is validated using events in two separate VRs: one in the OOT region, where $t_\text{cluster} < -12.5$\unit{ns}, and one in the in-time region, where the clusters that fail the cluster identification criteria, as defined in Section~\ref{ssec:strategy:single_csc} are selected, as shown in Table~\ref{tab:csc_bkg_validation}.

\begin{table}[!h]
  \centering
  \caption{
    Validation of the ABCD method for the single-CSC-cluster category in both VRs.
    The uncertainty in the prediction is the statistical uncertainty propagated from bins B, C, and D.
    The expected background event rate in bin A ($\lambda_\mathrm{A}$) and the background event rate in bins B, C, D, and A ($N_\mathrm{B}$, $N_\mathrm{C}$, $N_\mathrm{D}$, $N_\mathrm{A}$) are shown.
  }
  \label{tab:csc_bkg_validation}
  \begin{scotch}{l c c c D{+}{\,\pm\,}{3,3} c}
    Validation region  & $N_\mathrm{B}$ & $N_\mathrm{C}$ & $N_\mathrm{D}$ & \multicolumn{1}{c}{$\lambda_\mathrm{A}$} & $N_\mathrm{A}$ \\
    \hline
    Out-of-time region & 8              & 442            & 121            & 2.2+0.8       & 3              \\
    In-time region     & 8              & 317            & \phantom{1}87             & 2.2+0.8       & 2              \\
  \end{scotch}
\end{table}

\subsection{Single DT cluster}

{\tolerance=800
  The dominant backgrounds in the single-DT-cluster category are punch-through jets and low-\pt particles from pileup.
  To estimate the background from the low-\pt particles, the ABCD method is used.
  Like the single-CSC-cluster category, the same discriminating variables \deltaphi and \nhits are used, except the thresholds for the signal-like selection are $\deltaphi < 1$ and $\nhits > 100$.
  \par}

The background prediction is validated in a pileup-enriched VR, as shown in Table~\ref{tab:dt_bkg_validation_abcd}.
The pileup-enriched region is defined by inverting the loose identification criterion on the leading jet, and removing the RPC matching criteria, \deltaphi requirement, and filters that reject noncollision background.
To reduce the statistical uncertainties when estimating the background rates in bins C and D, we merge the MB3 and MB4 categories.
The final ABCD fit in the SR is also performed with those categories merged.

The punch-through jet background that is not accounted for by the ABCD method exhibits correlated behavior, which is highly enhanced in the signal region (bin A) and negligible in the other regions.
This additional background component is estimated by measuring the number of observed events in excess of the ABCD prediction in the region with the inner DT station hit veto inverted, and multiplying the result by $\epsilon/(1-\epsilon)$ where $\epsilon$ is the corresponding veto efficiency.
For clusters in MB2, only the MB1 veto is applied, so only the MB1 veto is inverted.
For clusters in MB3 and MB4, both MB1 and MB2 vetoes are inverted.
The number of excess events in the inverted region over the estimated background is measured to be $22 \pm 7$, $7 \pm 3$, and $2.0 \pm 1.7$ for clusters in MB2, MB3, and MB4, respectively.
The veto efficiency $\epsilon$ is measured to be $0.23 \pm 0.02$, $0.38 \pm 0.07$, $0.29 \pm 0.14$ for clusters in MB2, MB3, and MB4, respectively.
The statistical uncertainties in the measured veto efficiencies and the number of excess events in the inverted region are propagated as an additional systematic uncertainty on the background prediction.

The inner DT station (MB1 or MB1 plus MB2) hit veto efficiency is measured in a separate punch-through jet-enriched region, by selecting clusters that have a jet with $\pt > 10\GeV$ within $\Delta R < 0.4$.
The measured veto efficiency is fitted as a function of the matched jet \pt with the sum of an exponential and constant function for clusters from each station separately.
The veto efficiencies are then extrapolated by evaluating the fit function at 0\GeV to predict the veto efficiencies for clusters passing the jet veto.

The punch-through jet background prediction method is validated, as shown in Table~\ref{tab:singleDT_bkg_validation}, using a subset of the region with the inner DT station hit veto inverted by predicting background clusters that are matched to 2--5 MB1 or 2--5 MB2 hits, instead of $<$2 MB1 and $<$2 MB2 hits in the SR.

\begin{table}[h!]
  \centering
  \caption{
    Validation of the ABCD method for the single-DT-cluster category in a pileup-enriched region.
    The uncertainty in the prediction is the statistical uncertainty propagated from bins B, C, and D.
    Bins C and D for the MB3 and MB4 categories are combined to reduce the statistical uncertainty in the two regions.
    The final ABCD fit in the SR will also be performed with those bins combined.
  }

  \label{tab:dt_bkg_validation_abcd}
  \centering
  \begin{scotch}{l c c c D{+}{\,\pm\,}{3,3} c}
    Cluster station & $N_\mathrm{B}$ & $N_\mathrm{C}$      & $N_\mathrm{D}$      & \multicolumn{1}{c}{$\lambda_\mathrm{A}$} & $N_\mathrm{A}$ \\
    \hline
    MB2             & 2              & 130                 & 82                  & 1.3+0.9        & 3              \\
    MB3             & 1              & \multirow{2}{*}{\phantom{1}20} & \multirow{2}{*}{11} & 0.6+1.0        & 1              \\
    MB4             & 0              &                     &                     & 0.0+1.1        & 1              \\
  \end{scotch}
\end{table}

\begin{table}[h!]
  \centering
  \caption{
    Validation of the punch-through jet background prediction method for the single-DT-cluster category.
    The uncertainty in the prediction is the statistical uncertainty propagated from the extrapolated MB1/MB2 hit veto efficiency.
  }
  \label{tab:singleDT_bkg_validation}
  \begin{scotch}{l D{+}{\,\pm\,}{3,3} c}
    Cluster station & \multicolumn{1}{c}{$\lambda_\mathrm{A}$} & $N_\mathrm{A}$ \\
    \hline
    MB2             & 4.7+1.5        & 6              \\
    MB3             & 1.5+0.9        & 0              \\
    MB4             & 1.0+0.9        & 0              \\
  \end{scotch}
\end{table}

\section{Systematic uncertainties}
\label{sec:systematics}

The main source of background uncertainty is the statistical uncertainty of the ABCD method prediction from the limited data sample size of the background-enriched bins.
For the single-DT-cluster category, there is an additional systematic uncertainty for the punch-through jet background amounting to 32, 50, and 100\% for clusters in MB2, MB3, and MB4, respectively.
No additional background systematic uncertainties are assigned for the background predicted by the ABCD method because the ABCD background estimation method is validated, as detailed in Section~\ref{sec:background}. Background statistical uncertainty from the ABCD method is propagated to the SR prediction.

The dominant source of uncertainty in the signal prediction is missing higher-order QCD corrections that affect the Higgs boson \pt distribution.
This uncertainty is estimated through renormalization and factorization scale variations and amounts to 21\% for the gluon fusion production for all signal samples.
The other main sources of uncertainty include the signal modeling of the cluster reconstruction and selection criteria, as detailed in the following paragraphs, jet energy scale (3--6\%)~\cite{CMS:2016lmd}, PDFs (3\%), pileup modeling (2\%), integrated luminosity (1.6\%)~\cite{CMS-LUM-17-003,REFLUMI17,REFLUMI18}, \ptmiss trigger efficiency (5\% downward correction and 1\% uncertainty), and simulation sample size (3--5\%).

We studied the simulation modeling of the cluster reconstruction efficiency, cluster-level selections, and veto efficiencies to evaluate their effect on the signal prediction.
The accuracy of the simulation prediction for the cluster reconstruction efficiency relies on its ability to correctly model the response of the muon detectors in an environment with a large multiplicity of secondary particles.
This aspect is validated by measuring clusters produced in $\PZ\to\MM$ data events, in which one of the muons undergoes bremsstrahlung in the muon detectors and the associated photon produces an electromagnetic shower.
The discrepancy between data and simulation is taken as an additional systematic uncertainty in the cluster reconstruction and selection efficiencies.
The modeling of the veto efficiencies, including the jet, muon, ME1, MB1, and RPC hit vetoes, is determined from the simulation of jets and muons, the presence of pileup particles, and random noise.
The veto efficiencies are measured in data by randomly sampling the $(\eta, \phi)$ locations of clusters from the signal distribution and evaluating whether a jet, muon, or ME1, MB1, or RPC hit from $\PZ\to\MM$ events has been observed within $\Delta R<0.4$ of the cluster's location.
Differences between the measured efficiencies in data and in MC simulation are propagated as a systematic uncertainty.

For CSC clusters in the single-CSC-cluster category, the systematic uncertainty is dominated by the cluster reconstruction efficiency and the cluster identification efficiency.
This uncertainty is determined by measuring the difference in efficiencies in simulation and data, which amounts to an 8\% relative uncertainty.
Additionally, in signal simulations, the CSC hits are always assumed to be read out, while in the actual data acquisition, only those hits in a chamber that has at least three cathode hits and at least four anode hits at different CSC layers and match predefined hit patterns are read out.
This could lead to an underestimation of the efficiency of ME1/1 or ME1/2 vetoes in simulation, which were estimated to have an uncertainty of 1\%.

The signal loss from the vetoes is dominated by the muon veto.
In addition to the impact of muons from pileup, there are inefficiencies resulting from reconstructed muon segments produced by particles resulting from the LLP decay itself.
The simulation modeling of this effect is verified using a data control sample of clusters matched to trackless jets made to resemble the signal LLP decay by requiring the neutral energy fraction to be larger than 95\%.
A 10\% downward correction is applied to the signal efficiency to account for the simulation's mismodeling of the vetoes.

For CSC clusters in the double-cluster category, looser selections are applied, resulting in the systematic uncertainty being dominated by the cluster time spread requirement, which amounts to 10\% in the CSC-CSC category and 5\% in the DT-CSC category.
Furthermore, in the double-cluster category, the jet and muon vetoes are implemented with tighter identification criteria.
As a result, the effect of the presence of jets and muons is well modeled and no corrections and uncertainty are assigned.

For the DT clusters in the single-DT-cluster category, the systematic uncertainty is dominated by the cluster reconstruction efficiency, which is measured to be 15\%.
The MB1 and MB2 veto efficiencies are also measured separately by randomly sampling the locations of the clusters from the signal distribution and evaluating whether an MB1 or MB2 hit has been matched.
No correction or uncertainty is assigned for the MB2 veto.
A 10\% downward correction with a 7\% uncertainty is applied to the signal efficiency to account for additional noise MB1 hits in data.

For the DT clusters in the double-cluster category, the signal systematic uncertainty is dominated by the cluster reconstruction efficiency, which is measured to be 3\% and 1\% in the DT-DT and DT-CSC categories, respectively.

\section{Results and interpretation}
\label{sec:results}

In this section, we present the search results in each category and the interpretation of the combined results of all three categories in the twin Higgs and dark-shower models.

\subsection{Double-cluster category}

\begin{figure*}[h!]
  \centering
  \includegraphics[width=0.45\textwidth]{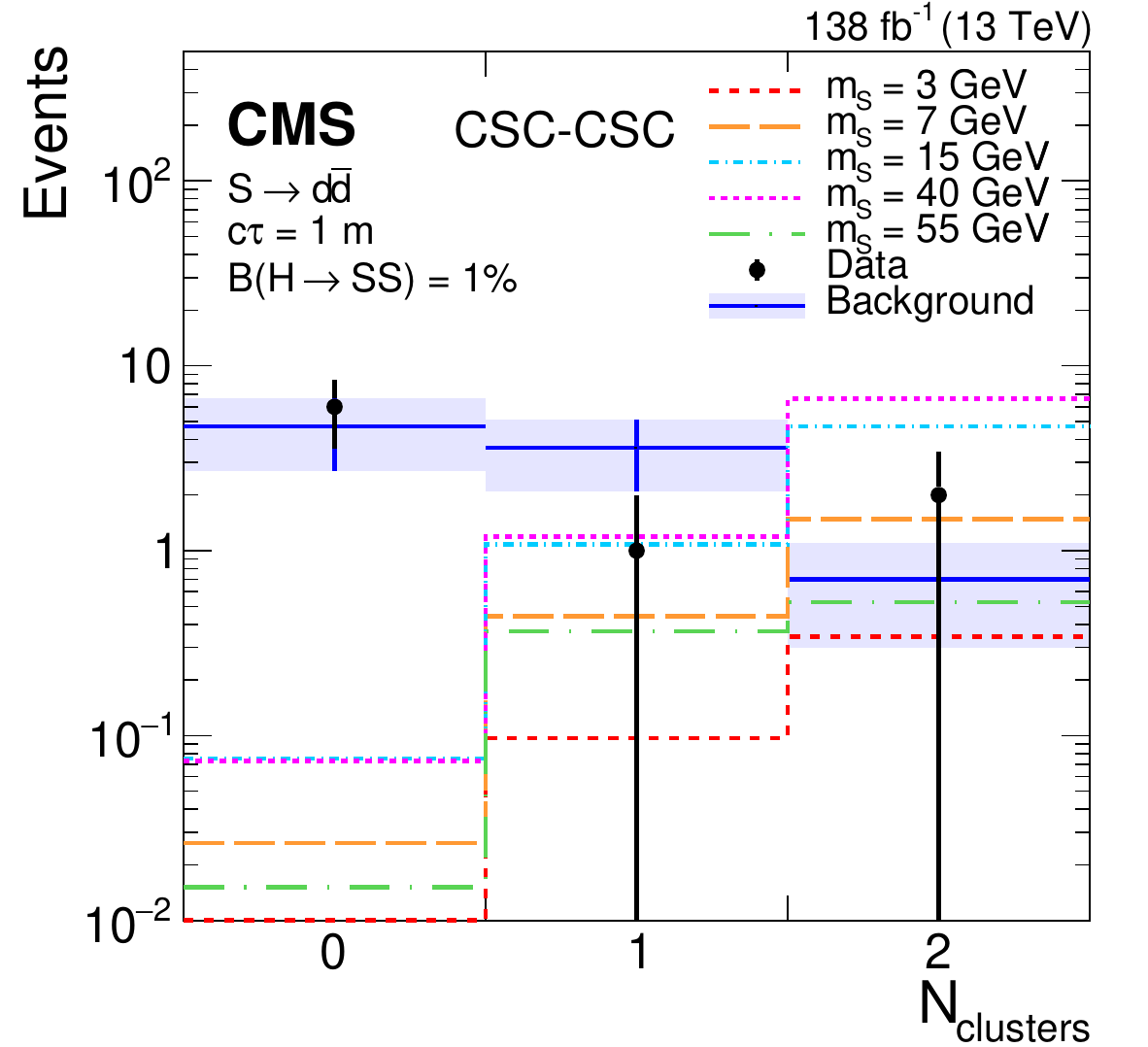}
  \includegraphics[width=0.45\textwidth]{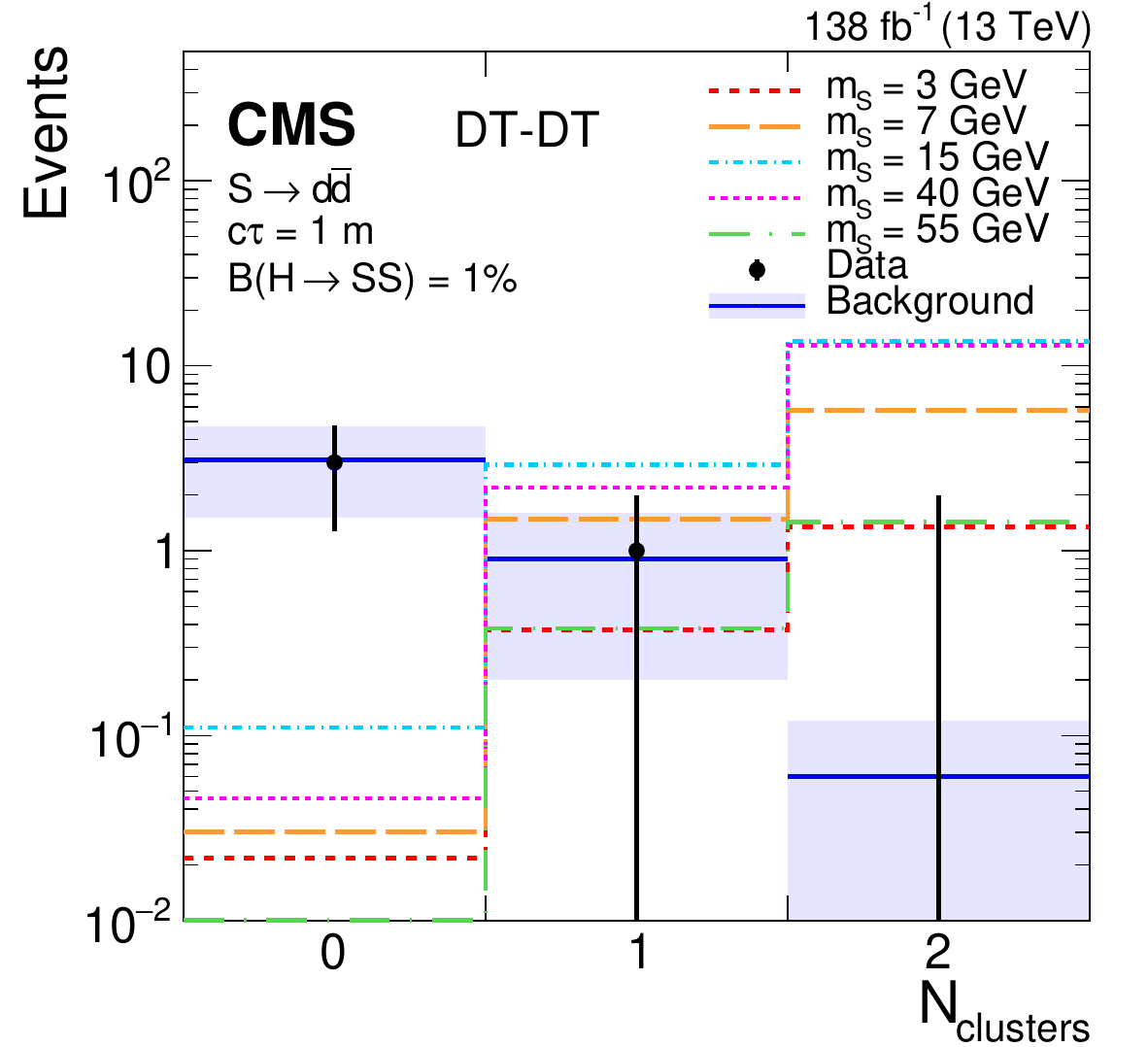}
  \includegraphics[width=0.45\textwidth]{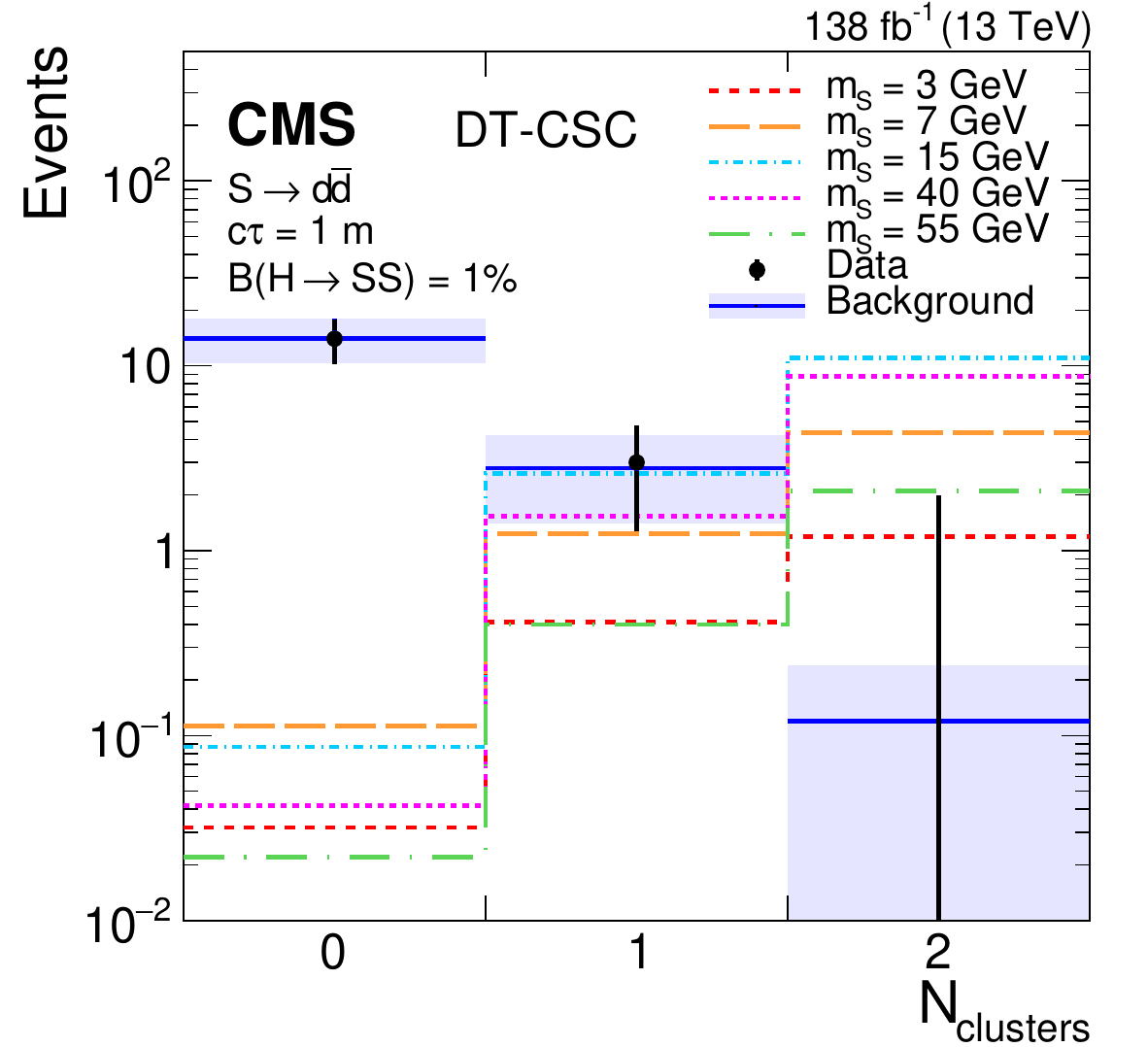}
  \caption{
    The signal (assuming $\brhtoss = 1\%$, $\PS \to \ddbar$, and $c\tau=1\unit{m}$), background, and data distributions of \nclusters passing the \nhits selection in the search region for CSC-CSC (upper left), DT-DT (upper right), and DT-CSC (lower) categories.
    The branching fraction \brhtoss has been set to a specific value purely to allow a visual comparison of the shapes of the signal and background distributions.
    The background prediction is obtained from the fit to the observed data assuming no signal contribution, and is shown in blue with the shaded region showing the fitted uncertainty.
    No events are observed in the rightmost bins of the DT-CSC and DT-DT categories, so only the associated uncertainty is shown.
  }
  \label{fig:double_result_nclusters}
\end{figure*}

The results of the search in the double-cluster category are shown in Table~\ref{tab:double_result}.
We observed no statistically significant deviation with respect to the SM prediction.
The signal and data distributions of \nclusters passing the \nhits selection are shown in Fig.~\ref{fig:double_result_nclusters}.

\begin{table*}[!h]
  \centering
  \caption{
    Number of predicted background and observed events in the double-cluster category.
    The background prediction is obtained from a fit to the observed data assuming no signal contribution.
  }
  \label{tab:double_result}
  \begin{scotch}{l l D{+}{\,\pm\,}{3,3} D{+}{\,\pm\,}{3,3} D{+}{\,\pm\,}{3,3} D{+}{\,\pm\,}{3,3} D{+}{\,\pm\,}{3,3}}
    & Category & \multicolumn{1}{c}{A}             & \multicolumn{1}{c}{B}             & \multicolumn{1}{c}{C}              & \multicolumn{1}{c}{D}             & \multicolumn{1}{c}{BD}            \\
    \hline
    \multirow{3}{*}{Background-only fit} & DT-DT    & 0.06+0.06 & \multicolumn{1}{c}{\NA}           & 3.1+1.6  & \multicolumn{1}{c}{\NA}           & 0.9+0.7   \\
    & CSC-CSC  & 0.7+0.4 & \multicolumn{1}{c}{\NA}           & 4.7+2.0  & \multicolumn{1}{c}{\NA}           & 3.6+1.5 \\
    & DT-CSC   & 0.12+0.12 & 1.9+1.2 & 14.1+3.8 & 0.9+0.7 & \multicolumn{1}{c}{\NA}           \\
    [\cmsTabSkip]
    \multirow{3}{*}{Observation}         & DT-DT    & \multicolumn{1}{c}{0}      & \multicolumn{1}{c}{\NA}           & \multicolumn{1}{c}{3}              & \multicolumn{1}{c}{\NA}           & \multicolumn{1}{c}{1}             \\
    & CSC-CSC  & \multicolumn{1}{c}{2}  & \multicolumn{1}{c}{\NA}           & \multicolumn{1}{c}{6}              & \multicolumn{1}{c}{\NA}           & \multicolumn{1}{c}{1}             \\
    & DT-CSC   & \multicolumn{1}{c}{0}  & \multicolumn{1}{c}{2}             & \multicolumn{1}{c}{14}             & \multicolumn{1}{c}{1}             & \multicolumn{1}{c}{\NA}           \\
  \end{scotch}
\end{table*}

\subsection{Single-CSC-cluster category}

\begin{figure*}[h!]
  \centering
  \includegraphics[width=0.45\textwidth]{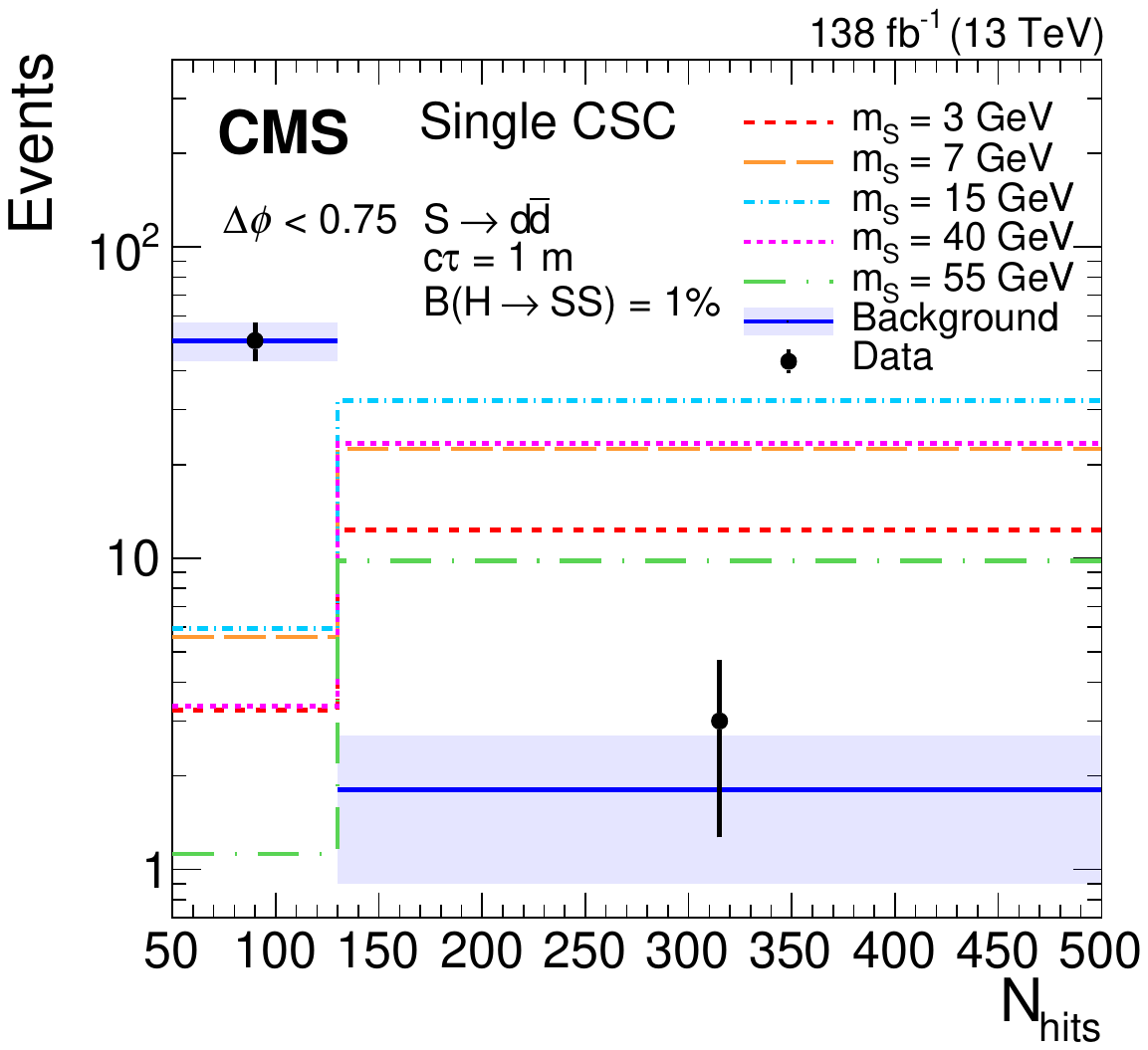}
  \includegraphics[width=0.45\textwidth]{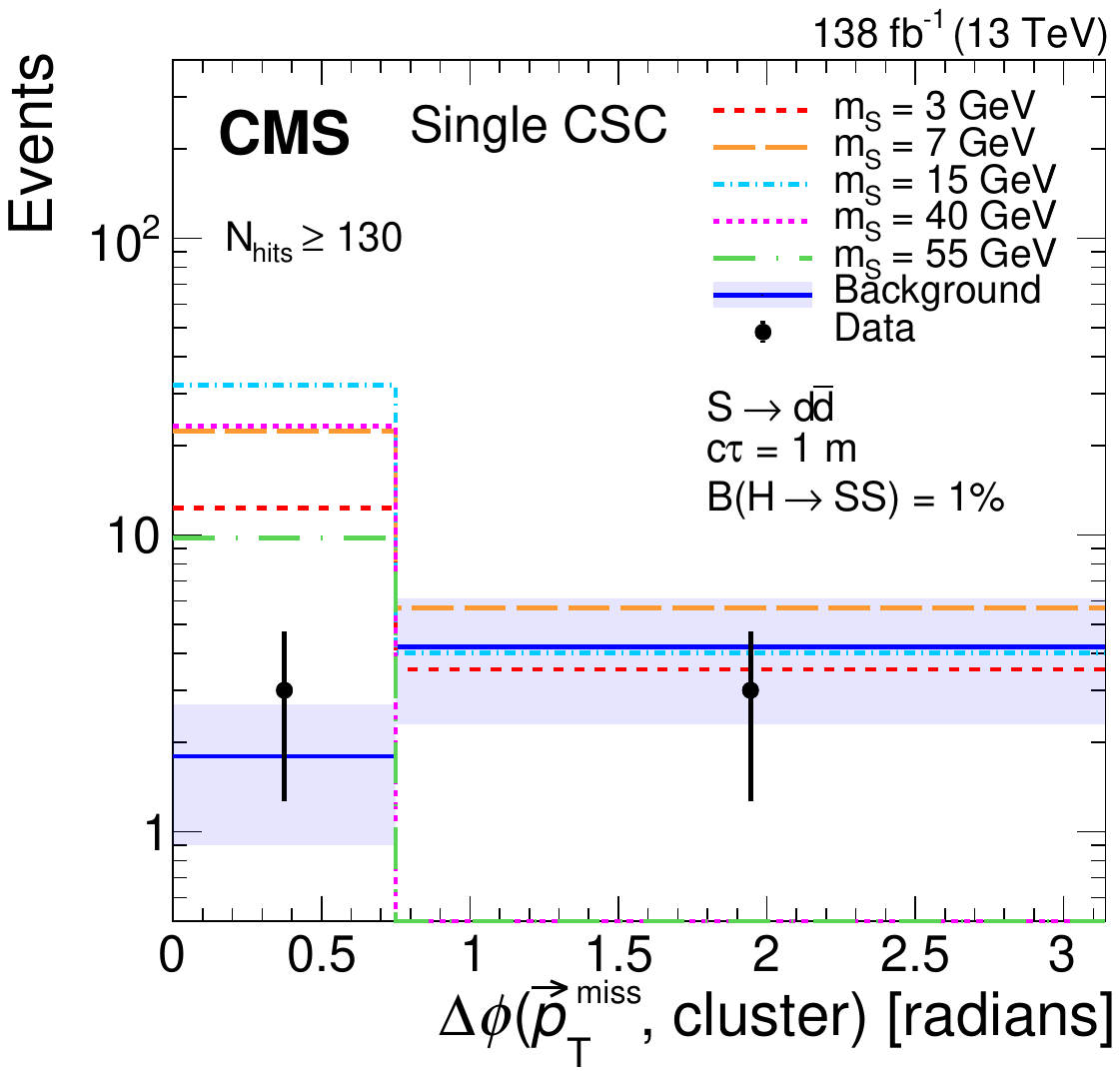}
  \caption{
    Distributions of \nhits (\cmsLeft) and \deltaphi (\cmsRight) in the search region of the single-CSC-cluster category.
    The background prediction is obtained from the fit to the observed data assuming no signal contribution, and is shown in blue with the shaded region showing the fitted uncertainty.
    The expected signal with $\brhtoss = 1\%$, $\PS\to\ddbar$, and $c\tau=1\unit{m}$ is shown for $m_\PS$ of 3, 7, 15, 40, and 55\GeV in various colors and dotted lines.
    The \nhits distribution includes only events in bins A and D, while the \deltaphi distribution includes only events in bins A and B.
    The rightmost bin in the \nhits distribution includes overflow events.
  }

  \label{fig:csc_result_nrechits_dphi}
\end{figure*}

The corresponding search result in the single-CSC-cluster category is shown in Table~\ref{tab:csc_result}.
No statistically significant deviation with respect to the SM prediction is observed.
The signal and data distributions of \nhits in bins A and D, and \deltaphi in bins A and B are shown in Fig.~\ref{fig:csc_result_nrechits_dphi}.

\begin{table}[!h]
  \centering
  \caption{
    Number of predicted background and observed events in the single-CSC-cluster category.
    The background prediction is obtained from a fit to the observed data assuming no signal contribution.
  }
  \label{tab:csc_result}
  \begin{scotch}{l D{+}{\,\pm\,}{3,3} D{+}{\,\pm\,}{3,3} D{+}{\,\pm\,}{3,3} D{+}{\,\pm\,}{3,3}}
    & \multicolumn{1}{c}{A}             & \multicolumn{1}{c}{B}             & \multicolumn{1}{c}{C}            & \multicolumn{1}{c}{D}          \\
    \hline
    Background-only fit & 1.8+0.8 & 4.2+1.7 & 120+11 & 51+7 \\
    [\cmsTabSkip]
    Observed            & \multicolumn{1}{c}{3}  & \multicolumn{1}{c}{3} & \multicolumn{1}{c}{121} &       \multicolumn{1}{c}{50}             \\
  \end{scotch}
\end{table}

\subsection{Single-DT-cluster category}

\begin{figure*}[h!]
  \centering
  \includegraphics[width=0.45\textwidth]{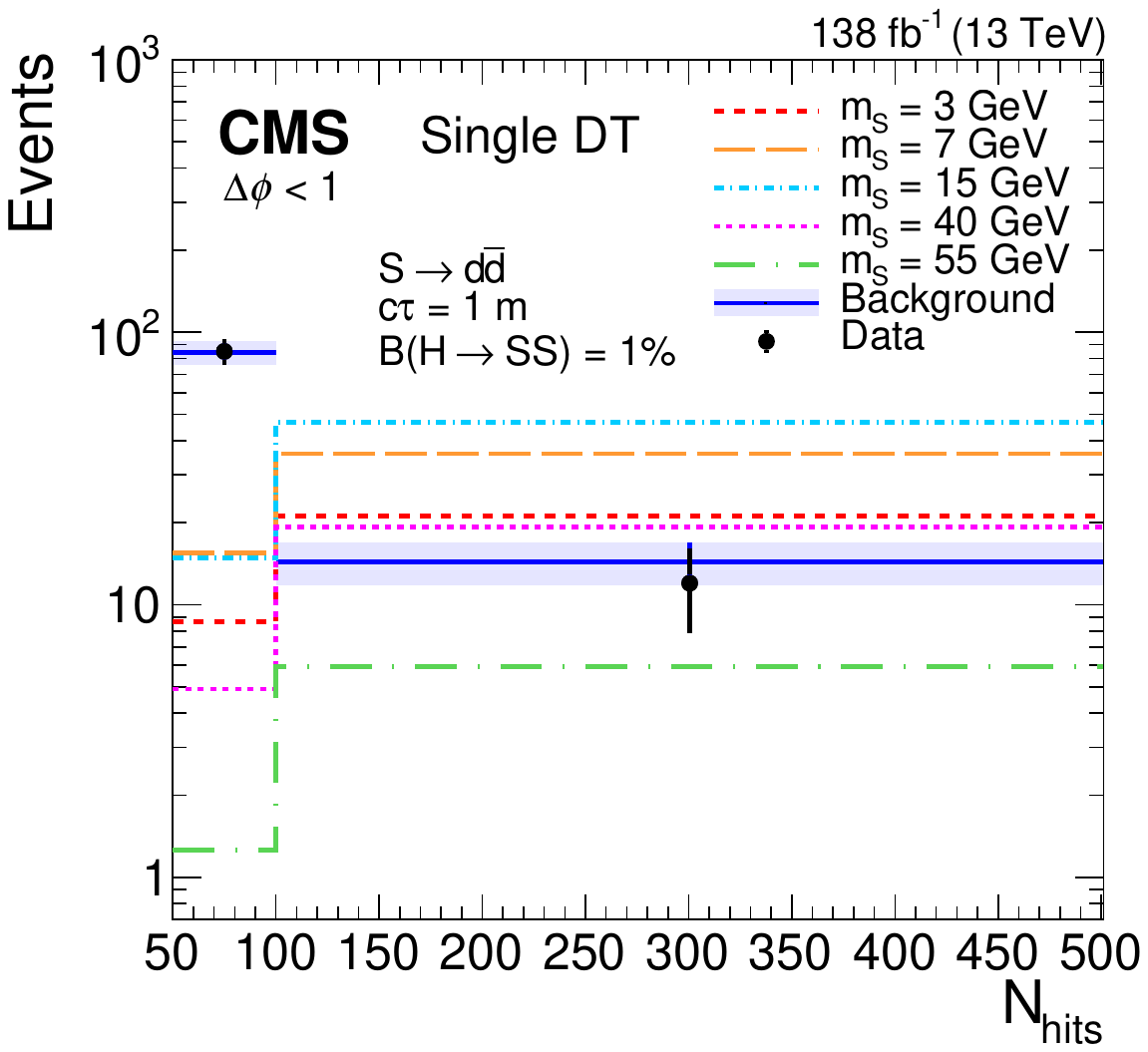}
  \includegraphics[width=0.45\textwidth]{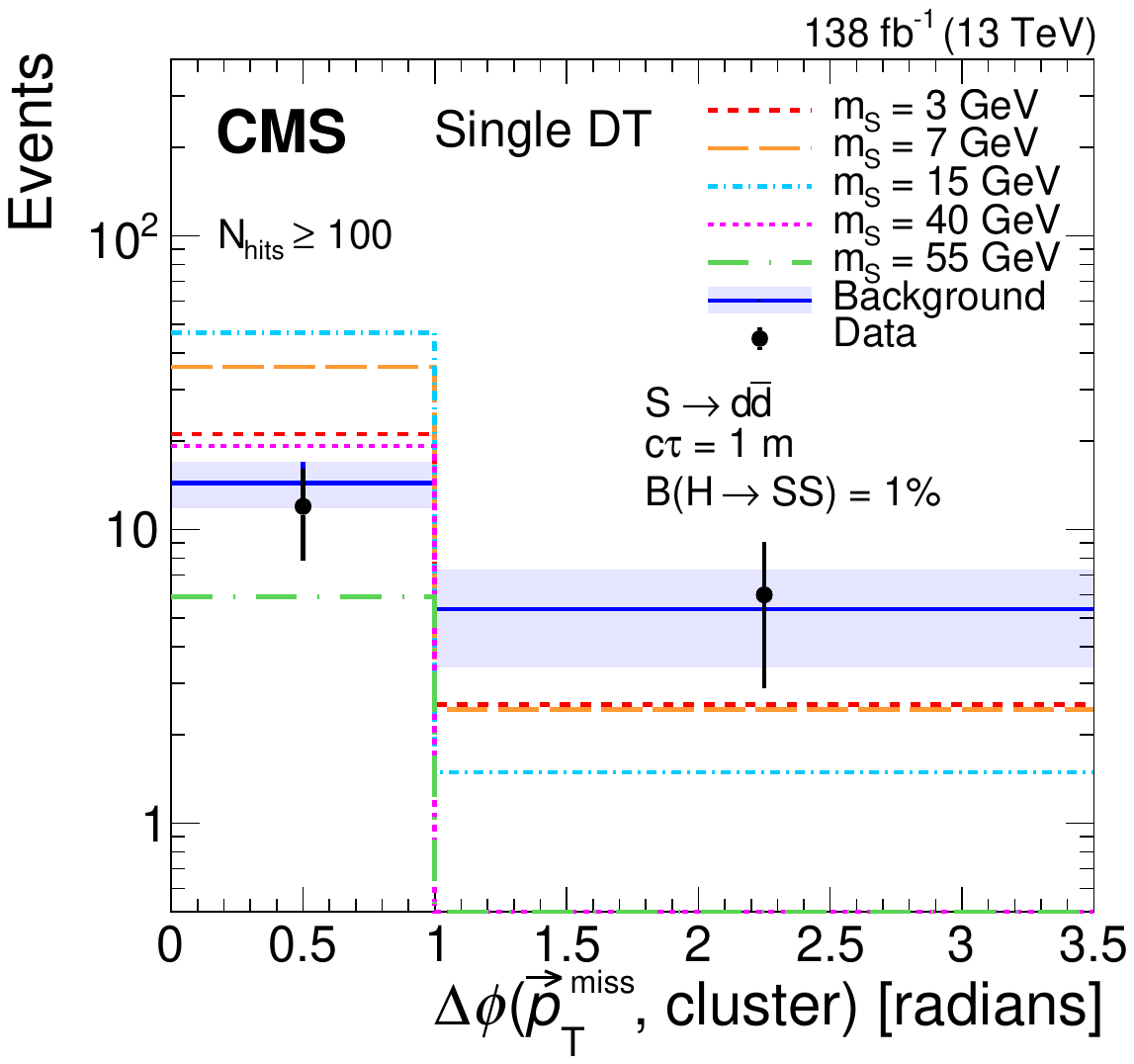}
  \caption{
    Distributions of \nhits (\cmsLeft) and \deltaphi (\cmsRight) in the search region of the single-DT-cluster category.
    The background prediction is obtained from the fit to the observed data assuming no signal contribution, and is shown in blue with the shaded region showing the fitted uncertainty.
    The expected signal with $\brhtoss = 1\%$, $\PS\to\ddbar$, and $c\tau=1\unit{m}$ is shown for $m_\PS$ of 3, 7, 15, 40, and 55\GeV in various colors and dotted lines.
    The \nhits distribution includes only events in bins A and D, while the \deltaphi one includes only events in bins A and B.
    The right-hand bin in the \nhits distribution includes overflow events.
  }
  \label{fig:dt_result_nrechits_dphi}
\end{figure*}

Using the methods detailed in Section~\ref{sec:background}, the background from punch-through jets and low-\pt pileup particles are estimated in the single-DT-cluster category.
The result of the search is shown in Table~\ref{tab:dt_result}.
We observed no statistically significant deviation with respect to the SM prediction.
The signal and data distributions of \nhits in bins A and D, and \deltaphi in bins A and B are shown in Fig.~\ref{fig:dt_result_nrechits_dphi}.

\begin{table*}[!h]
  \centering
  \caption{
    Number of predicted background and observed events in the single-DT-cluster category.
    The background prediction is obtained from a fit to the observed data assuming no signal contribution.
  }
  \label{tab:dt_result}
  \resizebox{\textwidth}{!}{
    \begin{scotch}{l l D{+}{\,\pm\,}{3,3} D{+}{\,\pm\,}{3,3} D{+}{\,\pm\,}{3,3} c c c}
      & \multirow{2}{*}{Category} & \multicolumn{1}{c}{A}             & \multicolumn{1}{c}{A}               & \multicolumn{1}{c}{A}             & \multirow{2}{*}{B} & \multirow{2}{*}{C}             & \multirow{2}{*}{D}             \\
      &                           & \multicolumn{1}{c}{(total)}       & \multicolumn{1}{c}{(punch-through)} & \multicolumn{1}{c}{(ABCD pred.)}  &                    &                                &                                \\
      \hline
      \multirow{3}{*}{Background-only fit}
      & MB2                       & 9.5+1.9 & 6.3+1.7   & 3.1+1.3 & 4.8\,$\pm$\,1.9      & 119\,$\pm$\,12               & 76.8\,$\pm$\,8.1                 \\
      & MB3                       & 3.7+1.5 & 3.1+1.1   & 0.6+1.1 & 0.5\,$\pm$\,0.5      & \multirow{2}{*}{\phantom{11}7\,$\pm$\,3\phantom{1}} & \multirow{2}{*}{\phantom{1}7.5\,$\pm$\,2.6} \\
      & MB4                       & 1.2+0.9 & 1.2+0.9   & 0.1+0.5 & 0.06\,$\pm$\,0.22    &                                &                                \\
      [\cmsTabSkip]
      \multirow{3}{*}{Observation}
      & MB2                       & 9             & \NA             & \NA           & \phantom{1.1}5\phantom{\,$\pm$\,1.11}                  & \phantom{1.}119\phantom{\,$\pm$\,11.1}                            & \phantom{1.}77\phantom{\,$\pm$\,11.1}                             \\
      & MB3                       & 1             & \NA             & \NA           & \phantom{1.1}1\phantom{\,$\pm$\,1.11}                   & \multirow{2}{*}{\phantom{111.}6\phantom{\,$\pm$\,11.1}}             & \multirow{2}{*}{\phantom{11.}8\phantom{\,$\pm$\,11.1}}             \\
      & MB4                       & 2             & \NA             & \NA           & \phantom{1.1}0\phantom{\,$\pm$\,1.11}                   &                                & \\
    \end{scotch}
  }
\end{table*}

\subsection{Interpretations}

In this section, we present the combination of the double-cluster, single-CSC-cluster, and single-DT-cluster categories and the interpretations in the context of the twin Higgs and dark-shower models.
The ABCD regions of the three categories are defined to be mutually exclusive and, in the combination, we fit the ABCD plane of each category simultaneously.

For fully hadronic decays of the LLP and in the region of $c\tau$ with the largest double-cluster acceptance, the double-cluster category is the most sensitive.
This region of $c\tau$ depends on the LLP mass.
Because of the better sensitivity of the double-cluster category, and to minimize changes to the previously published single-CSC-cluster category, overlapping events are kept in only one category with the following order of precedence: double-cluster, single-CSC-cluster, and single-DT-cluster category.
For electromagnetic and $\PGt$ decays of the LLP, the single-CSC-cluster category is the most sensitive for shorter $c\tau$, while for large $c\tau$ the single-DT-cluster category is the most sensitive.

All theoretical uncertainties assigned to signal simulations are fully correlated across categories.
Experimental uncertainties that are not related to the cluster, such as luminosity, jet energy scale, PDFs, and pileup modeling uncertainty, are fully correlated across categories.
Experimental uncertainties associated with cluster selections are assumed to be fully uncorrelated across categories.
All uncertainties are incorporated into the analysis via nuisance parameters and treated according to the frequentist paradigm.

We evaluate 95\% confidence level (\CL) upper limits on the branching fractions $\brhtoss$ and $\brhtopsipsi$ for both the twin Higgs and dark-shower models by performing both a background-only fit, as shown in Tables~\ref{tab:double_result}--\ref{tab:dt_result}, and a signal plus background fit, and using the modified frequentist criterion \CLs~\cite{JUNK1999435,Read_2002} with the profile likelihood ratio test statistic~\cite{CMS-NOTE-2011-005}.
The procedure has been validated by injecting a signal and verifying that an unbiased estimation of the signal strength is obtained.
The Higgs boson production and its couplings to the SM particles are taken from SM predictions.

We show the predicted number of signal events for the twin Higgs model in a few benchmark decay, mass, and lifetime scenarios in Table~\ref{tab:signal_yield}, selected to be most sensitive and relevant for the search.

\begin{table*}[!h]
  \centering
  \caption{
    Expected number of signal events in bin A for each category, for a few benchmark signal models assuming $\brhtoss = 1\%$.
  }
  \label{tab:signal_yield}
  \renewcommand{\arraystretch}{1.3}
  \begin{scotch}{l r r r r r}
    LLP decay mode, mass                         & \multirow{2}{*}{CSC-CSC} & \multirow{2}{*}{DT-DT} & \multirow{2}{*}{DT-CSC} & \multirow{2}{*}{Single CSC} & \multirow{2}{*}{Single DT} \\
    proper decay length                          &                          &                        &                         &                             &                            \\

    \hline
    $\ddbar$, 3\GeV, $c\tau = 1\unit{m}$         & 0.3                      & 1.3                    & 1.2                     & 12.3                        & 21.2                       \\
    $\ddbar$, 7\GeV, $c\tau = 1\unit{m}$         & 1.5                      & 5.7                    & 4.3                     & 22.5                        & 35.8                       \\
    $\ddbar$, 15\GeV, $c\tau = 1\unit{m}$        & 4.7                      & 13.6                   & 11.1                    & 32.0                        & 46.8                       \\
    $\ddbar$, 40\GeV, $c\tau = 1\unit{m}$        & 6.6                      & 12.9                   & 8.8                     & 23.4                        & 19.3                       \\
    $\ddbar$, 55\GeV, $c\tau = 1\unit{m}$        & 0.5                      & 1.4                    & 2.1                     & 9.8                         & 5.9                        \\
    [\cmsTabSkip]
    $\PGtp\PGtm$, 7\GeV, $c\tau = 1\unit{m}$     & 0.6                      & 1.8                    & 1.6                     & 14.2                        & 22.5                       \\
    $\PGtp\PGtm$, 15\GeV, $c\tau = 1\unit{m}$    & 1.7                      & 5.2                    & 3.9                     & 20.1                        & 28.9                       \\
    $\PGtp\PGtm$, 40\GeV, $c\tau = 1\unit{m}$    & 3.3                      & 4.5                    & 3.3                     & 21.3                        & 17.0                       \\
    $\PGtp\PGtm$, 55\GeV, $c\tau = 1\unit{m}$    & 0.3                      & 0.9                    & 1.0                     & 10.6                        & 6.0                        \\
    [\cmsTabSkip]
    $\Pgpz\Pgpz$, 0.4\GeV, $c\tau = 0.1\unit{m}$ & 0.1                      & 0.4                    & 0.4                     & 6.8                         & 19.2                       \\
    $\Pgpz\Pgpz$, 1\GeV, $c\tau = 0.1\unit{m}$   & 0.4                      & 1.3                    & 1.1                     & 11.6                        & 30.7\\
  \end{scotch}
\end{table*}

\begin{figure*}[htb!]
  \centering
  \includegraphics[width=0.45\textwidth]{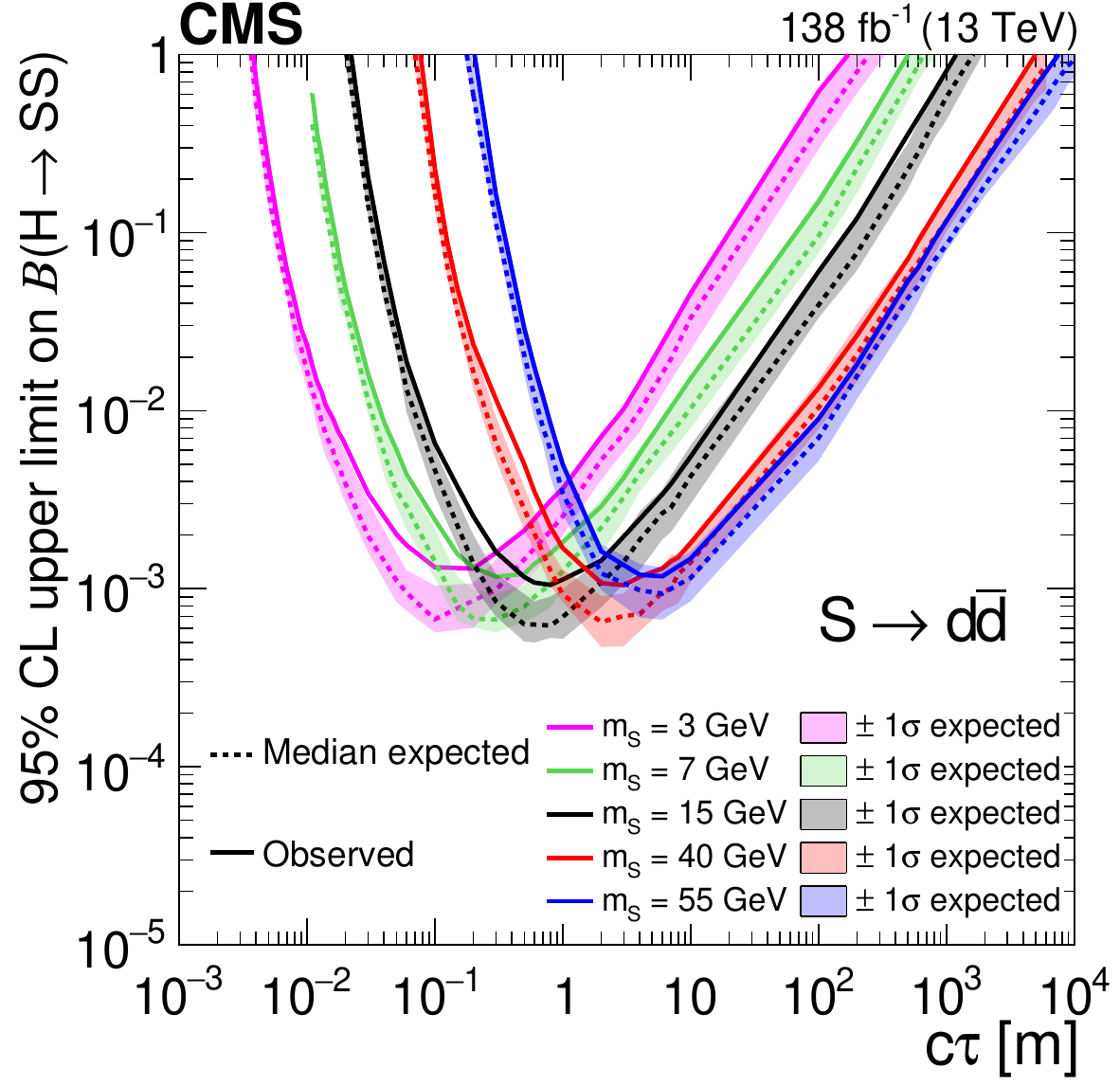}
  \includegraphics[width=0.45\textwidth]{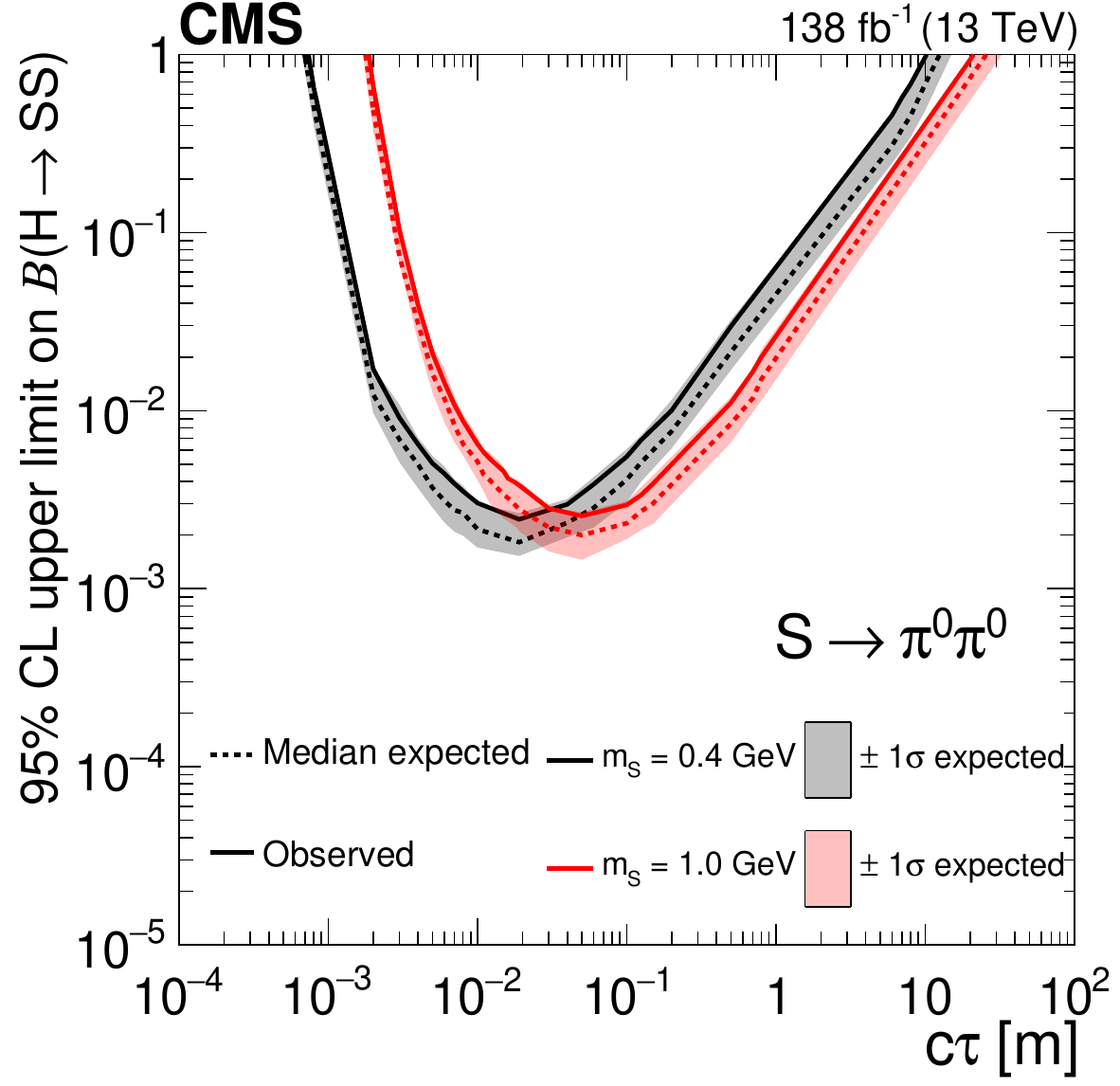}
  \includegraphics[width=0.45\textwidth]{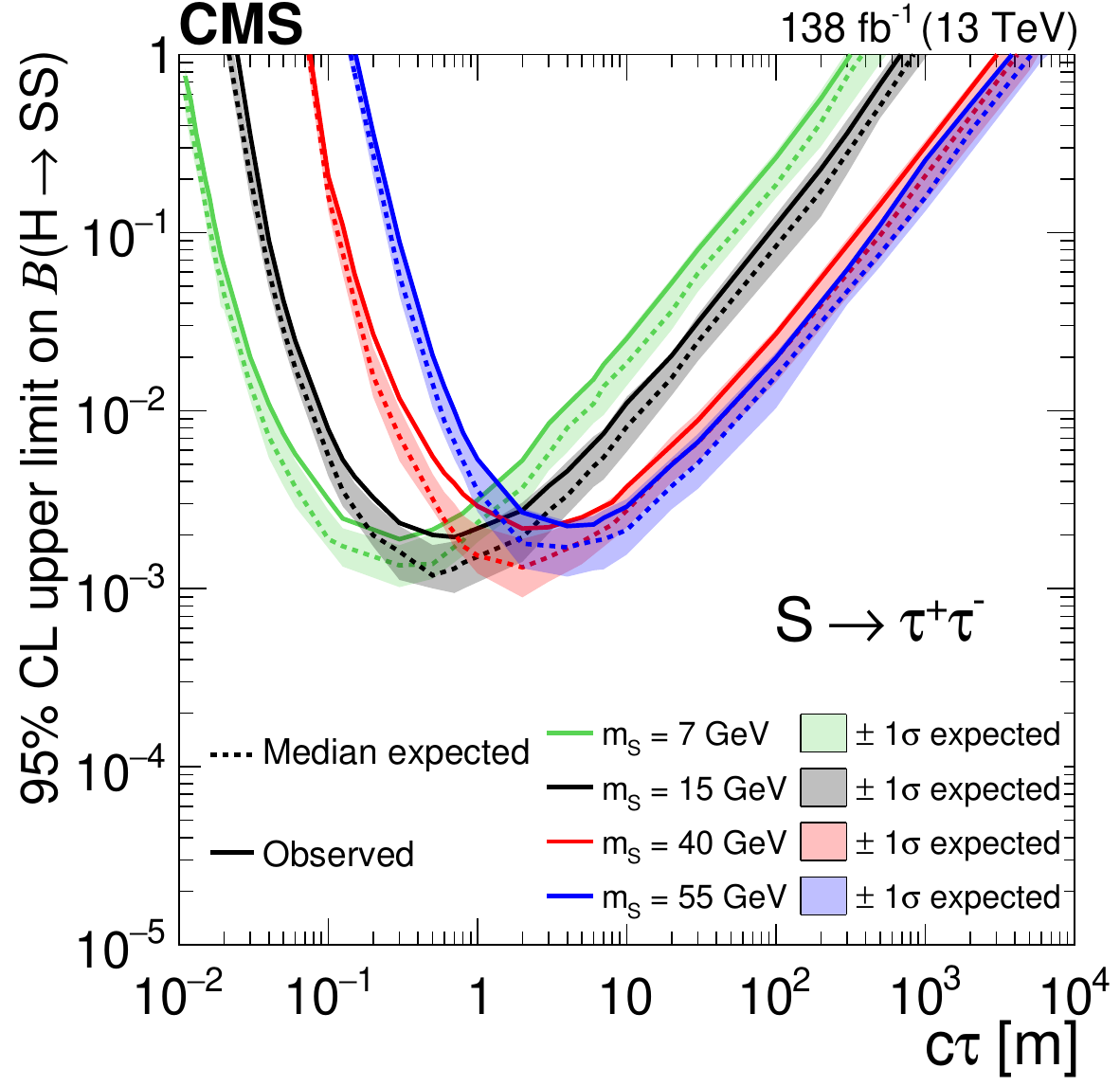}
  \caption{
    The 95\% \CL expected (dotted curves) and observed (solid curves) upper limits on the branching fraction $\brhtoss$ as functions of $c\tau$ for the $\PS\to\ddbar$ (upper left), $\PS\to\Pgpz\Pgpz$ (upper right), and $\PS\to\PGtp\PGtm$ (lower) decay modes.
    The exclusion limits are shown for different mass hypotheses.
  }
  \label{fig:twin_higgs_limits_1d}
\end{figure*}

\begin{figure*}[h!]
  \centering
  \includegraphics[width=0.45\textwidth]{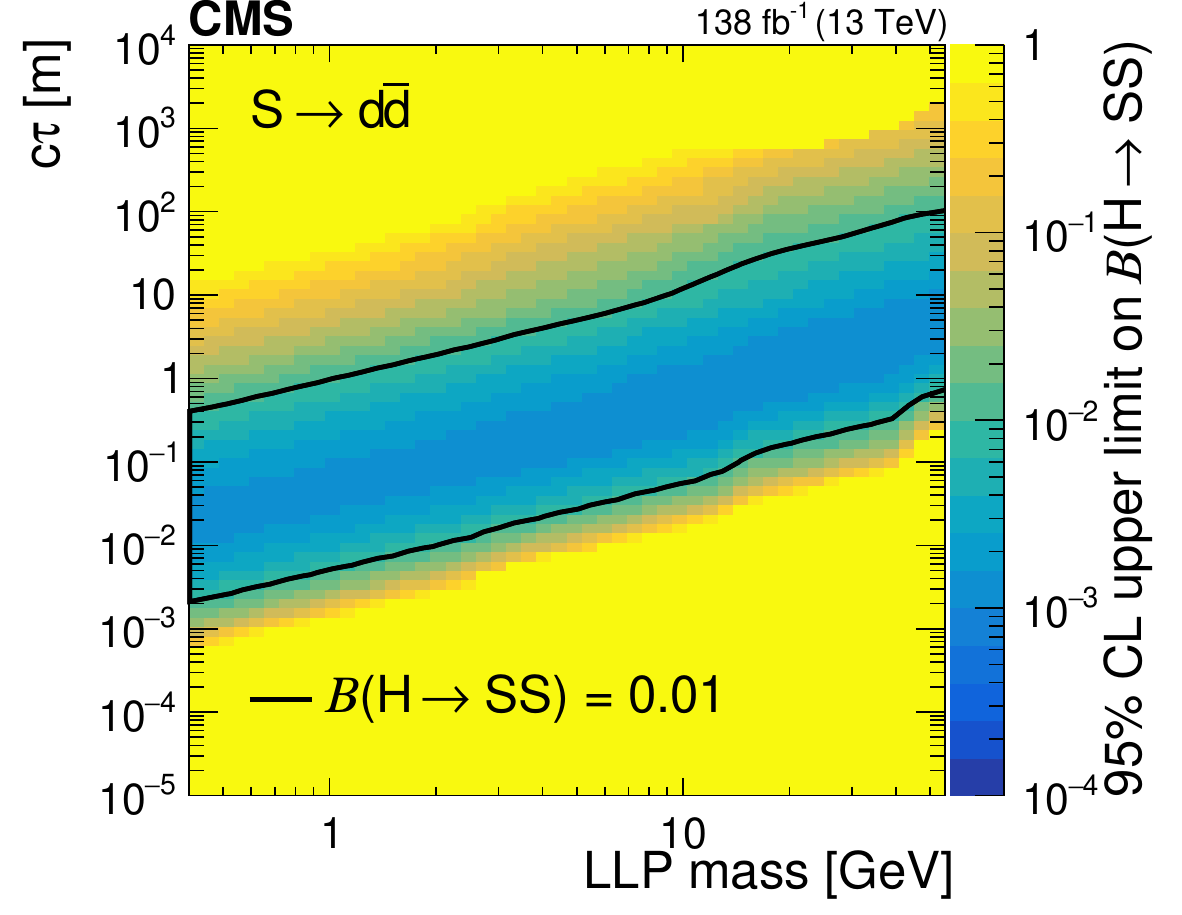}
  \includegraphics[width=0.45\textwidth]{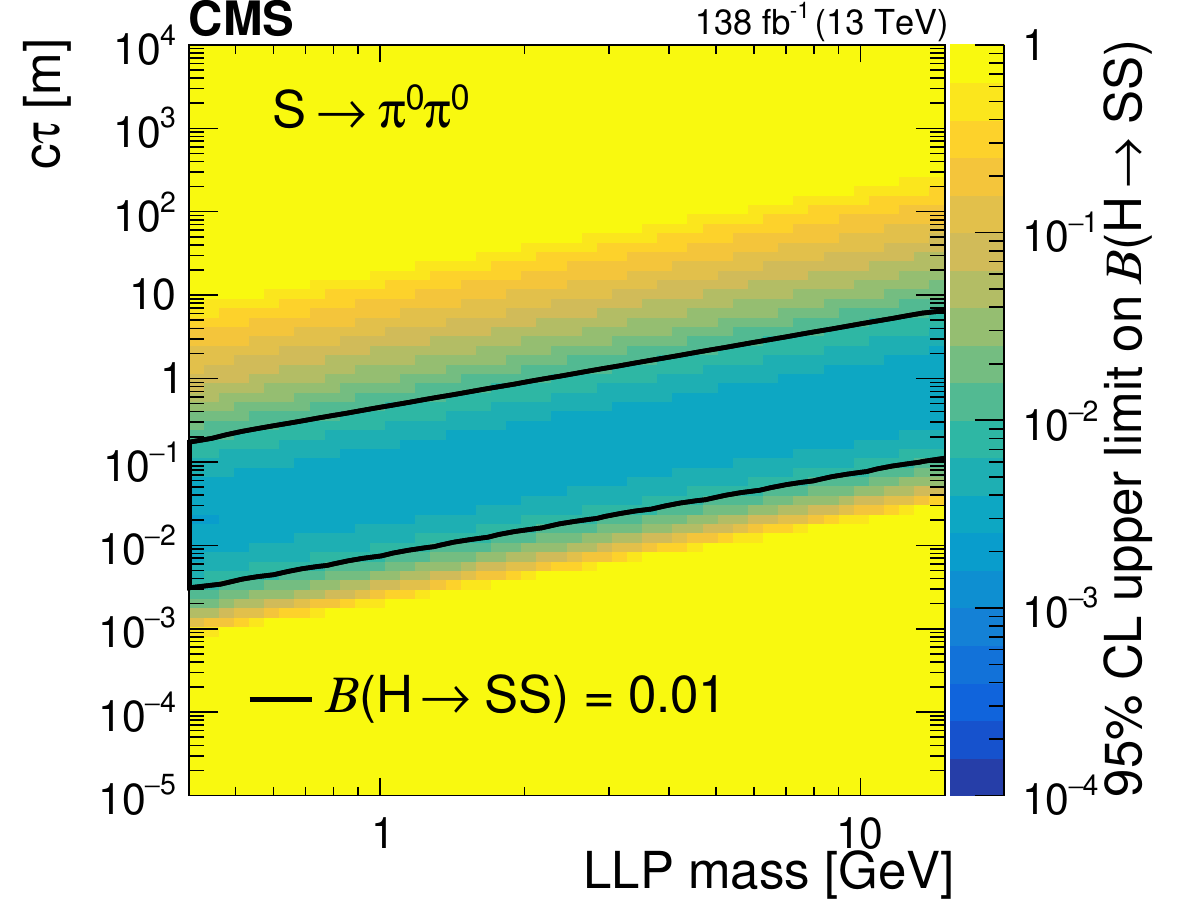}
  \includegraphics[width=0.45\textwidth]{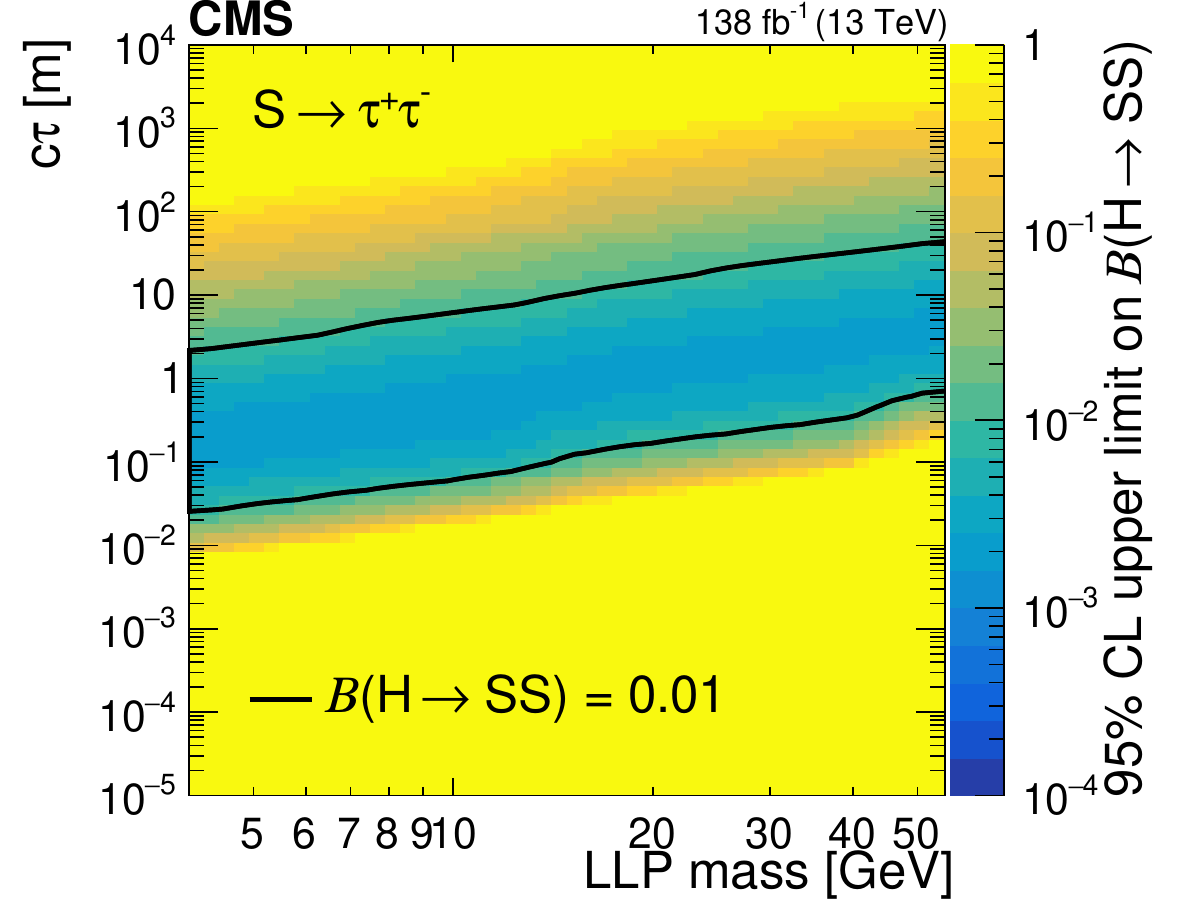}
  \caption{
    The 95\% \CL observed upper limits on the branching fraction $\brhtoss$ as functions of mass and $c\tau$ for the $\PS\to\ddbar$ (upper left), $\PS\to\Pgpz\Pgpz$ (upper right), and $\PS\to\PGtp\PGtm$ (lower) decay modes.
    The area inside the solid contours represents the region for which the limit is smaller than 0.01.
  }
  \label{fig:twin_higgs_limits_2d}
\end{figure*}

\begin{figure*}[h!]
  \centering
  \includegraphics[width=0.45\textwidth]{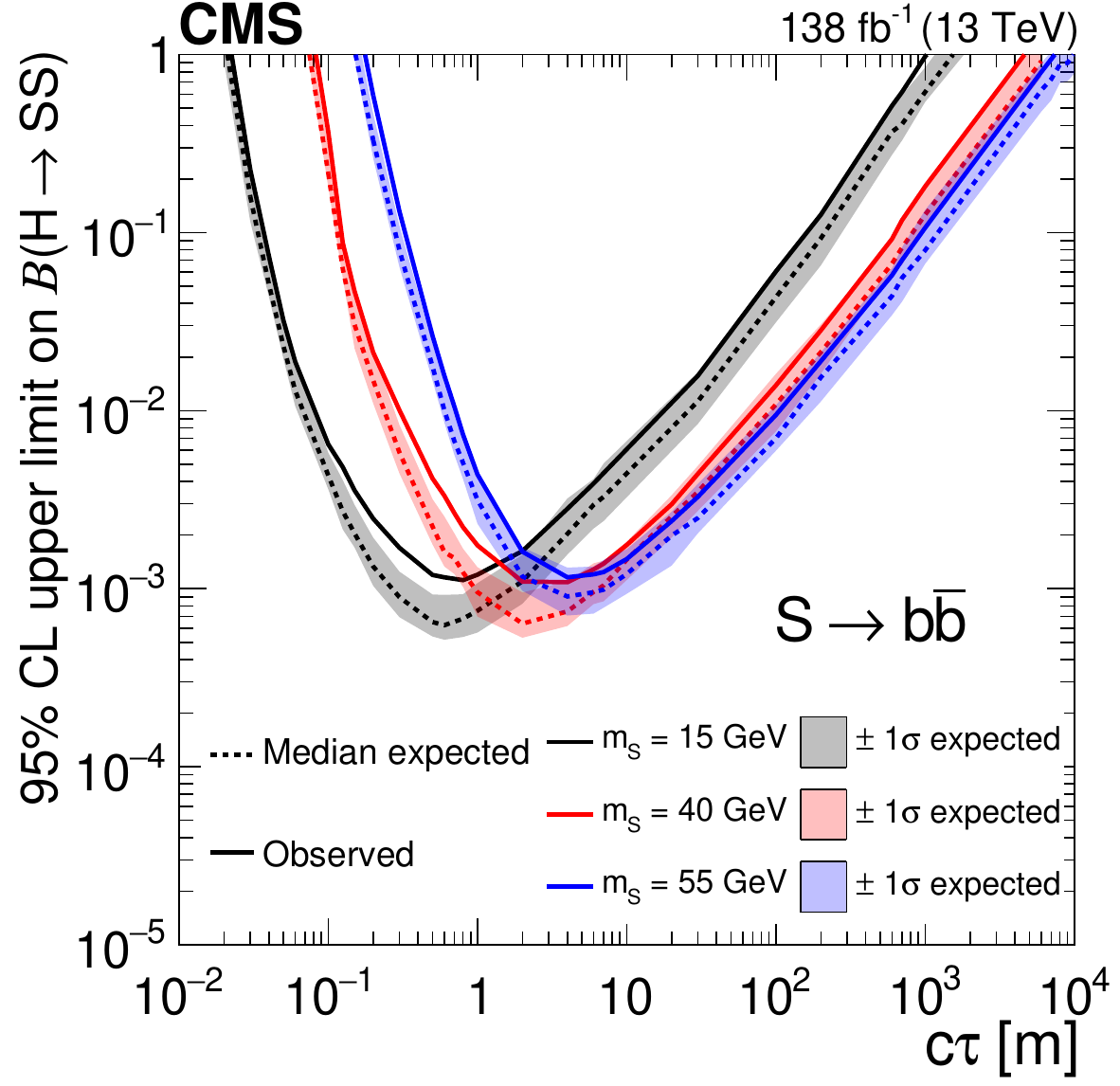}
  \includegraphics[width=0.45\textwidth]{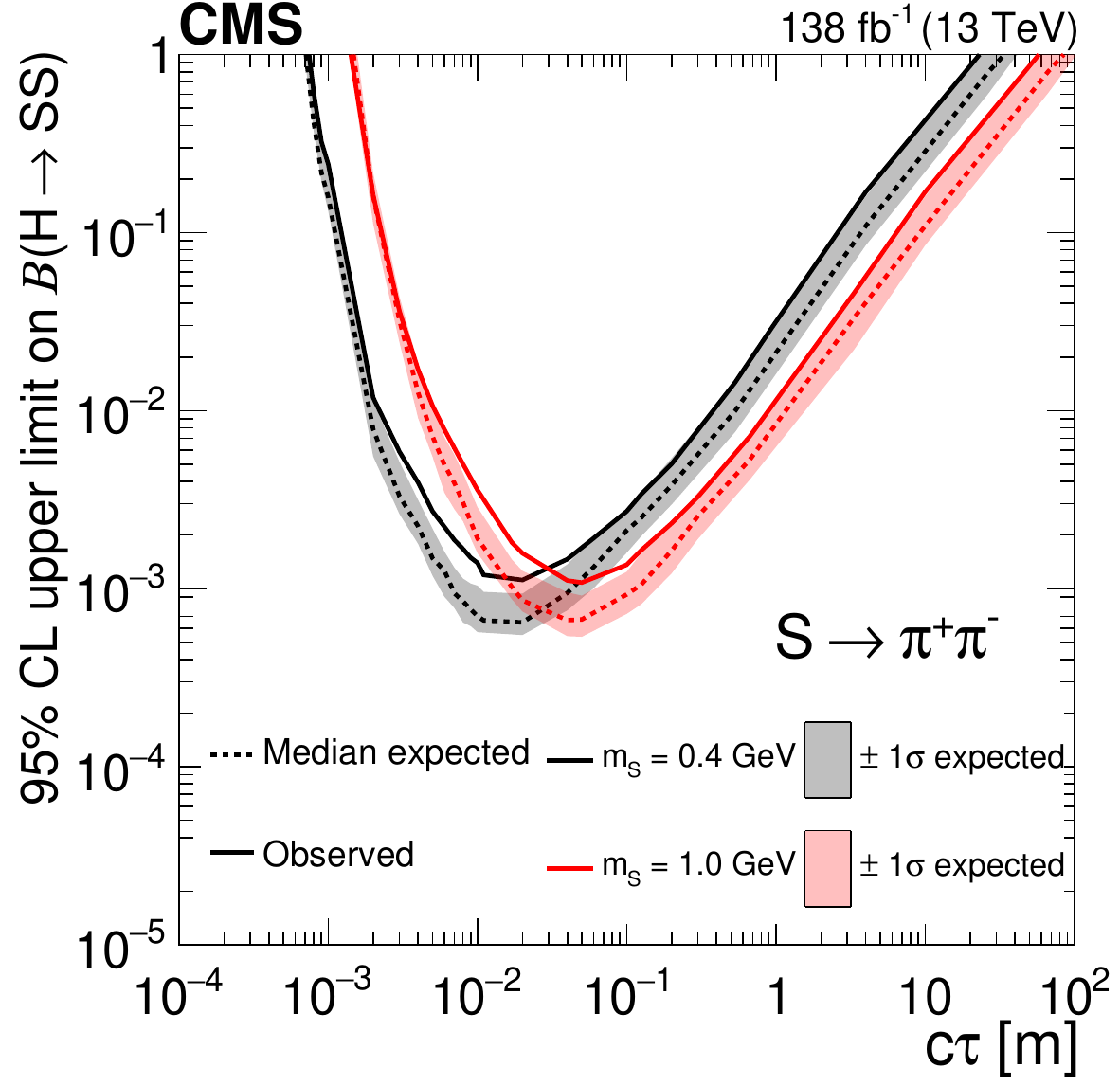}
  \includegraphics[width=0.45\textwidth]{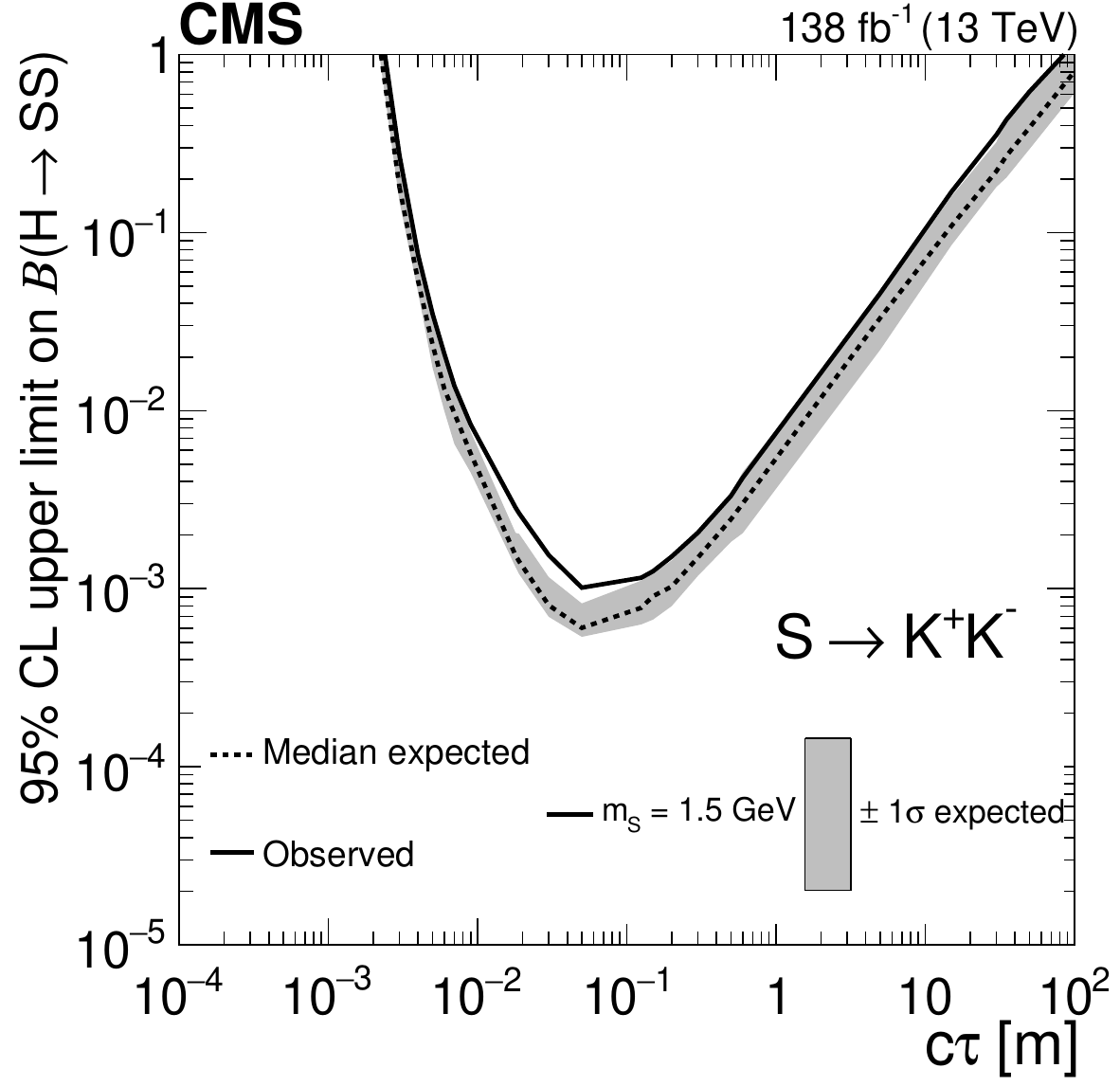}
  \includegraphics[width=0.45\textwidth]{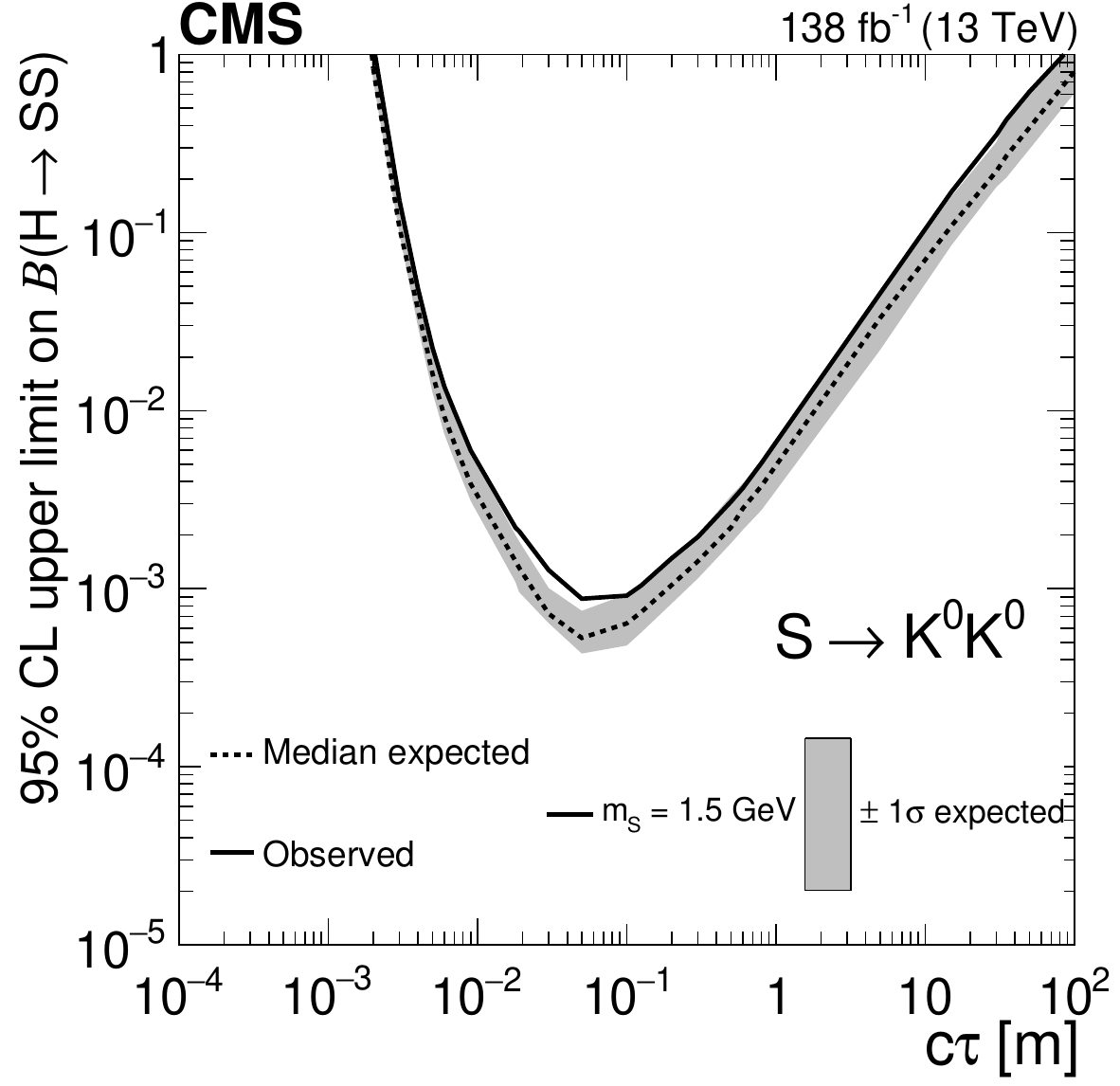}
  \includegraphics[width=0.45\textwidth]{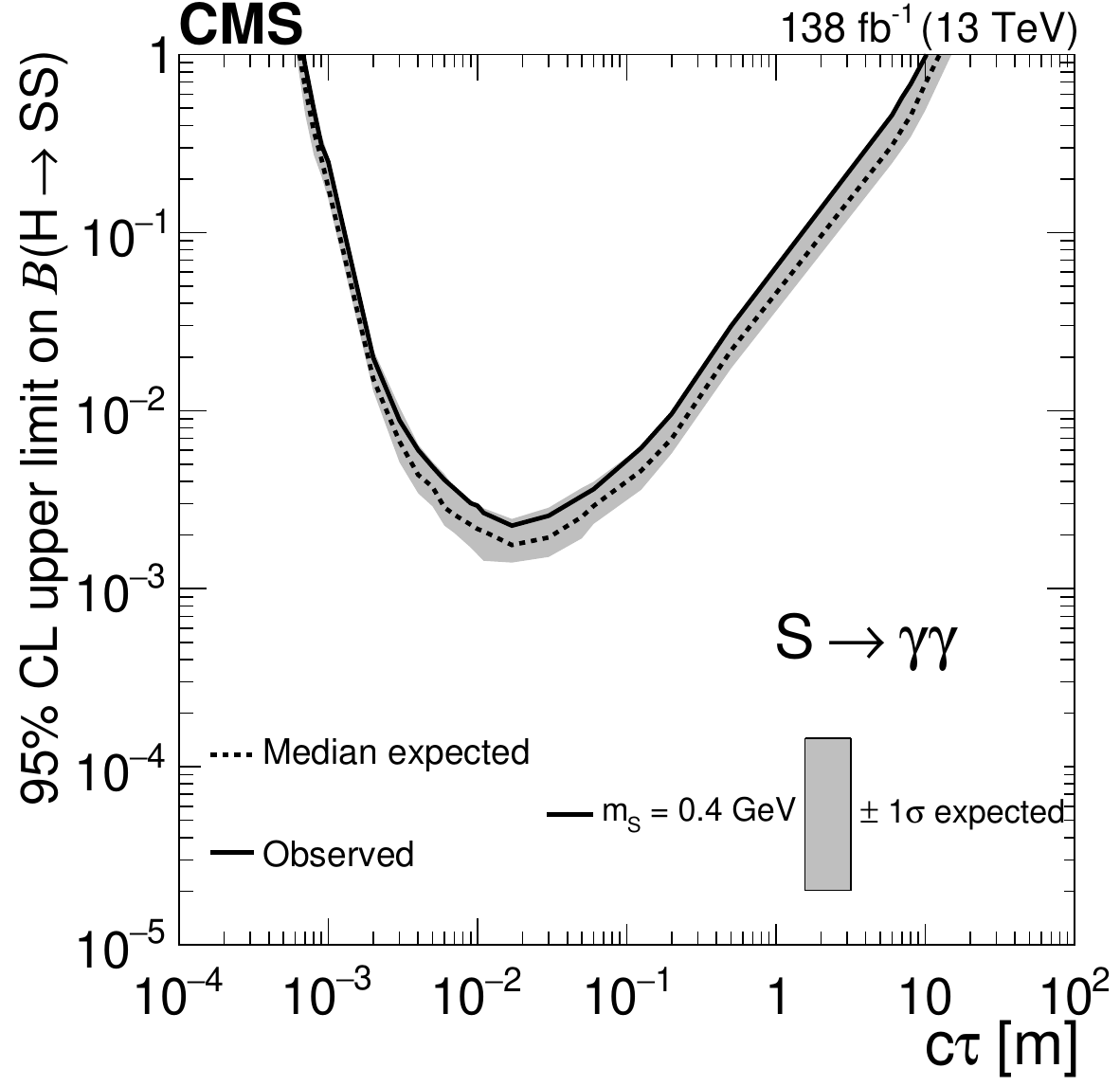}
  \includegraphics[width=0.45\textwidth]{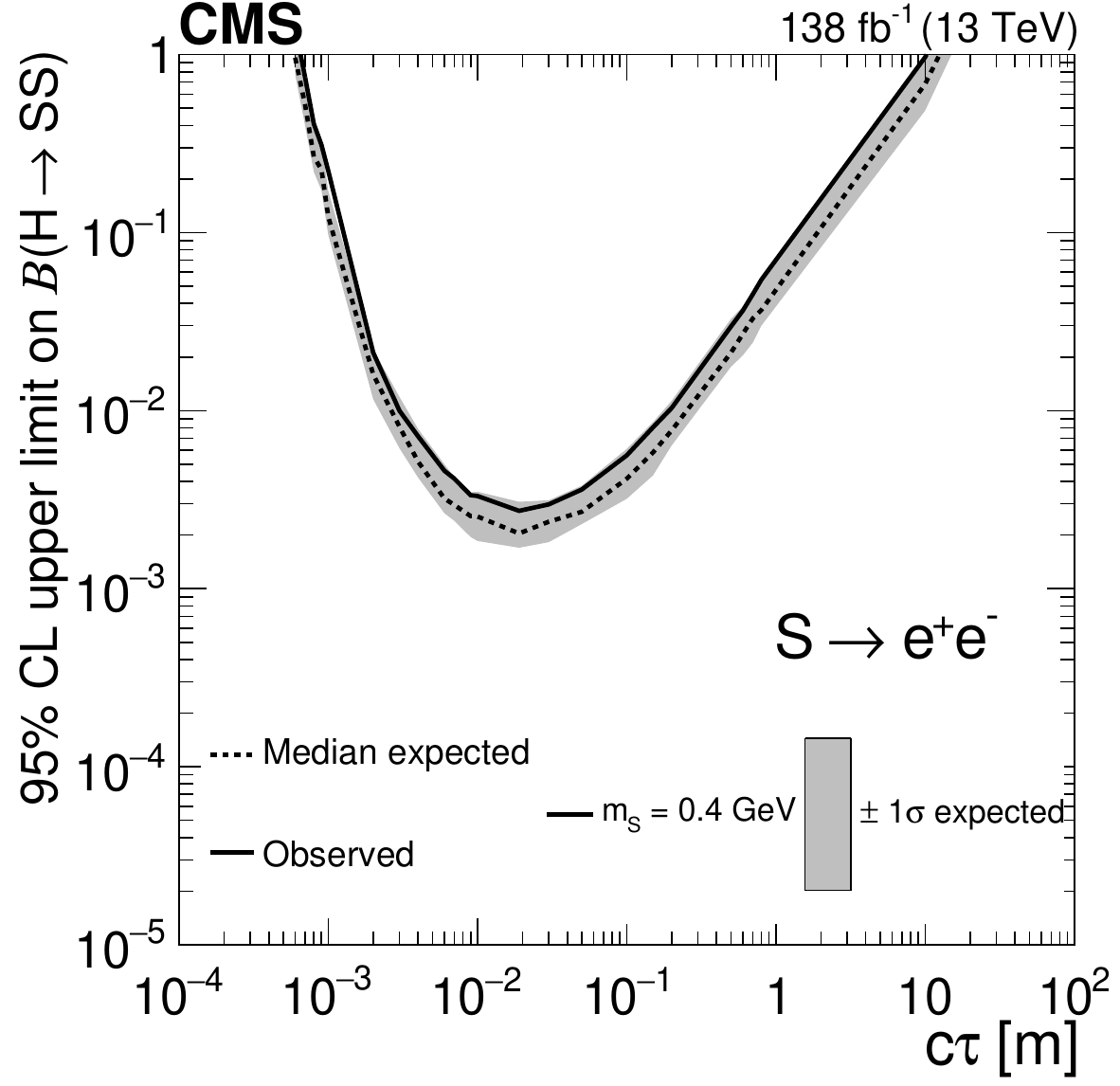}
  \caption{
    The 95\% \CL expected (dotted curves) and observed (solid curves) upper limits on the branching fraction $\brhtoss$ as functions of $c\tau$ for the $\PS\to\bbbar$ (upper left), $\PS\to\Pgpp\Pgpm$ (upper right), $\PS\to \PKpKm$ (middle left), $\PS\to \PKzKz$ (middle right), $\PS\to \Pgg\Pgg$ (lower left), and $\PS\to \EE$ (lower right) decay modes.
    The exclusion limits are shown for different mass hypotheses.
  }
  \label{fig:twin_higgs_limits_1d_additional}
\end{figure*}

\begin{figure*}[h!]
  \centering
  \includegraphics[width=0.45\textwidth]{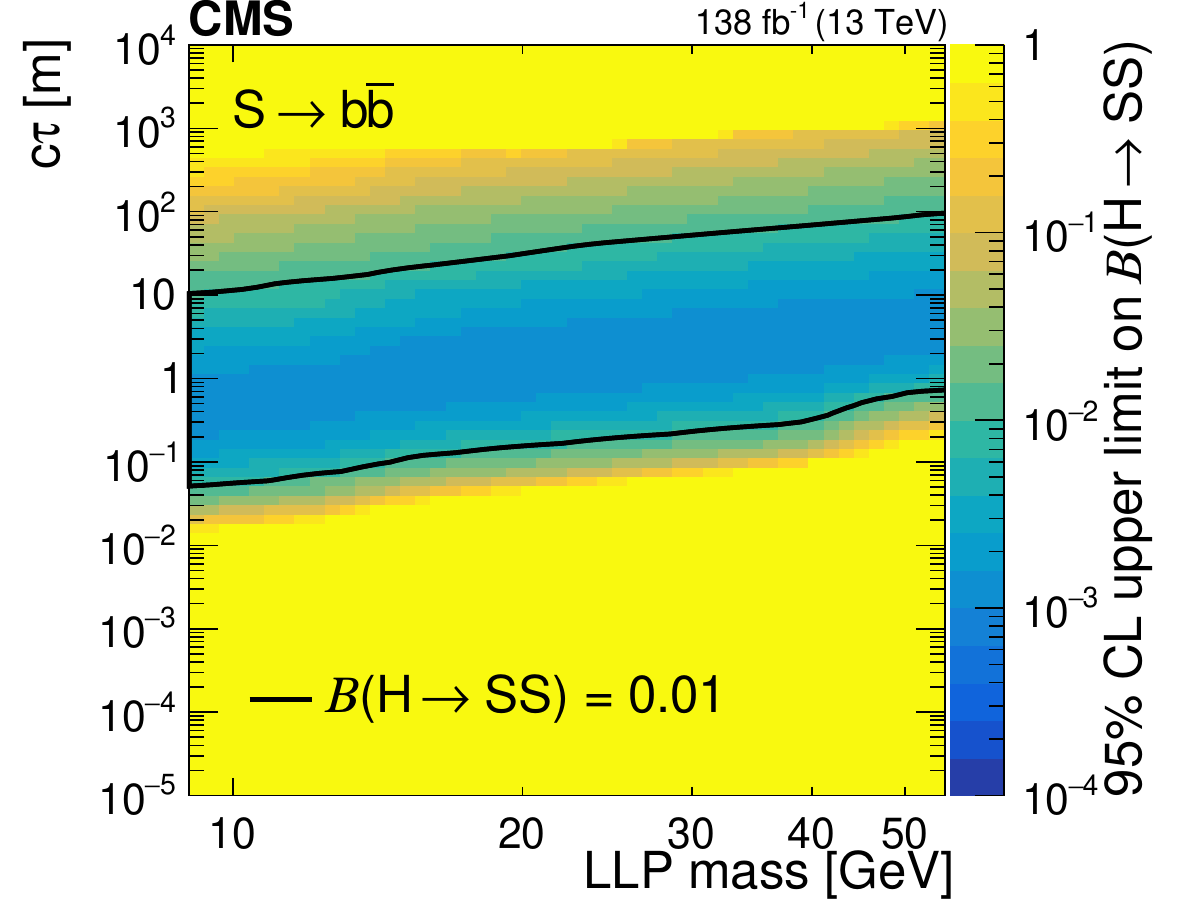}
  \includegraphics[width=0.45\textwidth]{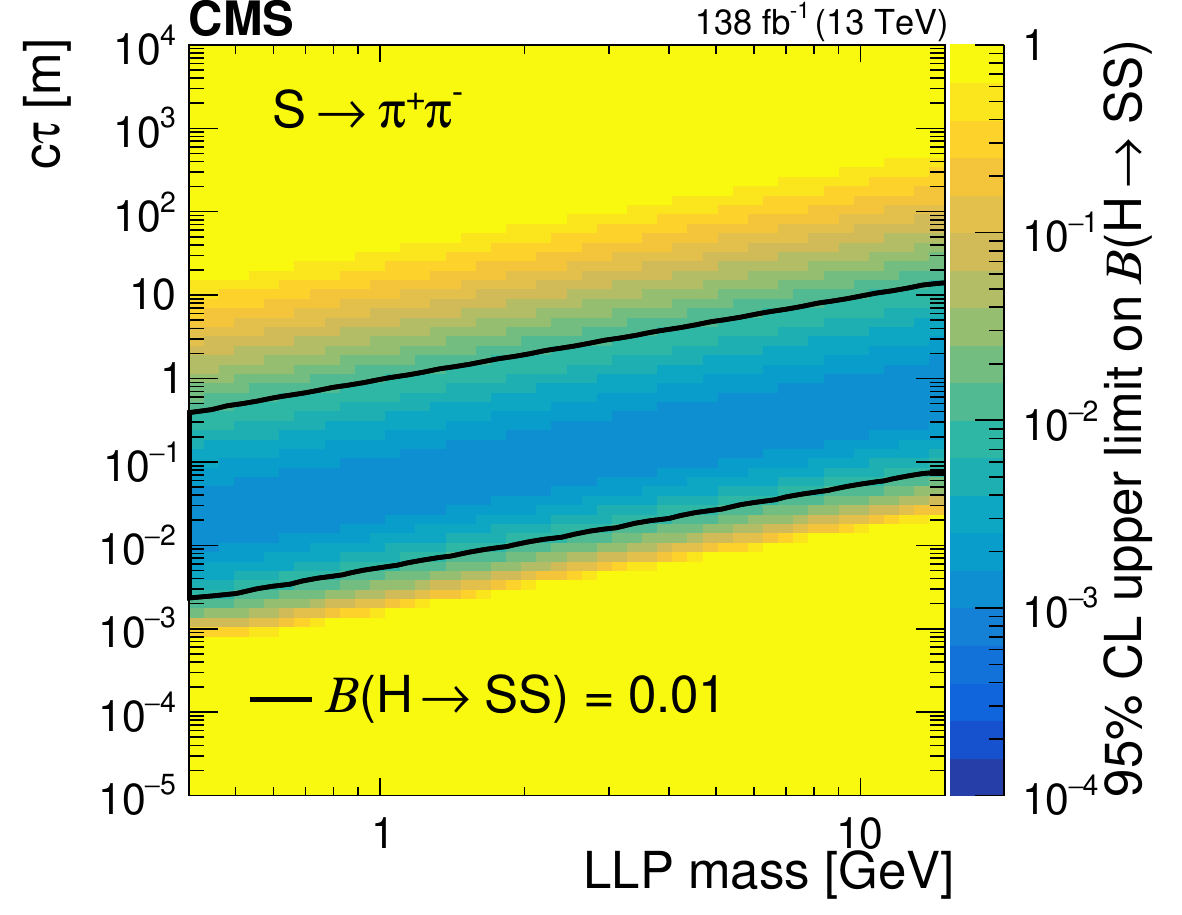}
  \includegraphics[width=0.45\textwidth]{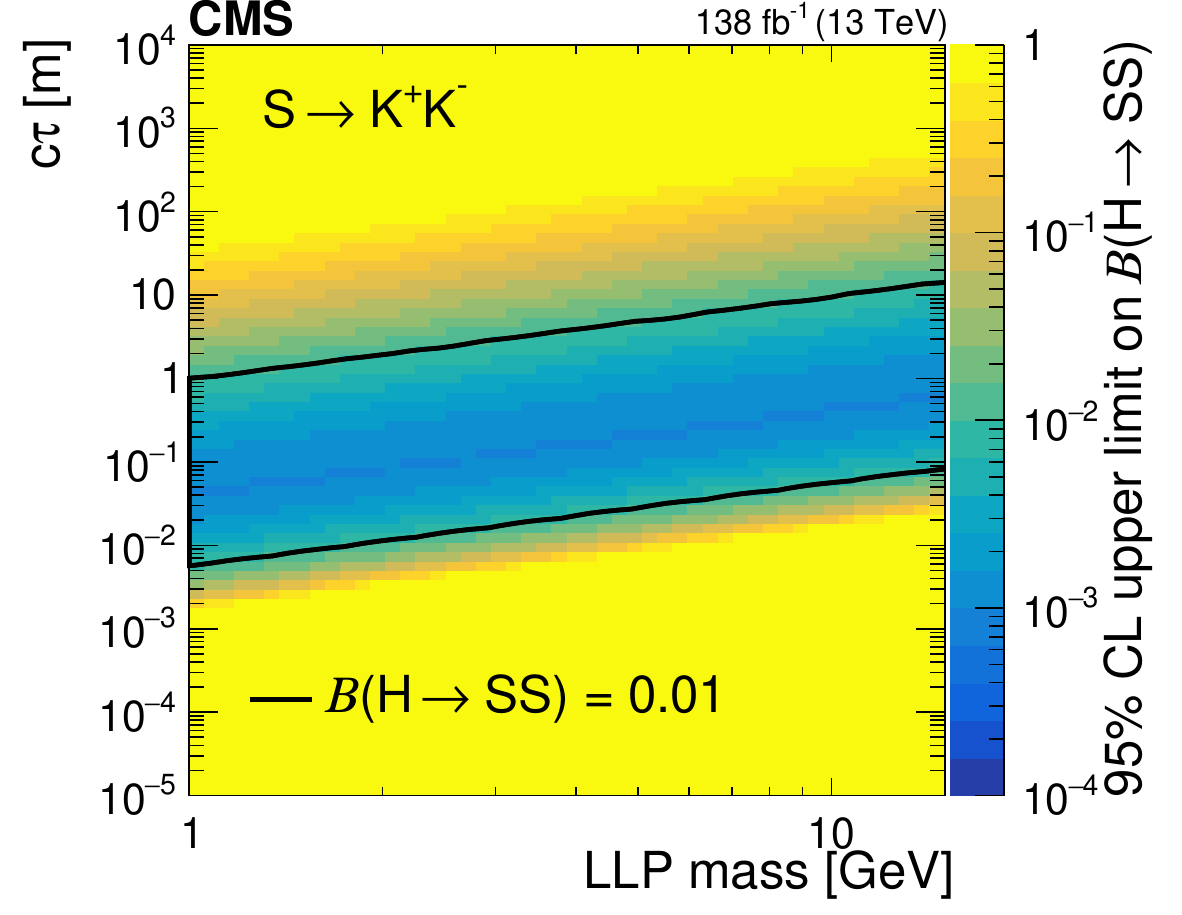}
  \includegraphics[width=0.45\textwidth]{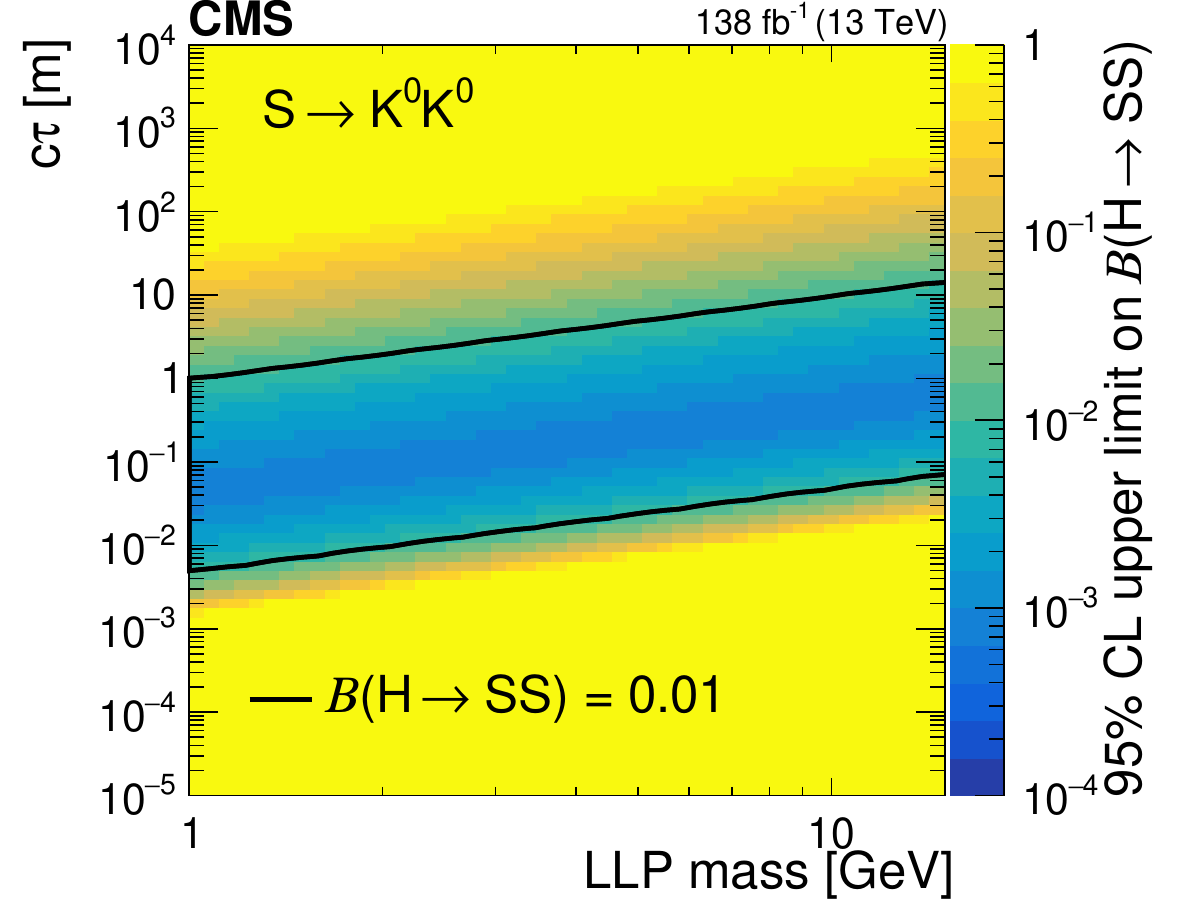}
  \includegraphics[width=0.45\textwidth]{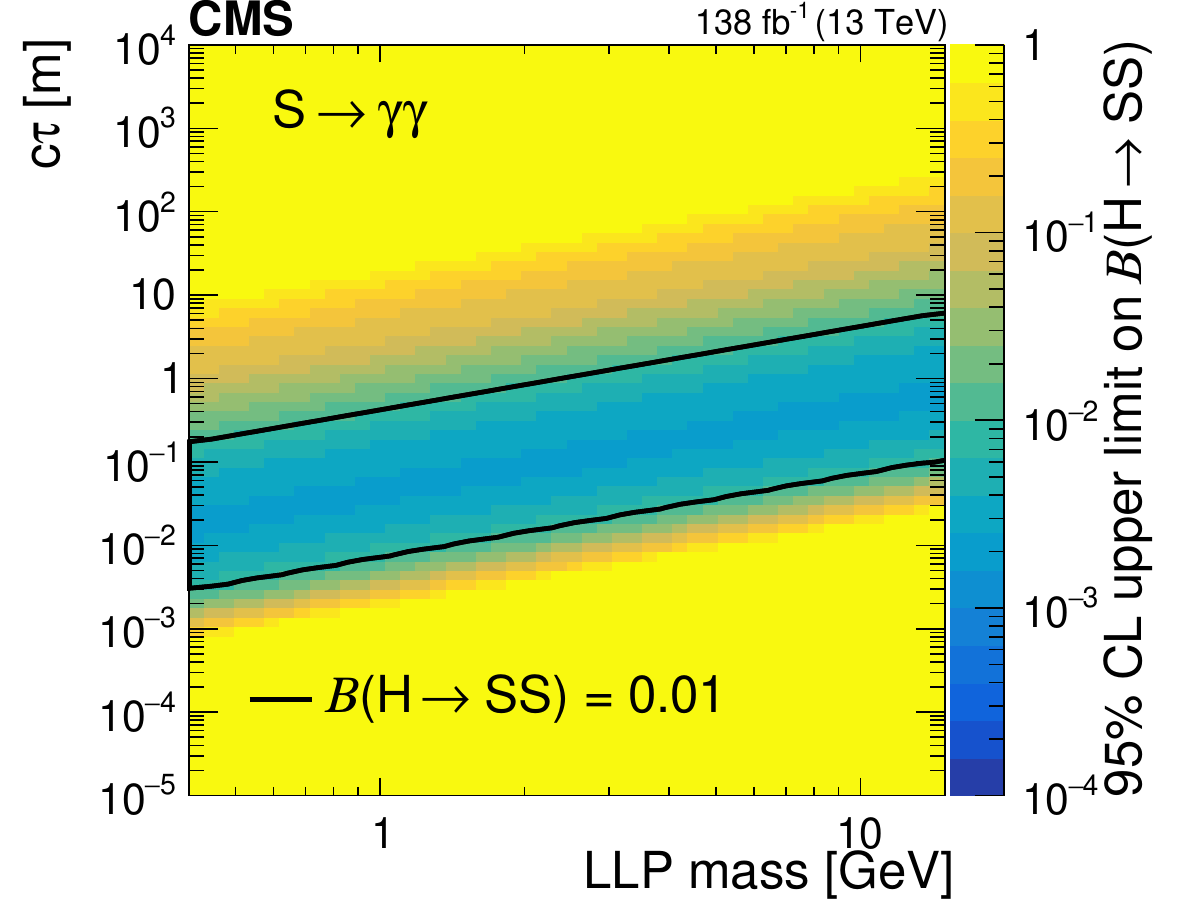}
  \includegraphics[width=0.45\textwidth]{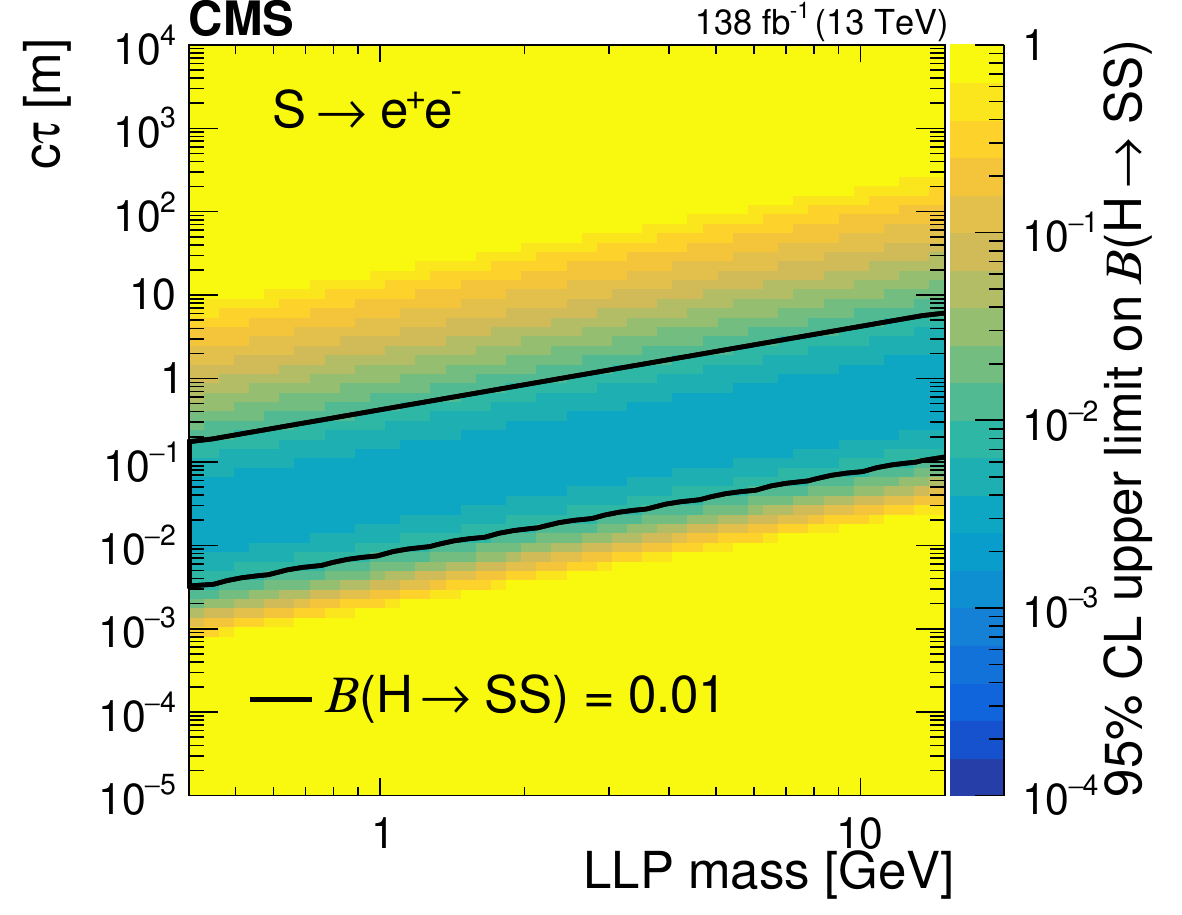}
  \caption{
    The 95\% \CL observed upper limits on the branching fraction $\brhtoss$ as functions of mass and $c\tau$ for the $\PS\to\bbbar$ (upper left), $\PS\to\Pgpp\Pgpm$ (upper right), $\PS\to \PKpKm$ (middle left), $\PS\to \PKzKz$ (middle right), $\PS\to\Pgg\Pgg$ (lower left), $\PS\to \EE$ (lower right) decay modes.
    The area inside the solid contours represents the region for which the limit is smaller than 0.01.
  }
  \label{fig:twin_higgs_limits_2d_additional}
\end{figure*}

The upper limits for the twin Higgs model are shown in Fig.~\ref{fig:twin_higgs_limits_1d} for the $\PS\to\ddbar$, $\PS\to\Pgpz\Pgpz$, and $\PS\to\PGtp\PGtm$ decay modes, as functions of $c\tau$ for a selection of $m_\PS$ values.
Figure~\ref{fig:twin_higgs_limits_2d} shows these limits as a function of $m_\PS$ and $c\tau$ and extend as low as $m_\PS = 0.4\GeV$.
The sensitivity for 55\GeV is slightly lower than for the other mass points because the two LLPs in the event are geometrically closer to each other at higher masses, so when both decay in the muon detectors, they are likely to merge to create one muon detector shower, instead of two distinct clusters, lowering the sensitivity of the double-cluster category.
Additional upper limits for other decay modes, including $\PS\to\bbbar$, $\PS\to\Pgpp\Pgpm$, $\PS\to\PK\PK$, $\PS\to\bbbar$, $\PS\to\EE$, and $\PS\to\Pgg\Pgg$, are shown in Fig.~\ref{fig:twin_higgs_limits_1d_additional} as a function of $c\tau$ and in Fig.~\ref{fig:twin_higgs_limits_2d_additional} as a function of both $m_\PS$ and $c\tau$.
Owing to the differences in geometric acceptance, the sensitivity at shorter proper decay lengths is dominated by the single-CSC-cluster category, at longer proper decay lengths by the single-DT-cluster category, and at intermediate proper decay lengths, where the analysis is most sensitive, by the double-cluster category.
The addition of the single-DT and double-cluster categories improves upon the previous result based on only CSC clusters~\cite{CMS:2021juv} by a factor of 2 for $c\tau$ above 0.2, 0.5, 2.0, and 18.0\unit{m} for LLP masses 7, 15, 40, and 55\GeV, respectively.
The limits for electromagnetic decay modes are slightly less stringent, because the efficiency for reconstructing the corresponding MDS object is smaller.

\begin{figure*}[h!]
  \centering
  \includegraphics[width=0.6\textwidth]{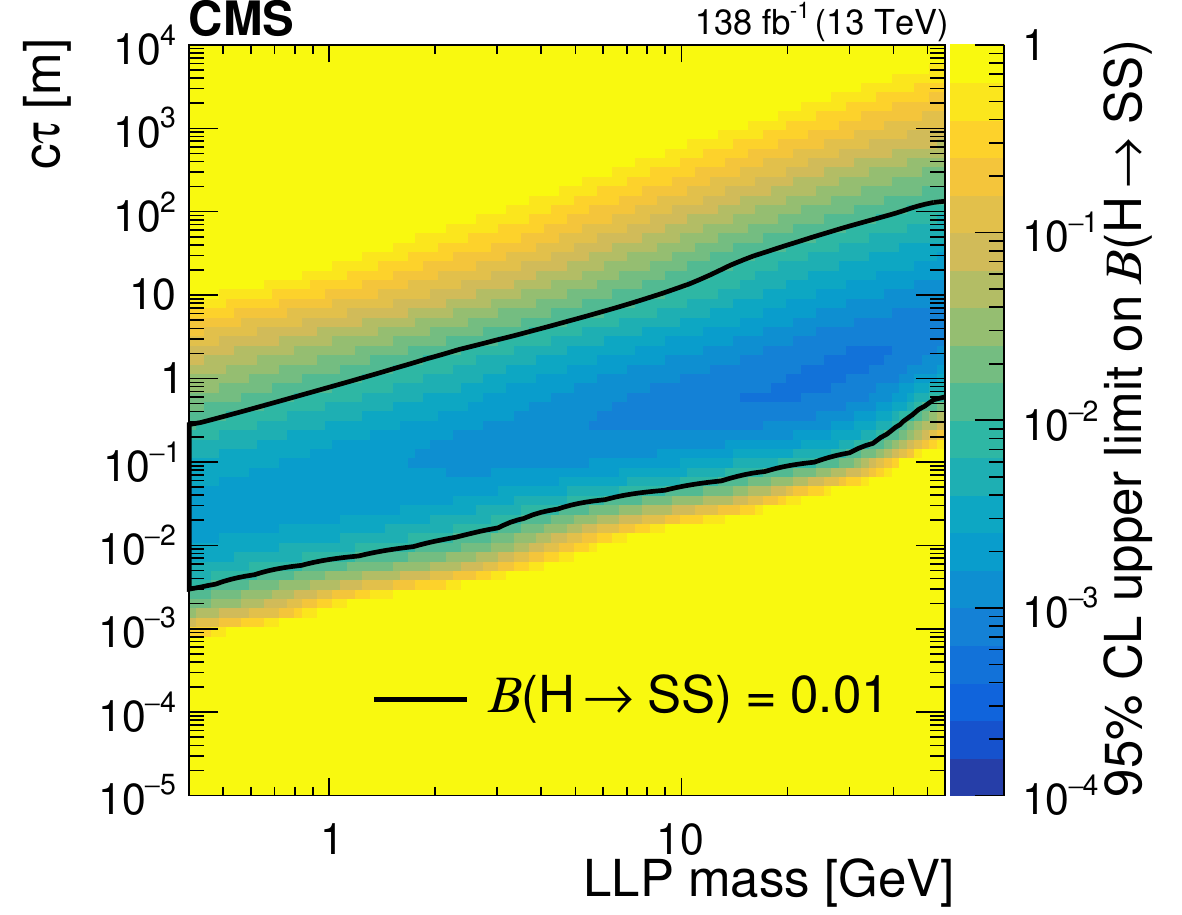}
  \caption{
    The 95\% \CL observed upper limits on the branching fraction $\brhtoss$ as a function of mass and $c\tau$, assuming the branching fractions for \PS are identical to those of a Higgs boson evaluated at $m_\PS$~\cite{GERSHTEIN2021136758}.
    The area inside the solid contours represents the region for which the limit is smaller than 0.01.
  }
  \label{fig:twin_higgs_limits_2d_mixbr}
\end{figure*}

We also consider an explicit scenario where the branching fractions of the \PS decays are identical to those of a Higgs boson evaluated at $m_\PS$~\cite{GERSHTEIN2021136758}.
Figure~\ref{fig:twin_higgs_limits_2d_mixbr} shows the upper limits as functions of both mass and $c\tau$ using the branching fractions for \PS calculated in Ref.~\cite{GERSHTEIN2021136758}.
This search provides the most stringent constraint on $\brhtoss$ for proper decay lengths in the range of 0.04--0.40\unit{m} and above 4\unit{m}, of 0.3--0.9\unit{m} and above 3\unit{m}, and of above 0.8\unit{m} for an LLP mass of 15, 40, and 55\GeV, respectively.
For LLP masses below 10\GeV, this search provides the most stringent constraints on LLPs decaying to particles other than muons.

\begin{figure*}[h!]
  \centering
  \includegraphics[width=0.5\textwidth]{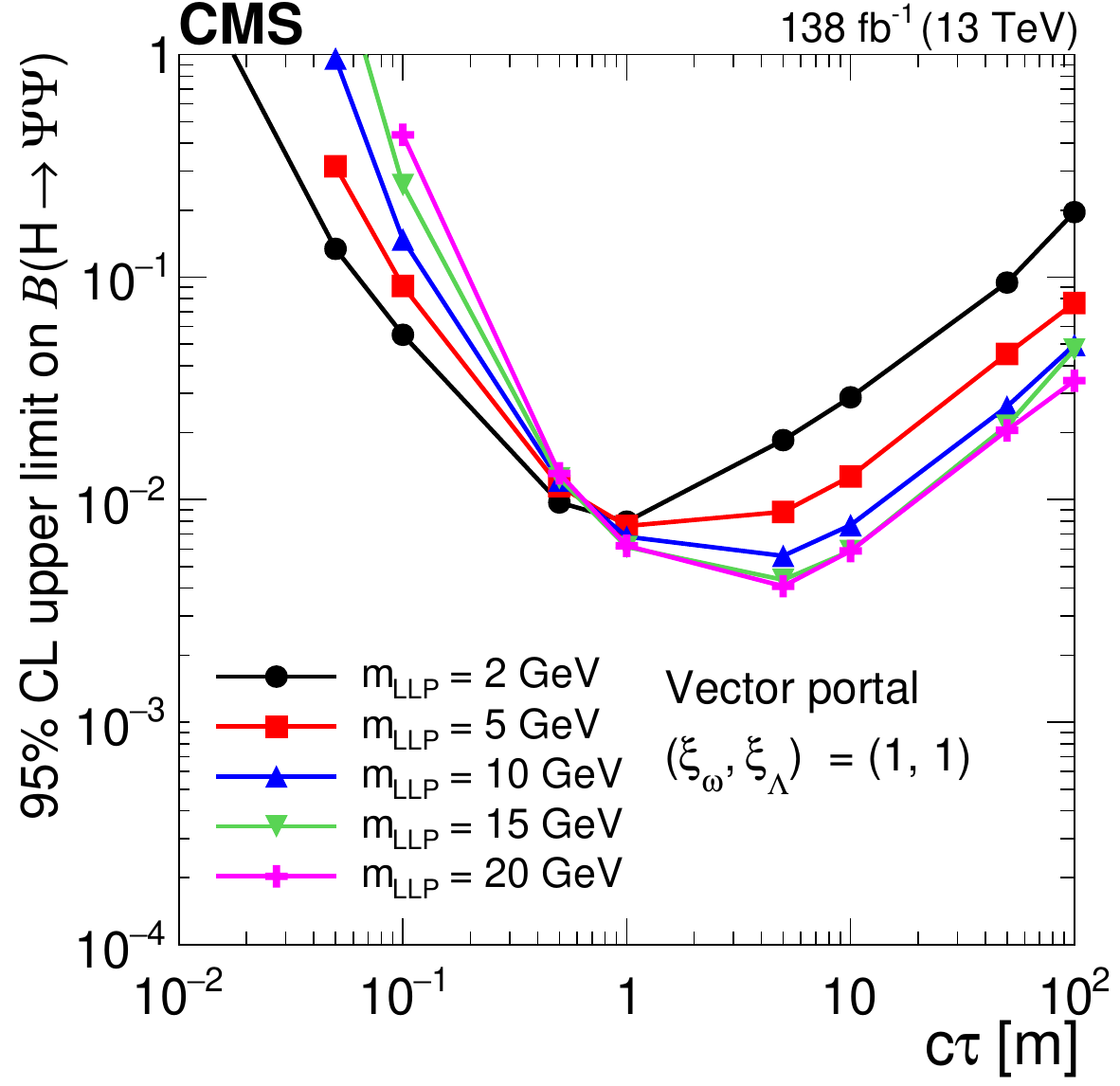}
  \caption{
    The 95\% \CL observed upper limits on the branching fraction $\brhtopsipsi$ as functions of $c\tau$ for the dark shower vector portal assuming $(\xi_{\omega}$, $\xi_{\Lambda})  = (1, 1)$.
    The exclusion limits are shown for different LLP mass hypotheses.
    The limits are calculated only at the proper decay lengths indicated by the markers and the lines connecting the markers are linear interpolations.
    The plot is restricted to physical $\brhtopsipsi$ values of below 1.
  }
  \label{fig:limit_darkshower_vector}
\end{figure*}

\begin{figure*}[h!]
  \centering
  \includegraphics[width=0.45\textwidth]{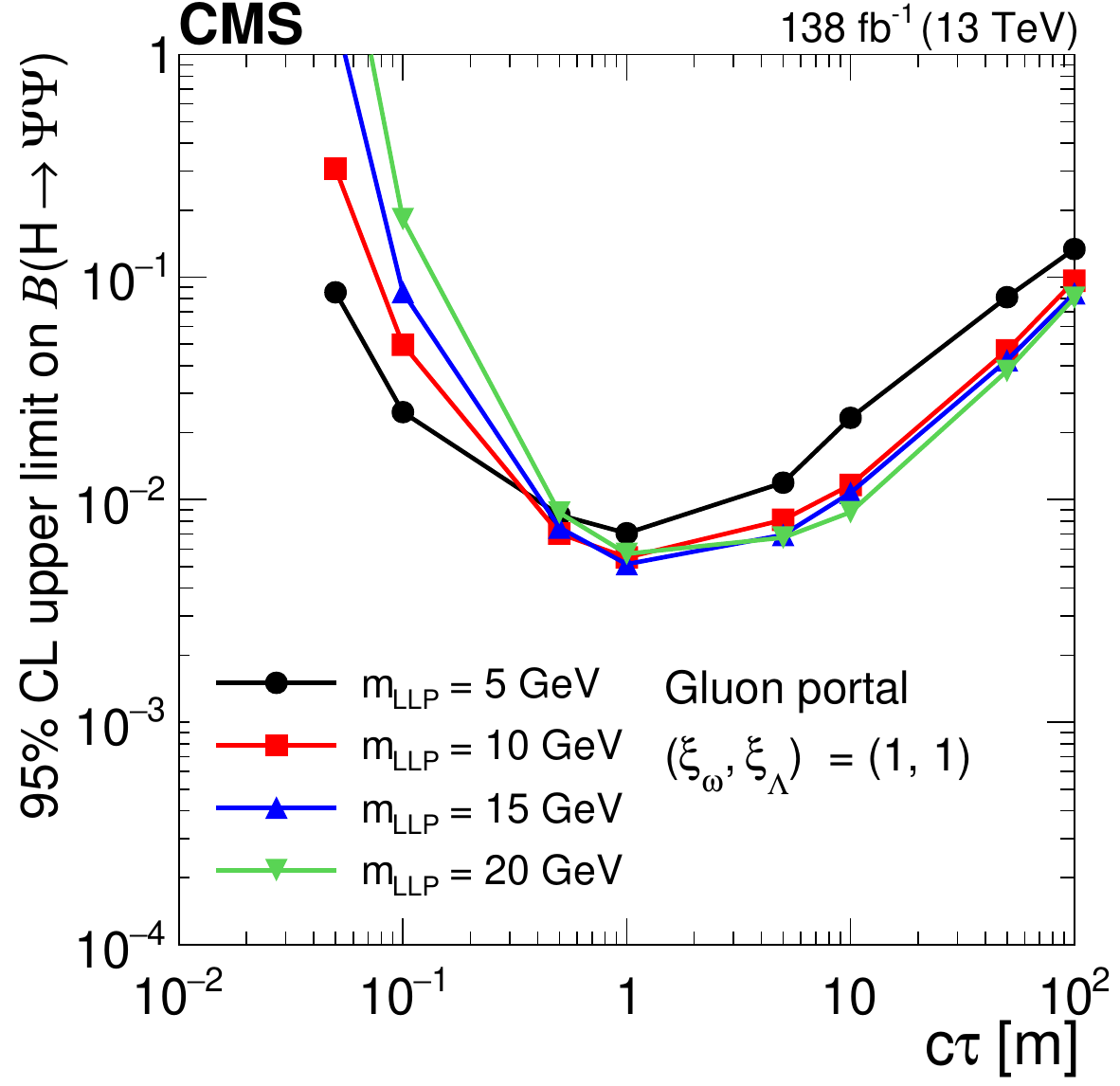}
  \includegraphics[width=0.45\textwidth]{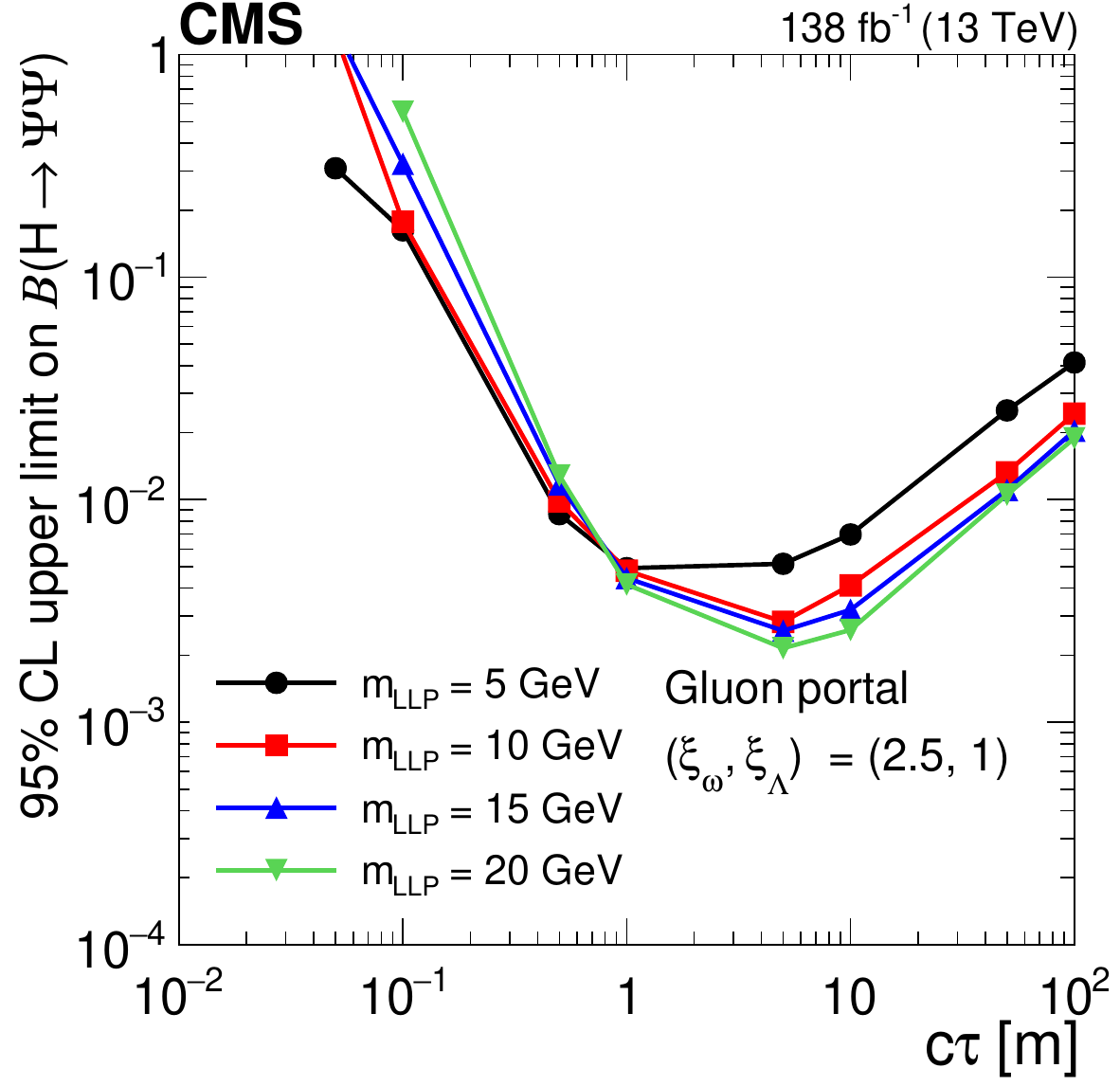}
  \includegraphics[width=0.45\textwidth]{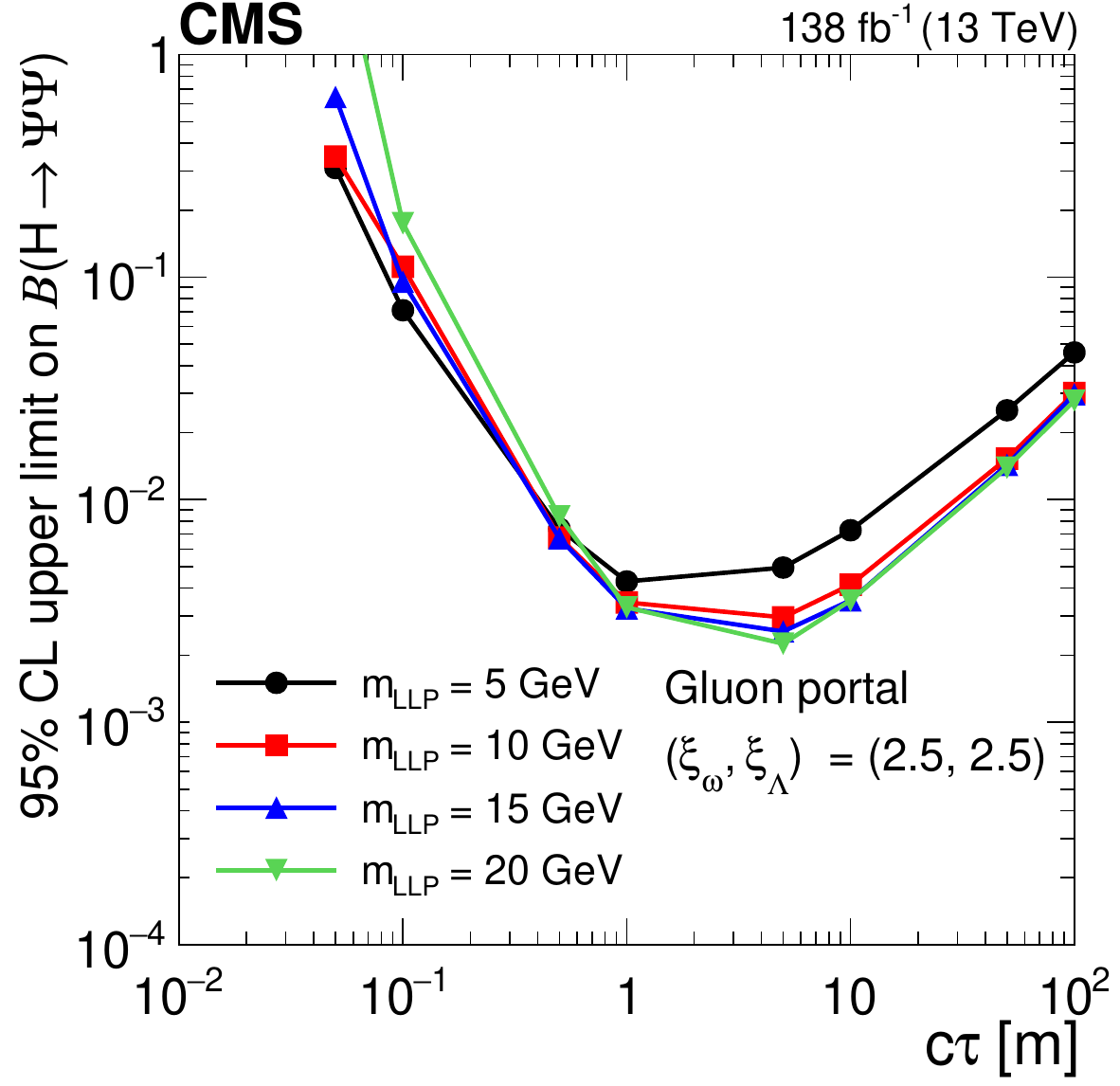}
  \caption{
    The 95\% \CL observed upper limits on the branching fraction $\brhtopsipsi$ as functions of $c\tau$ for the dark shower gluon portal, assuming $(\xi_{\omega}$, $\xi_{\Lambda})  = (1, 1)$ (upper left), $(\xi_{\omega}$, $\xi_{\Lambda})  = (2.5, 1.0)$ (upper right), and $(\xi_{\omega}$, $\xi_{\Lambda})  = (2.5, 2.5)$ (lower).
    The exclusion limits are shown for different mass hypotheses.
    The limits are calculated only at the proper decay lengths indicated by the markers and the lines connecting the markers are linear interpolations.
    The plot is restricted to physical $\brhtopsipsi$ values of below 1.
  }
  \label{fig:limit_darkshower_gluon}
\end{figure*}

\begin{figure*}[h!]
  \centering
  \includegraphics[width=0.45\textwidth]{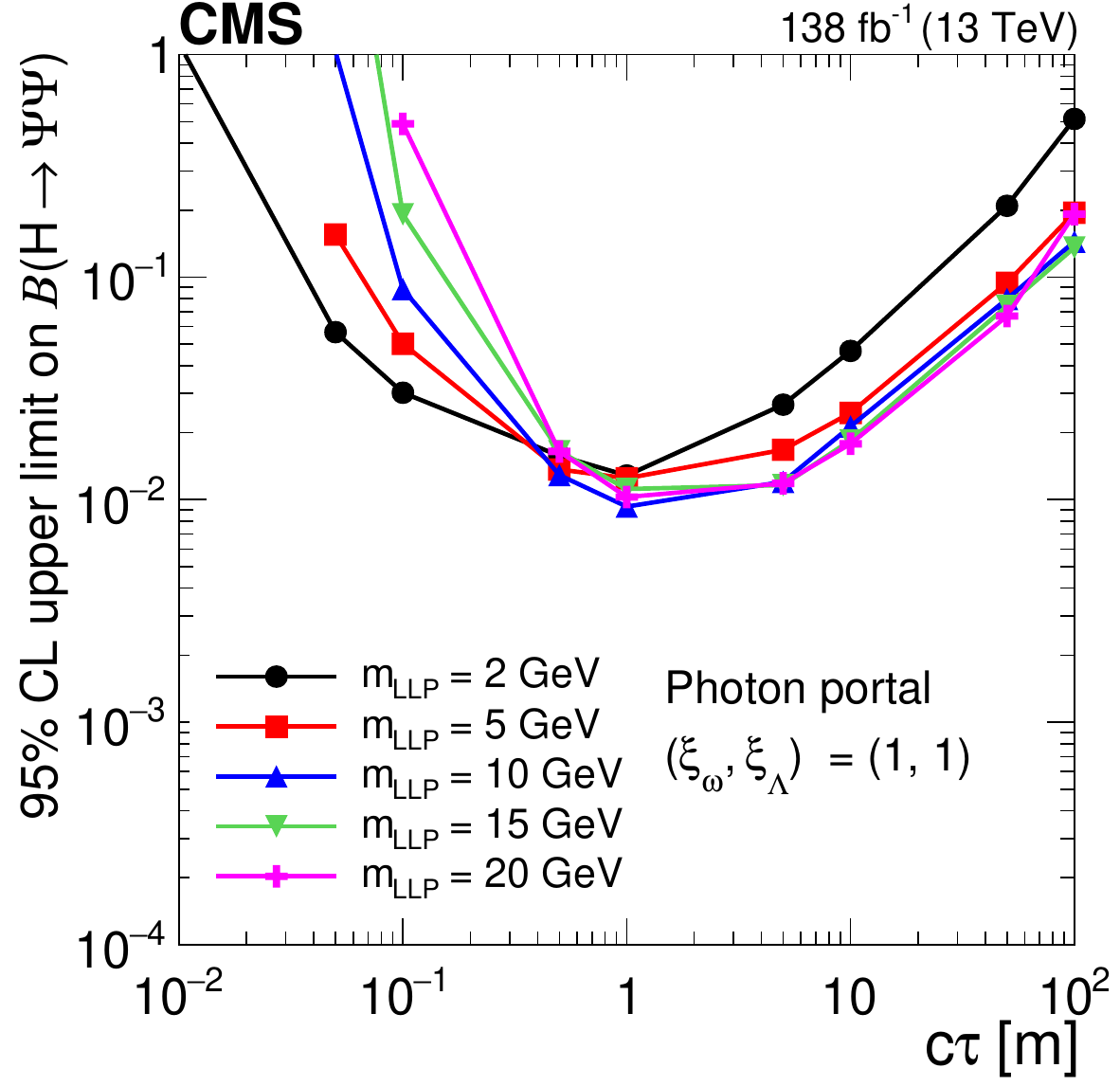}
  \includegraphics[width=0.45\textwidth]{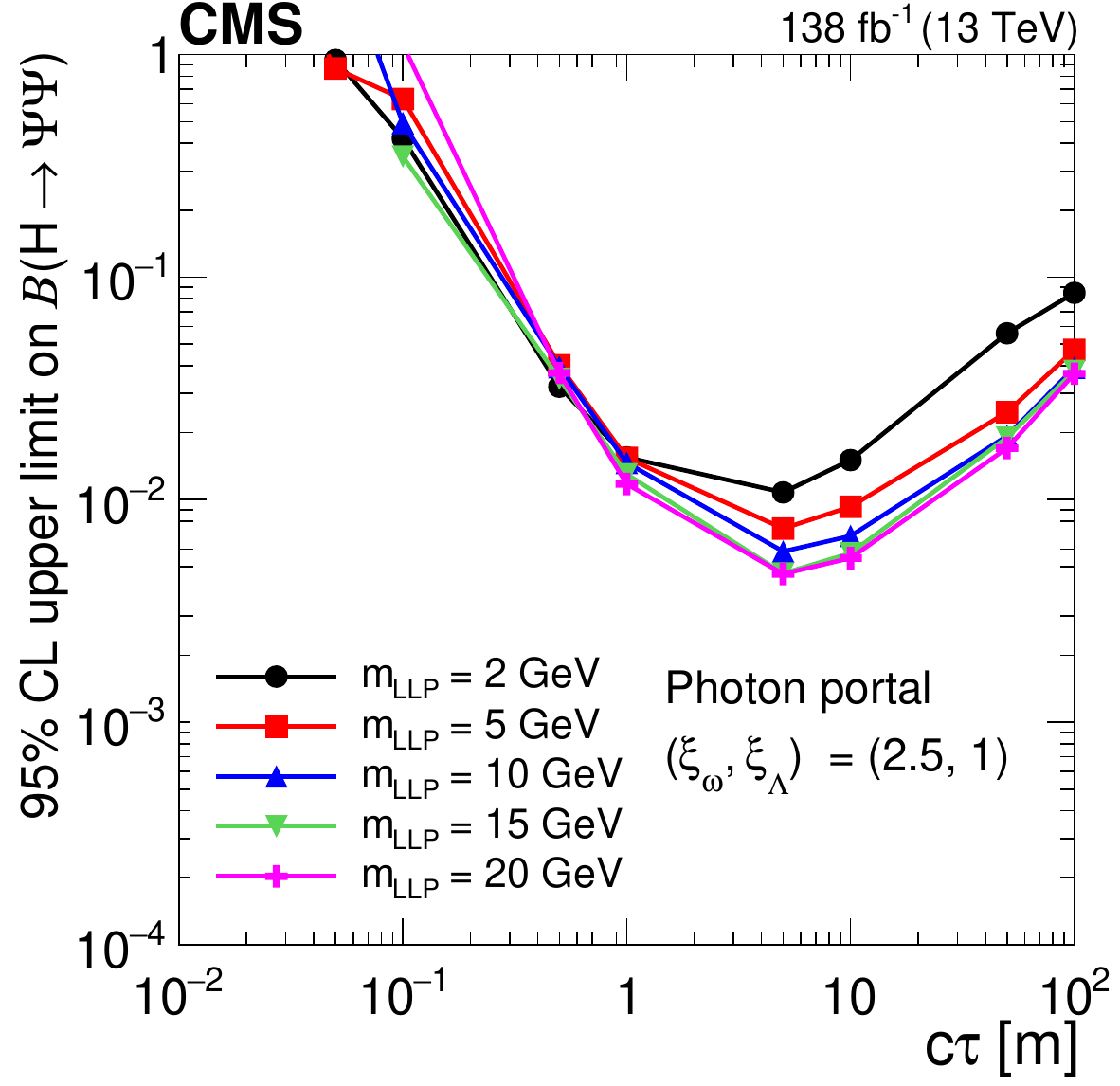}
  \includegraphics[width=0.45\textwidth]{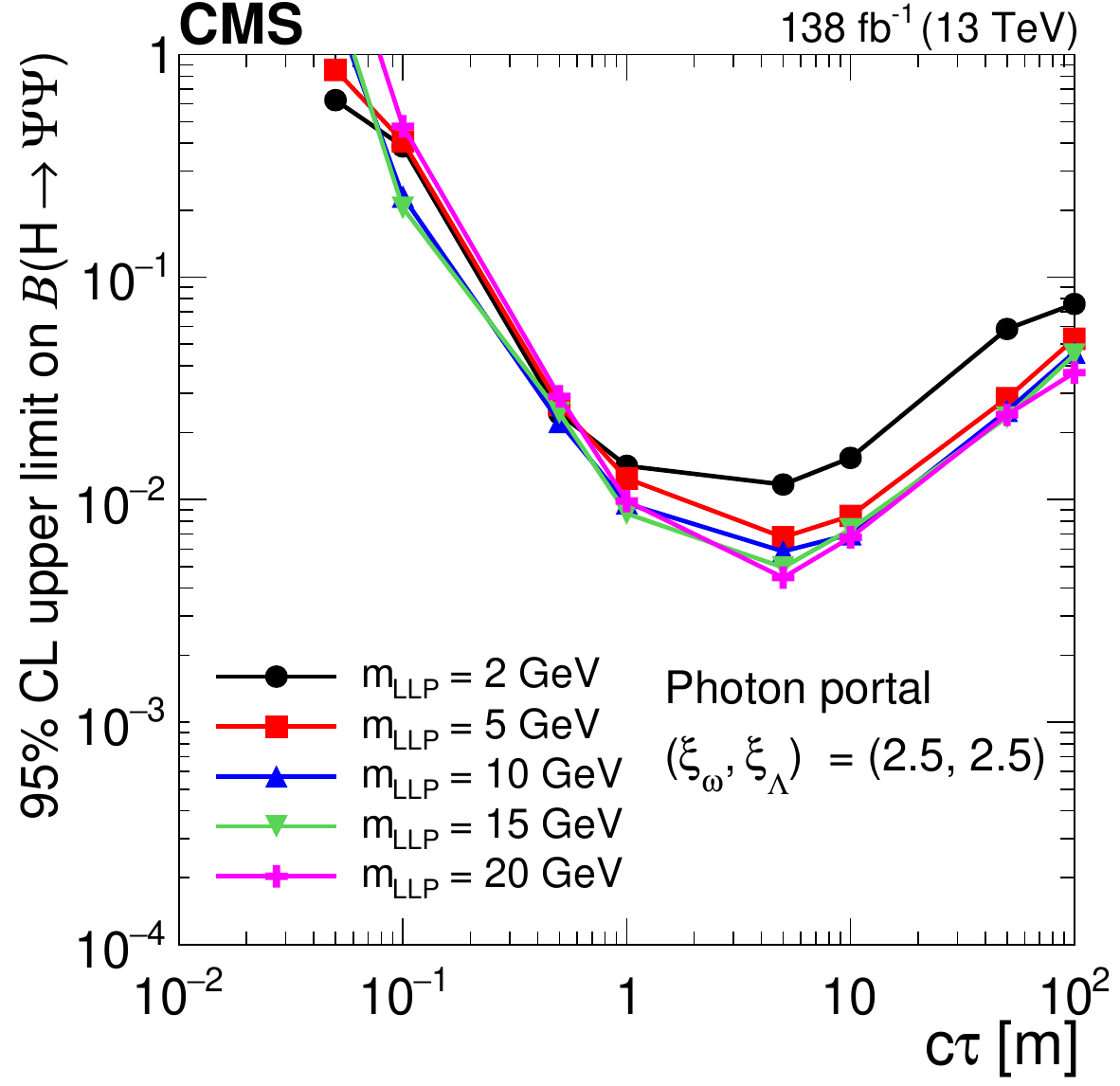}
  \caption{
    The 95\% \CL observed upper limits on the branching fraction $\brhtopsipsi$ as functions of $c\tau$ for the dark shower photon portal, assuming $(\xi_{\omega}$, $\xi_{\Lambda})  = (1, 1)$ (upper left), $(\xi_{\omega}$, $\xi_{\Lambda})  = (2.5, 1.0)$ (upper right), and $(\xi_{\omega}$, $\xi_{\Lambda})  = (2.5, 2.5)$ (lower).
    The exclusion limits are shown for different mass hypotheses.
    The limits are calculated only at the proper decay lengths indicated by the markers and the lines connecting the markers are linear interpolations.
    The plot is restricted to physical $\brhtopsipsi$ values of below 1.
  }
  \label{fig:limit_darkshower_photon}
\end{figure*}

\begin{figure*}[h!]
  \centering
  \includegraphics[width=0.45\textwidth]{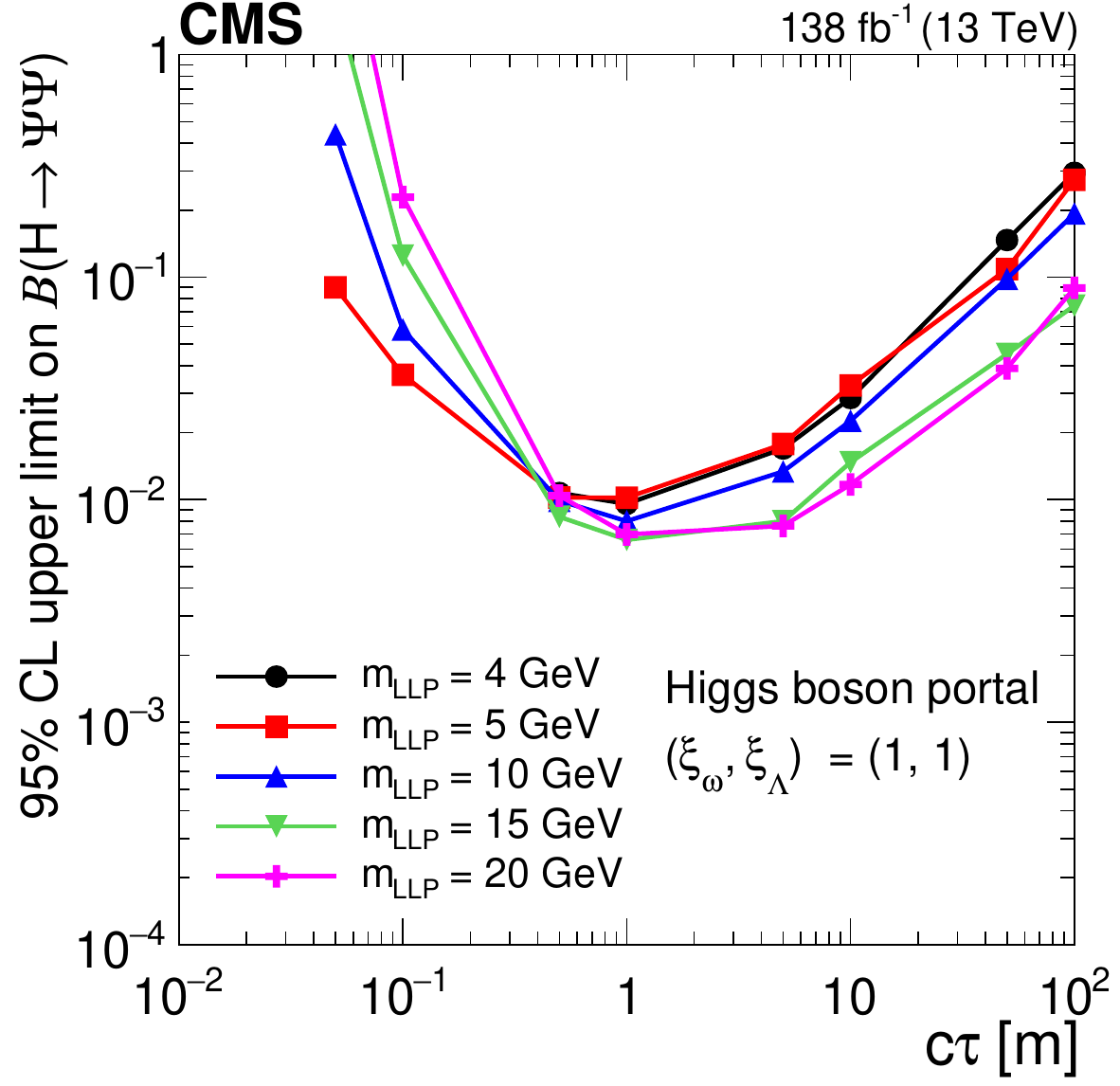}
  \includegraphics[width=0.45\textwidth]{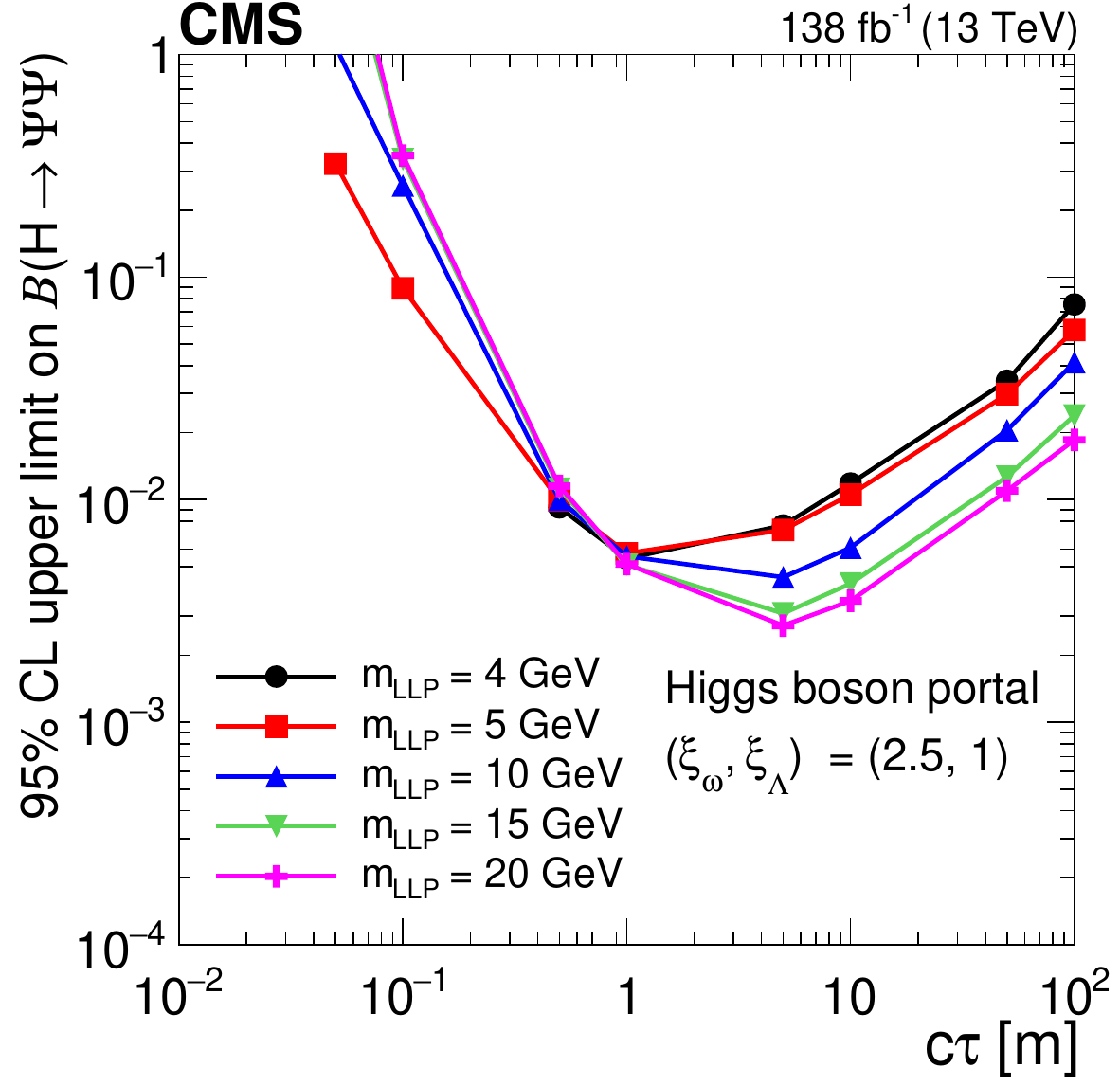}
  \includegraphics[width=0.45\textwidth]{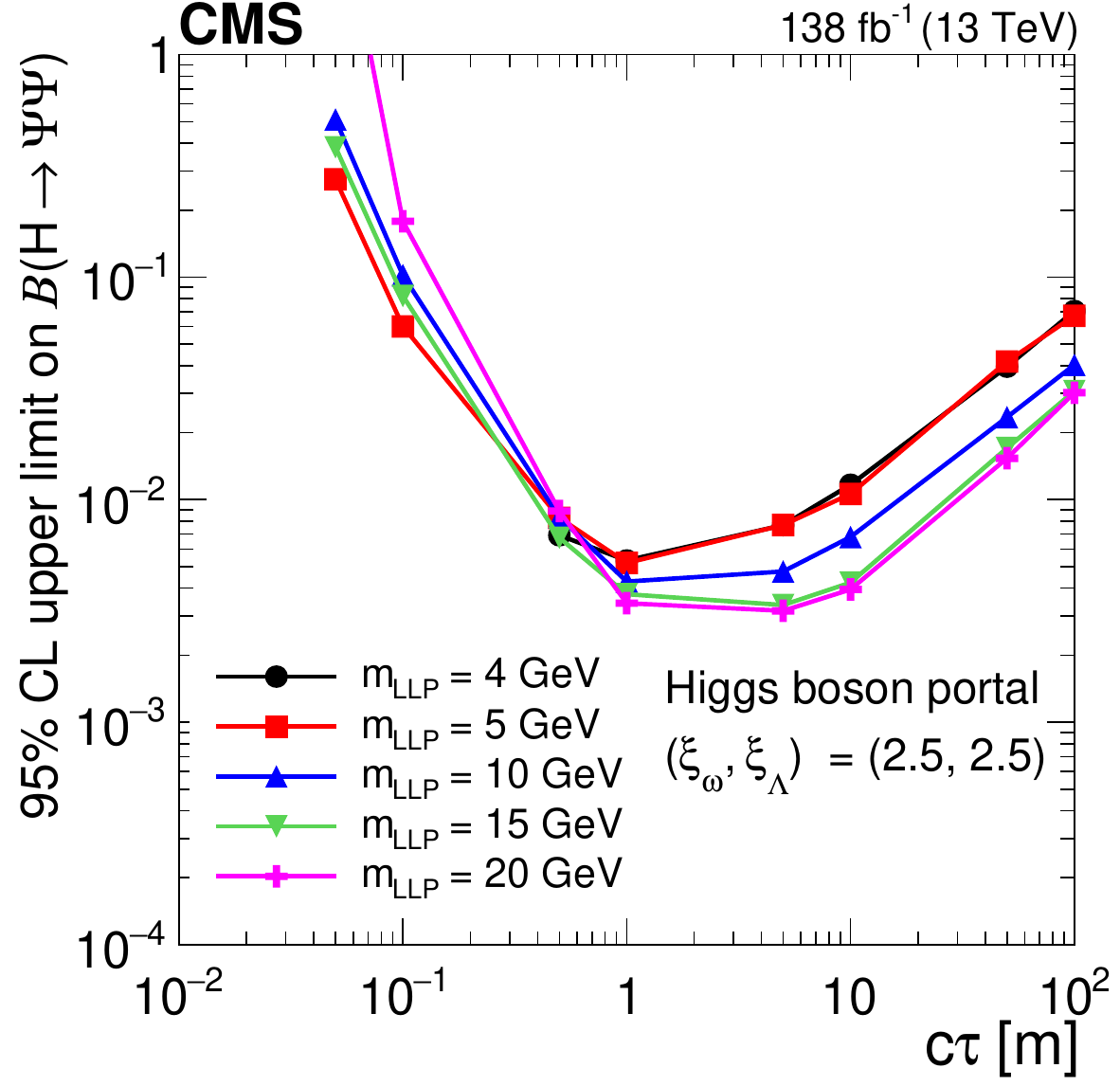}
  \caption{
    The 95\% \CL observed upper limits on the branching fraction $\brhtopsipsi$ as functions of $c\tau$ for the dark shower Higgs boson portal, assuming $(\xi_{\omega}$, $\xi_{\Lambda})  = (1, 1)$ (upper left), $(\xi_{\omega}$, $\xi_{\Lambda})  = (2.5, 1.0)$ (upper right), and $(\xi_{\omega}$, $\xi_{\Lambda})  = (2.5, 2.5)$ (lower).
    The exclusion limits are shown for different mass hypotheses.
    The limits are calculated only at the proper decay lengths indicated by the markers and the lines connecting the markers are linear interpolations.
    The plot is restricted to physical $\brhtopsipsi$ values of below 1.
  }
  \label{fig:limit_darkshower_higgs}
\end{figure*}

\begin{figure*}[h!]
  \centering
  \includegraphics[width=0.45\textwidth]{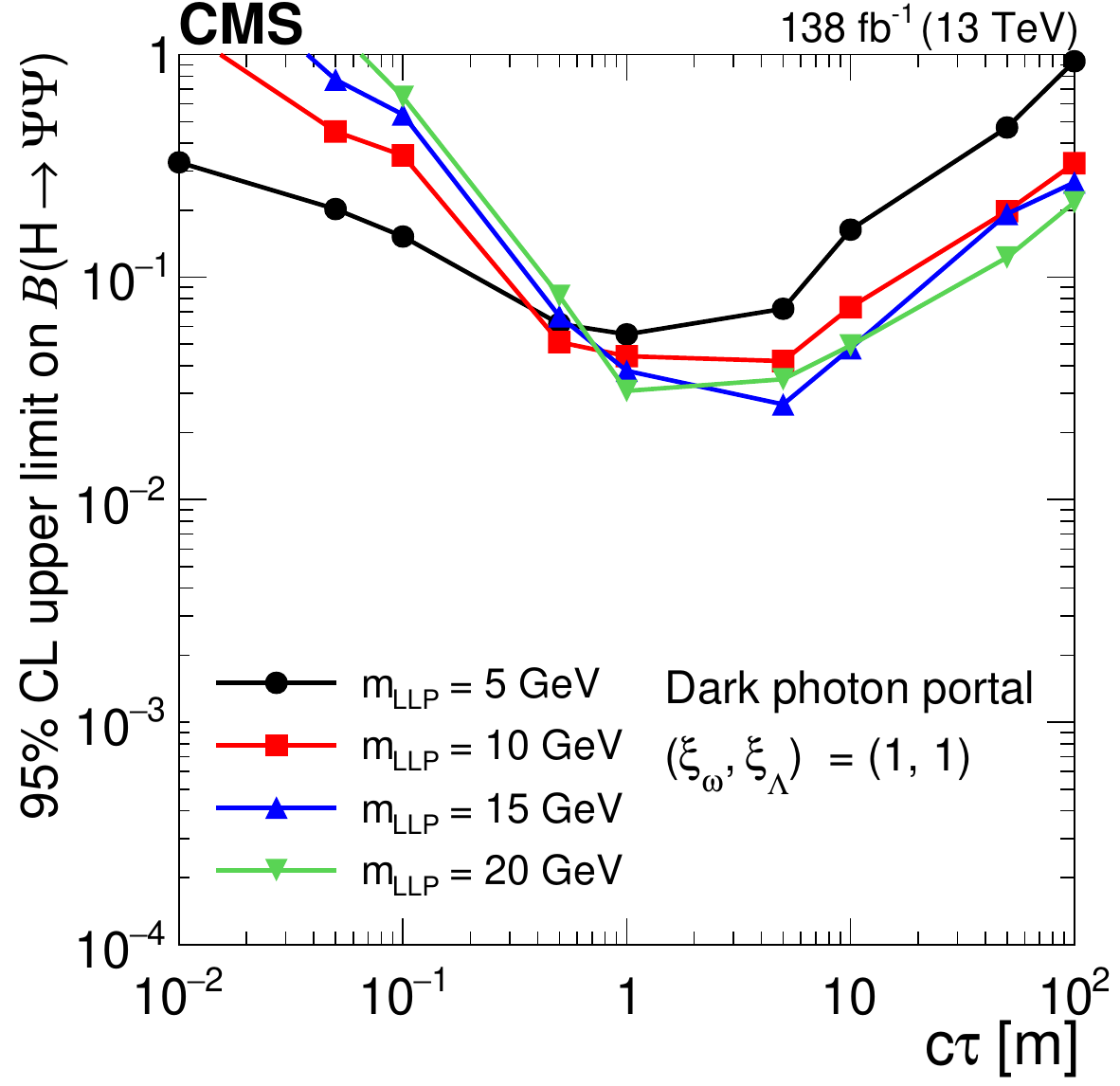}
  \includegraphics[width=0.45\textwidth]{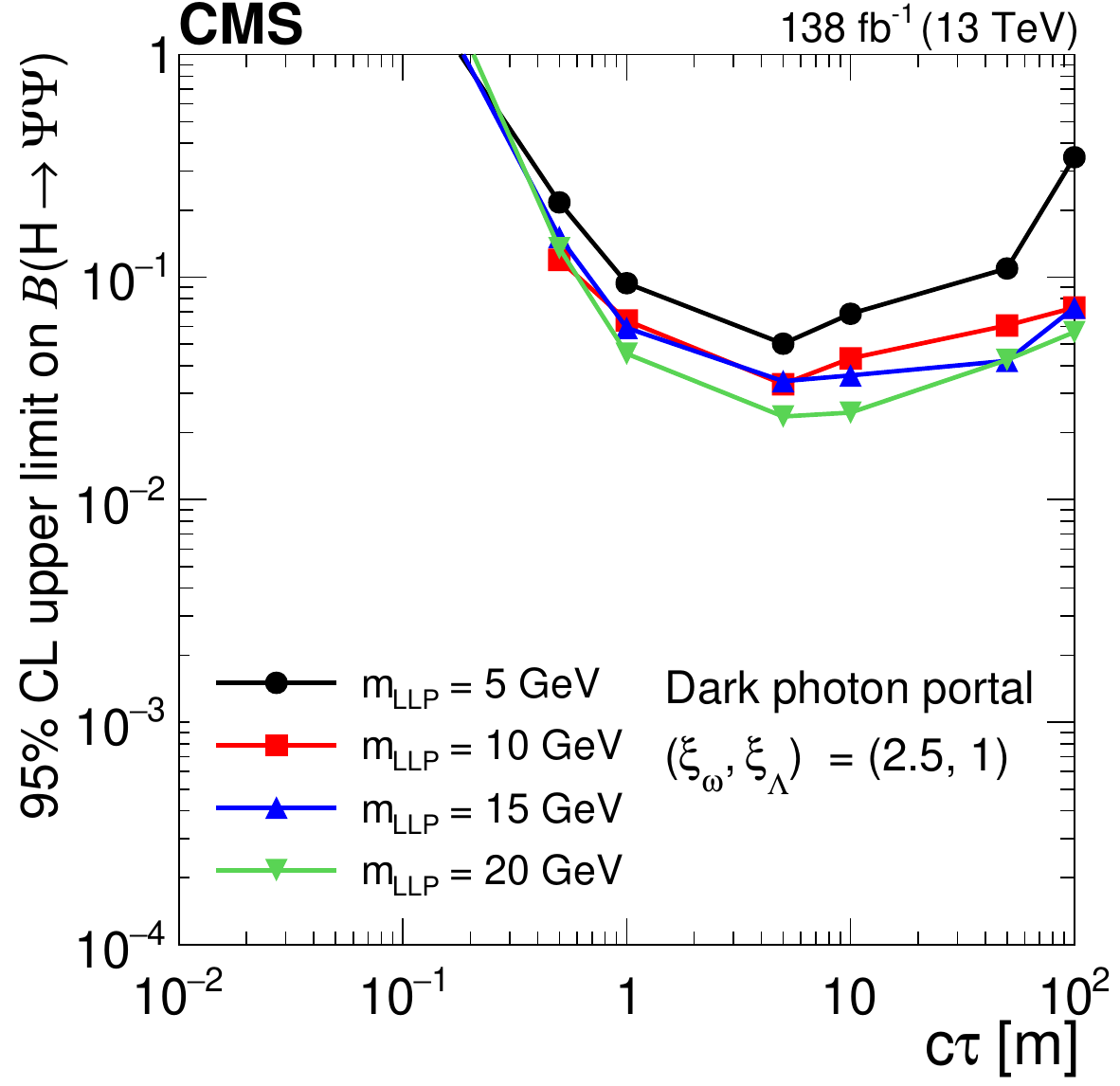}
  \includegraphics[width=0.45\textwidth]{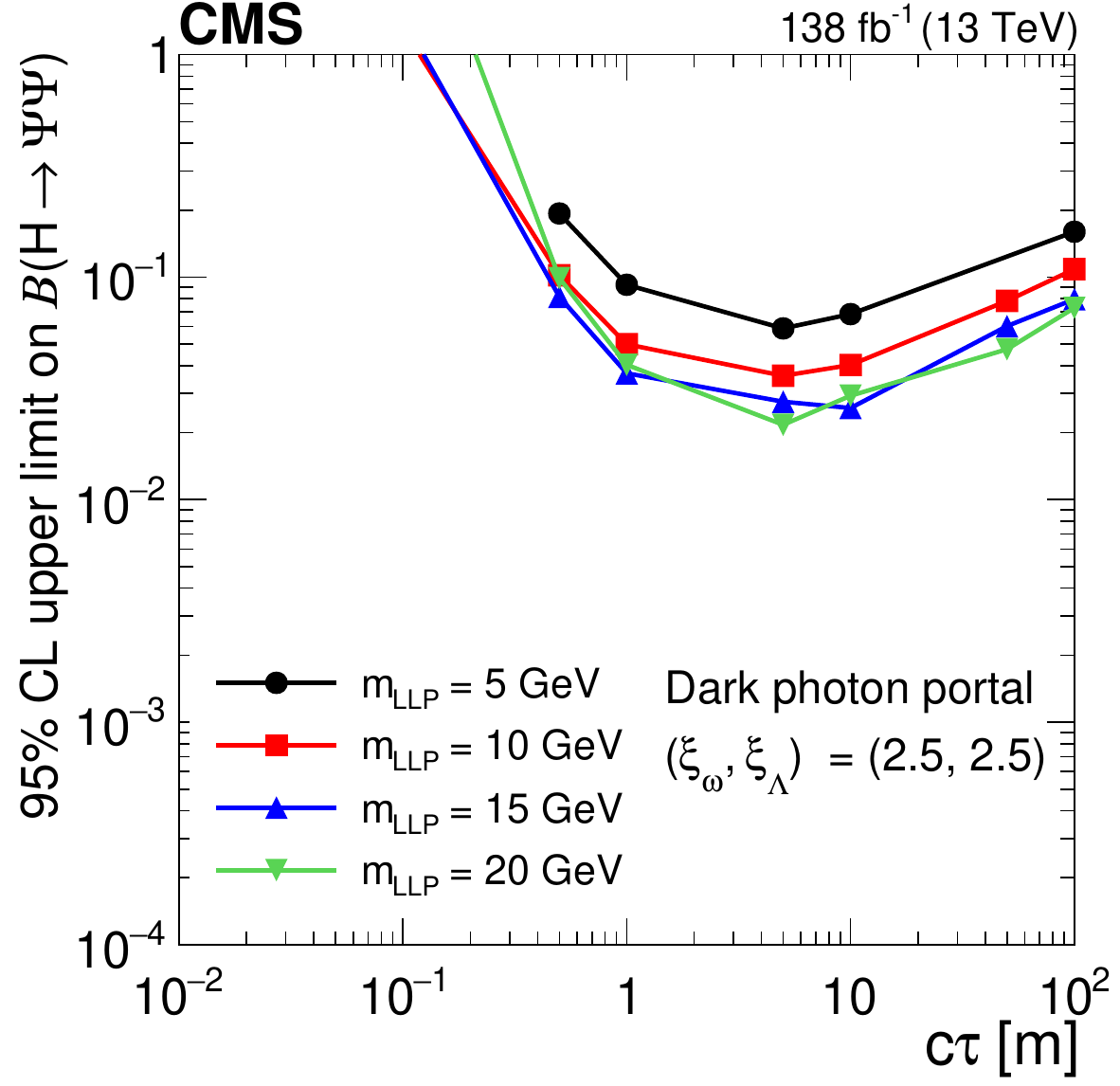}
  \caption{
    The 95\% \CL observed upper limits on the branching fraction $\brhtopsipsi$ as functions of $c\tau$ for the dark shower dark-photon portal, assuming $(\xi_{\omega}$, $\xi_{\Lambda})  = (1, 1)$ (upper left), $(\xi_{\omega}$, $\xi_{\Lambda})  = (2.5, 1.0)$ (upper right), and $(\xi_{\omega}$, $\xi_{\Lambda})  = (2.5, 2.5)$ (lower).
    The exclusion limits are shown for different mass hypotheses.
    The limits are calculated only at the proper decay lengths indicated by the markers and the lines connecting the markers are linear interpolations.
    The plot is restricted to physical $\brhtopsipsi$ values of below 1.
  }
  \label{fig:limit_darkshower_darkphoton}
\end{figure*}

Finally, for the dark-shower models the observed upper limits for the vector, gluon, photon, Higgs boson and dark-photon portals are shown in Fig.~\ref{fig:limit_darkshower_vector},~\ref{fig:limit_darkshower_gluon},~\ref{fig:limit_darkshower_photon},~\ref{fig:limit_darkshower_higgs},~and~\ref{fig:limit_darkshower_darkphoton}, respectively, as a function of $c\tau$ for a selection of LLP mass and $(\xi_{\omega}, \xi_{\Lambda})$ values.
This search sets the first LHC limits on models of dark showers produced via Higgs boson decay, and is sensitive to the branching fraction of the Higgs boson to dark quarks as low as $2 \times 10^{-3}$.

\clearpage
\section{Summary}
\label{sec:summary}

Data from proton-proton collisions at $\sqrt{s} = 13\TeV$ recorded by the CMS experiment in 2016--2018, corresponding to an integrated luminosity of 138\fbinv, have been used to conduct the first search that uses both the barrel and endcap CMS muon detectors to detect hadronic and electromagnetic showers from long-lived particle (LLP) decays.
Based on this unique detector signature, the search is largely model-independent, with sensitivity to a broad range of LLP decay modes and masses extending below the \GeVns scale.
With the excellent shielding provided by the inner CMS detector, the CMS magnet and its steel flux-return yoke, the background is suppressed to a low level and a search for both single and pairs of LLPs decays is possible.

No significant deviation from the standard model background is observed.
The most stringent LHC constraints to date are set on the branching fraction of the Higgs boson to LLPs with masses below 10\GeV and decaying to particles other than muons.
For higher LLP masses the search provides the most stringent branching fraction limits for specific proper decay lengths: 0.04--0.40\unit{m} and above 5\unit{m} for 15\GeV LLP; 0.3--0.9\unit{m} and above 3\unit{m} for 40\GeV LLP; and above 0.9\unit{m} for 55\GeV LLP.

Finally, the first LHC limits on models of dark showers produced via \PH decay are set, and constrain branching fractions of the \PH decay to dark quarks as low as $2 \times 10^{-3}$ at 95\% confidence level.

\begin{acknowledgments}
  We congratulate our colleagues in the CERN accelerator departments for the excellent performance of the LHC and thank the technical and administrative staffs at CERN and at other CMS institutes for their contributions to the success of the CMS effort. In addition, we gratefully acknowledge the computing centers and personnel of the Worldwide LHC Computing Grid and other centers for delivering so effectively the computing infrastructure essential to our analyses. Finally, we acknowledge the enduring support for the construction and operation of the LHC, the CMS detector, and the supporting computing infrastructure provided by the following funding agencies: SC (Armenia), BMBWF and FWF (Austria); FNRS and FWO (Belgium); CNPq, CAPES, FAPERJ, FAPERGS, and FAPESP (Brazil); MES and BNSF (Bulgaria); CERN; CAS, MoST, and NSFC (China); MINCIENCIAS (Colombia); MSES and CSF (Croatia); RIF (Cyprus); SENESCYT (Ecuador); ERC PRG, RVTT3 and TK202 (Estonia); Academy of Finland, MEC, and HIP (Finland); CEA and CNRS/IN2P3 (France); SRNSF (Georgia); BMBF, DFG, and HGF (Germany); GSRI (Greece); NKFIH (Hungary); DAE and DST (India); IPM (Iran); SFI (Ireland); INFN (Italy); MSIP and NRF (Republic of Korea); MES (Latvia); LAS (Lithuania); MOE and UM (Malaysia); BUAP, CINVESTAV, CONACYT, LNS, SEP, and UASLP-FAI (Mexico); MOS (Montenegro); MBIE (New Zealand); PAEC (Pakistan); MES and NSC (Poland); FCT (Portugal); MESTD (Serbia); MCIN/AEI and PCTI (Spain); MOSTR (Sri Lanka); Swiss Funding Agencies (Switzerland); MST (Taipei); MHESI and NSTDA (Thailand); TUBITAK and TENMAK (Turkey); NASU (Ukraine); STFC (United Kingdom); DOE and NSF (USA).
   
  \hyphenation{Rachada-pisek} Individuals have received support from the Marie-Curie program and the European Research Council and Horizon 2020 Grant, contract Nos.\ 675440, 724704, 752730, 758316, 765710, 824093, and COST Action CA16108 (European Union); the Leventis Foundation; the Alfred P.\ Sloan Foundation; the Alexander von Humboldt Foundation; the Science Committee, project no. 22rl-037 (Armenia); the Belgian Federal Science Policy Office; the Fonds pour la Formation \`a la Recherche dans l'Industrie et dans l'Agriculture (FRIA-Belgium); the Agentschap voor Innovatie door Wetenschap en Technologie (IWT-Belgium); the F.R.S.-FNRS and FWO (Belgium) under the ``Excellence of Science -- EOS" -- be.h project n.\ 30820817; the Beijing Municipal Science \& Technology Commission, No. Z191100007219010 and Fundamental Research Funds for the Central Universities (China); the Ministry of Education, Youth and Sports (MEYS) of the Czech Republic; the Shota Rustaveli National Science Foundation, grant FR-22-985 (Georgia); the Deutsche Forschungsgemeinschaft (DFG), under Germany's Excellence Strategy -- EXC 2121 ``Quantum Universe" -- 390833306, and under project number 400140256 - GRK2497; the Hellenic Foundation for Research and Innovation (HFRI), Project Number 2288 (Greece); the Hungarian Academy of Sciences, the New National Excellence Program - \'UNKP, the NKFIH research grants K 124845, K 124850, K 128713, K 128786, K 129058, K 131991, K 133046, K 138136, K 143460, K 143477, 2020-2.2.1-ED-2021-00181, and TKP2021-NKTA-64 (Hungary); the Council of Science and Industrial Research, India; ICSC -- National Research Center for High Performance Computing, Big Data and Quantum Computing, funded by the EU NexGeneration program (Italy); the Latvian Council of Science; the Ministry of Education and Science, project no. 2022/WK/14, and the National Science Center, contracts Opus 2021/41/B/ST2/01369 and 2021/43/B/ST2/01552 (Poland); the Funda\c{c}\~ao para a Ci\^encia e a Tecnologia, grant CEECIND/01334/2018 (Portugal); the National Priorities Research Program by Qatar National Research Fund;  MCIN/AEI/10.13039/501100011033, ERDF ``a way of making Europe", and the Programa Estatal de Fomento de la Investigaci{\'o}n Cient{\'i}fica y T{\'e}cnica de Excelencia Mar\'{\i}a de Maeztu, grant MDM-2017-0765 and Programa Severo Ochoa del Principado de Asturias (Spain); the Chulalongkorn Academic into Its 2nd Century Project Advancement Project, and the National Science, Research and Innovation Fund via the Program Management Unit for Human Resources \& Institutional Development, Research and Innovation, grant B37G660013 (Thailand); the Kavli Foundation; the Nvidia Corporation; the SuperMicro Corporation; the Welch Foundation, contract C-1845; and the Weston Havens Foundation (USA).  
\end{acknowledgments}

\bibliography{auto_generated}

\clearpage

\cleardoublepage \appendix\section{The CMS Collaboration \label{app:collab}}\begin{sloppypar}\hyphenpenalty=5000\widowpenalty=500\clubpenalty=5000
\cmsinstitute{Yerevan Physics Institute, Yerevan, Armenia}
{\tolerance=6000
A.~Hayrapetyan, A.~Tumasyan\cmsAuthorMark{1}\cmsorcid{0009-0000-0684-6742}
\par}
\cmsinstitute{Institut f\"{u}r Hochenergiephysik, Vienna, Austria}
{\tolerance=6000
W.~Adam\cmsorcid{0000-0001-9099-4341}, J.W.~Andrejkovic, T.~Bergauer\cmsorcid{0000-0002-5786-0293}, S.~Chatterjee\cmsorcid{0000-0003-2660-0349}, K.~Damanakis\cmsorcid{0000-0001-5389-2872}, M.~Dragicevic\cmsorcid{0000-0003-1967-6783}, P.S.~Hussain\cmsorcid{0000-0002-4825-5278}, M.~Jeitler\cmsAuthorMark{2}\cmsorcid{0000-0002-5141-9560}, N.~Krammer\cmsorcid{0000-0002-0548-0985}, A.~Li\cmsorcid{0000-0002-4547-116X}, D.~Liko\cmsorcid{0000-0002-3380-473X}, I.~Mikulec\cmsorcid{0000-0003-0385-2746}, J.~Schieck\cmsAuthorMark{2}\cmsorcid{0000-0002-1058-8093}, R.~Sch\"{o}fbeck\cmsorcid{0000-0002-2332-8784}, D.~Schwarz\cmsorcid{0000-0002-3821-7331}, M.~Sonawane\cmsorcid{0000-0003-0510-7010}, S.~Templ\cmsorcid{0000-0003-3137-5692}, W.~Waltenberger\cmsorcid{0000-0002-6215-7228}, C.-E.~Wulz\cmsAuthorMark{2}\cmsorcid{0000-0001-9226-5812}
% ...
\par}
\vskip\cmsinstskip
\dag:~Deceased\\
$^{1}$Also at Yerevan State University, Yerevan, Armenia\\
$^{2}$Also at TU Wien, Vienna, Austria\\
% ...
\end{sloppypar}
\end{document}